\documentclass[sn-mathphys,Numbered]{sn-jnl}

\usepackage{graphicx}%
\usepackage{multirow}%
\usepackage{amsmath,amssymb,amsfonts}%
\usepackage{amsthm}%
\usepackage{mathrsfs}%
\usepackage[title]{appendix}%
\usepackage{xcolor}%
\usepackage{textcomp}%
\usepackage{manyfoot}%
\usepackage{booktabs}%
\usepackage{algorithm}%
\usepackage{algorithmicx}%
\usepackage{algpseudocode}%
\usepackage{listings}%
\usepackage[width=.75\textwidth]{caption}
\DeclareFontFamily{OT1}{pzc}{}
\DeclareFontShape{OT1}{pzc}{m}{it}%
{<-> s * [1.15] pzcmi7t}{}
\DeclareMathAlphabet{\mathpzc}{OT1}{pzc}{m}{it}

\theoremstyle{thmstyleone}%
\theoremstyle{thmstyletwo}%

\theoremstyle{thmstylethree}%

\begin{document}

\title[Pseudoscalar Mesons and Emergent Mass]{$\,$\\[-6ex]\hspace*{\fill}{\normalsize{\sf\emph{Preprint no}.\
NJU-INP 084/24}}\\[1ex]Pseudoscalar Mesons and Emergent Mass}


\author*[1]{\fnm{Kh\'epani} \sur{Raya}}\email{khepani.raya@dci.uhu.es}

\author[1,2]{\fnm{Adnan} \sur{Bashir}}\email{adnan.bashir@umich.mx}

\author[3]{\fnm{Daniele} \sur{Binosi}}\email{binosi@ectstar.eu}

\author[4,5]{\fnm{Craig D.} \sur{Roberts}}\email{cdroberts@nju.edu.cn}

\author[1]{\fnm{Jos\'e} \sur{Rodr\'iguez-Quintero}}\email{jose.rodriguez@dfaie.uhu.es}


\affil[1]{\orgdiv{Dpto. Ciencias Integradas}, \orgname{Centro de Estudios Avanzados en Fis., Mat. y Comp., Fac. Ciencias Experimentales, Universidad de Huelva}, \orgaddress{\city{Huelva}, \postcode{21071}, \country{Spain}}}

\affil[2]{\orgdiv{Instituto de F\'isica y Matem\'aticas}, \orgname{Universidad
Michoacana de San Nicol\'as de Hidalgo}, \orgaddress{\city{Morelia, Michoac\'an}, \postcode{58040}, \country{M\'exico}}}

\affil[3]{\orgdiv{European Centre for Theoretical Studies in Nuclear Physics and Related Areas}, \orgaddress{Villa Tambosi, Strada delle Tabarelle 286, \postcode{I-38123}, \city{Villazzano (TN)}, \country{Italy}}}

\affil[4]{\orgdiv{School of Physics}, \orgaddress{Nanjing University, \city{Jiangsu}, \postcode{210093},  \country{China}}}

\affil[5]{\orgdiv{Institute for Nonperturbative Physics}, \orgaddress{Nanjing University, \city{Jiangsu}, \postcode{210093},  \country{China}}}


\abstract{Despite its role in the continuing evolution of the Universe, only a small fraction of the mass of visible material can be attributed to the Higgs boson alone. 
The overwhelmingly dominant share may/should arise from the strong interactions that act in the heart of nuclear matter; namely, those described by quantum chromodynamics. 
This contribution describes how studying and explaining the attributes of pseudoscalar mesons can open an insightful window onto understanding the origin of mass in the Standard Model and how these insights inform our knowledge of hadron structure. 
The survey ranges over  distribution amplitudes and functions, electromagnetic and gravitational form factors, light-front wave functions, and generalized parton distributions. Advances made using continuum Schwinger function methods and their relevance for experimental efforts are highlighted.}





\maketitle

\section{Introduction}\label{sec1}
When studying the many facets of the Universe, one must be aware of the length/energy scales involved.
For instance, the laws that seem to govern everyday phenomena might not be applicable at astronomical or subatomic scales.
This apparent shift in perspective is probably only a reflection of our current limitations in comprehending the Universe.
Despite the challenges, we have been able to separate the mass-energy budget of the universe into three sources: dark energy, dark matter and visible matter \cite{Huterer:2017buf}.
While dark energy and dark matter contribute the largest portion ($71\%$ and $24\%$, respectively) \cite{Seife:2003abc}, their effects on our daily life and accurate description are practically unimportant.
Visible matter, on the other hand, constitutes only $5$\% of the whole; yet, it is the source of almost everything that is tangible.
In terms of fundamental particles and their interactions, this small percentage is described by the so-called Standard Model (SM) of particle physics \cite{Glashow:1961tr, Weinberg:1967tq, Salam:1968rm, Politzer:1973fx, Gross:1973id}.
The SM is remarkably successful in explaining observations; so much so that there are few empirical indications that it needs improvement -- see, \emph{e.g}., Refs.\,\cite{Aoyama:2020ynm, Crivellin:2023zui}.
Thus, before attempting to go \emph{beyond}, it is worth resolving outstanding issues within the SM.

One such challenge is to explain the origin of visible mass.
In this connection, the Higgs boson typically comes to mind \cite{Englert:2014zpa, Higgs:2014aqa}.
However, although it is vital to the character of our Universe, the Higgs only contributes about $1$-$2\%$ of the visible mass.
The remaining $\gtrsim 98$\% owes to strong interactions in the SM, \emph{i.e}., quantum chromodynamics (QCD) \cite{Marciano:1979wa, Marciano:1977su}, through a dynamical source that is now often dubbed emergent hadron mass (EHM).
These same strong interactions are also responsible for the formation of protons and neutrons (nucleons) and, from them, the atomic nuclei that constitute almost the entirety of visible material.

QCD is the Poincar\'e-invariant quantum non-Abelian gauge field theory that describes the color-charge interactions between gluon and quark partons.
However, these partons cannot be studied in isolation.
Insofar as QCD is concerned, only compact (fm-size) color-neutral objects can be detected.
One says that the the gluons and quarks are confined, but the meaning of this statement is vigorously debated \cite[Sec.\,5]{Ding:2022ows}: it can be argued that confinement and EHM are two sides of the same coin.  
Our perspective stresses a dynamical picture of confinement, wherein nonperturbative interactions drive changes in the analytic structure of colored Schwinger functions that ensure the absence of color-carrying objects from the Hilbert space of observable states \cite[Sec.\,5]{Ding:2022ows}.  
Such changes are evident in both continuum and lattice analyses -- see, e.g., Refs.\,\cite{Gao:2017uox, Binosi:2019ecz, Falcao:2020vyr, Boito:2022rad}.

Being the most abundant and stable hadron in the Universe, the proton has played a major role in the scientific endeavor.
It debuted a little over a century ago, in Rutherford's experiments that exposed the substructure of the atom \cite{RutherfordI, RutherfordII, RutherfordIII, RutherfordIV}; and later in the 1950s, when it was revealed that the proton also has a finite size \cite{Hofstadter:1956qs}.
Subsequently, deep inelastic scattering measurements on the proton provided empirical evidence for the existence of quarks \cite{Breidenbach:1969kd}.

\begin{figure}[t]%
\centering
\begin{tabular}{c}
$\rule[0cm]{1.5cm}{0cm} p \rule[0cm]{1.1cm}{0cm} \rho \rule[0cm]{1.05cm}{0cm} \pi \rule[0cm]{0.9cm}{0cm} K \rule[0cm]{0.9cm}{0cm} \eta_c \rule[0cm]{0.9cm}{0cm} \eta_b$
\\
\includegraphics[width=0.70\textwidth]{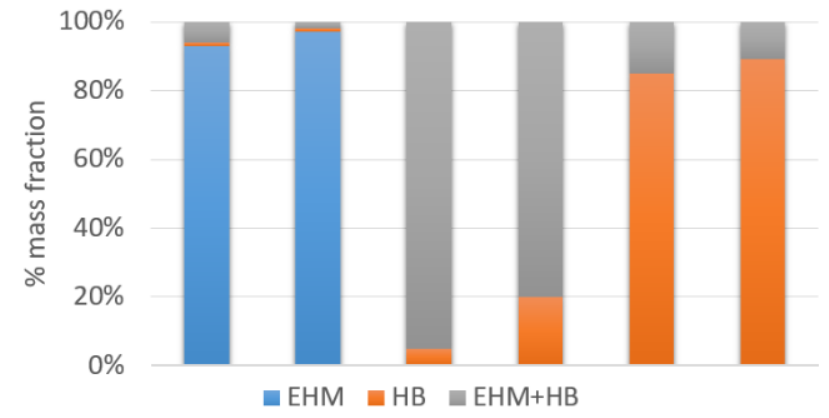}
\end{tabular}
\caption{Mass budgets of the proton,  $\rho$ meson and different pseudoscalars. Three types of contributions play a role: EHM, HB and the EHM + HB interference. Clearly, proton and $\rho$ are dominated by EHM (above $90$\%), heavy quarkonia by HB ($85-90$\%), and finally, pions and kaons by the interplay between EHM and HB mass generation ($95$\% and $80$\%, respectively).
These budgets are drawn at a resolving scale of $\zeta=2\,$GeV, but the images are little changed if one chooses instead to use renormalization-point-independent quantities.
\label{fig:MassBuds}}
\end{figure}

In addressing EHM and confinement, the proton has also been prominent.
For instance, it is argued that the proton mass, $m_p \approx 1 \,\text{GeV}$, about 2000-times that of the electron, is a natural \emph{mass} scale for visible matter \cite{Roberts:2021xnz, Roberts:2022rxm}.
To this value, by itself, the Higgs boson (HB) only contributes $\approx 1$\%, a fact highlighted in Fig.\,\ref{fig:MassBuds}.
The largest fraction, some $94$\%, is a definitive expression of EHM; and the remaining $\sim 5$\% results from constructive EHM + HB interference \cite{Roberts:2016vyn}.

The proton size, $r_p \approx 0.8\,\text{fm}$ \cite{Gao:2021sml, Cui:2022fyr}, is also crucial.
It defines the characteristic compactness scale for QCD's colour-neutral bound states and thereby represents a natural \emph{length} for confinement.
Understanding the origin of proton mass and size, and how different SM mechanisms affect its properties, are primary goals of modern science \cite{Belle-II:2018jsg, Aguilar:2019teb, Yuan:2019zfo, Brodsky:2020vco, Chen:2020ijn, Anderle:2021wcy, Arrington:2021biu, AbdulKhalek:2021gbh, Quintans:2022utc, Amoroso:2022eow, Carman:2023zke}.

In contrast to protons, $\pi$- and $K$-mesons, pions and kaons, are light, despite also being hadron bound states, and somewhat more compact \cite{Cui:2022fyr}.
At some level of approximation, they carry the strong force between nucleons \cite{Yukawa:1935xg}; and the existence of our known Universe requires that they be light compared to $m_p$.
At first glance, pions and kaons are the simplest QCD bound states; in fact, they are often drawn as two-body systems, \textit{viz}.\ a quark and antiquark, somehow held together.
However, this picture is simplistic \cite{Horn:2016rip, Roberts:2021nhw}.

In reality, pions and kaons are Nature's most fundamental Nambu-Goldstone (NG) bosons, which emerge as a consequence of dynamical chiral symmetry breaking (DCSB) in the SM \cite{Maris:1997hd, Maris:1997tm}.
DCSB is a corollary of EHM.
In the absence of HB couplings into QCD, \textit{i.e}., in the chiral limit, $\pi$- and $K$-mesons would be massless and structurally indistinguishable.

Restoring HB couplings, pions and kaons acquire their unusually low masses: $m_\pi \approx 0.15\,m_p$ and $m_K \approx 0.53\,m_p$; and exhibit structural dissimilarities.
As displayed in Fig.\,\ref{fig:MassBuds}, alone, HB couplings into QCD generate only a small fraction of the $\pi$ and $K$ masses.
In these cases, however, the bulk of the bound-state masses are generated by constructive EHM + HB interference.
This contrasts starkly with the $\rho$-meson: constituted from the same valence degrees-of-freedom as the pion, $m_\rho \approx 0.82 \, m_p \approx 5.6 \, m_\pi$ and the associated mass budget is qualitatively indistinguishable from that of the proton -- Fig.\,\ref{fig:MassBuds}.

It should now be clear that in attempting to discover the origin of visible mass, a focus solely on proton structure is inadequate.
It is imperative to broaden the goal and provide a simultaneous and unifying understanding of Nature's most fundamental Nambu-Goldstone bosons, \textit{viz}.\ pions and kaons.
Furthermore, every pseudoscalar meson would be a Nambu-Goldstone boson if it were not for HB couplings into QCD.
This means that exploring and explaining the properties of the entire collection of pseudoscalar mesons offers a unique opportunity for elucidating the observable expressions and impacts of Nature's two known mass generating mechanisms and the interference between them -- compare, \textit{e.g}., Figs.\,1 and 18 in Ref.\,\cite{Ding:2022ows}.
No wonder, then, that with high-energy, high-luminosity facilities becoming a reality, much experimental and theoretical attention is shifting toward the study of pseudoscalar mesons \cite{Belle-II:2018jsg, Aguilar:2019teb, Yuan:2019zfo, Brodsky:2020vco, Chen:2020ijn, Anderle:2021wcy, Arrington:2021biu, AbdulKhalek:2021gbh, Quintans:2022utc}.

Hereafter, we sketch possibilities offered by studies of pseudoscalar mesons to provide insights into the emergence of hadron mass and structure.
The vehicle for this discussion is provided by continuum Schwinger function methods (CSMs), an approach that has delivered significant progress, especially in the past decade
\cite{Roberts:2015lja, Eichmann:2016yit, Burkert:2017djo, Fischer:2018sdj, Qin:2020rad, Roberts:2020hiw, Roberts:2021nhw, Binosi:2022djx, Roberts:2022rxm, Ding:2022ows, Ferreira:2023fva, Deur:2023dzc}.
The manuscript is organized as follows:
Sec.\,\ref{Sec:EHM} discusses emergent phenomena in QCD and their study using CSMs;
pseudoscalar meson distribution amplitudes and functions, and their connection with EHM are addressed in Sec.\,\ref{Sec:DAsDFs};
a complementary perspective, obtained via electromagnetic and gravitational form factors, is provided in Sec.\,\ref{Sec:FFs};
Sec.\,\ref{Sec:GPDs} follows, using light-front wave functions (LFWFs) and generalized parton distributions (GPDs) to address some questions concerning the three-dimensional structure of pseudoscalar mesons;
and finally, a perspective is drawn in Sec.\,\ref{Sec:Conclusions}.

\section{Emergent phenomena in QCD}
\label{Sec:EHM}

Recall the QCD Lagrangian density: 
\begin{eqnarray}
    \mathcal{L}_{\text{QCD}}&=& \bar{q}_f[\gamma\cdot \partial + i g\frac{1}{2} \lambda^a \gamma \cdot A^a + m_f] q_f + \frac{1}{4}G_{\mu\nu}^a G_{\mu\nu}^a \,,\\
    G_{\mu\nu}^a &=& \partial_\mu A_\nu^a+\partial_\nu A_\mu^a - \underline{g f^{abc} A_\mu^b A_\nu^c}\,,
\end{eqnarray}
here $q_f$ denote the $f$-flavored quark fields, with current quark masses $m_f$; $A_\mu^a$ are the gluon fields and $\lambda^a$ the generators of SU(3) in the fundamental representation; and $g$ is the \emph{unique} QCD coupling \cite{Deur:2023dzc}.
Plainly, $\mathcal{L}_{\text{QCD}}$ looks very similar to the Lagrangian density of quantum electrodynamics (QED), except for the underlined piece.
This special term gives QCD its non-Abelian character and underlies asymptotic freedom \cite{Politzer:2005kc, Gross:2005kv, Wilczek:2005az}.
It also plays a key role in all non-perturbative facets of QCD, including confinement and EHM.

\begin{figure}[t]%
\centering
\begin{tabular}{lr}
\includegraphics[width=0.47\textwidth]{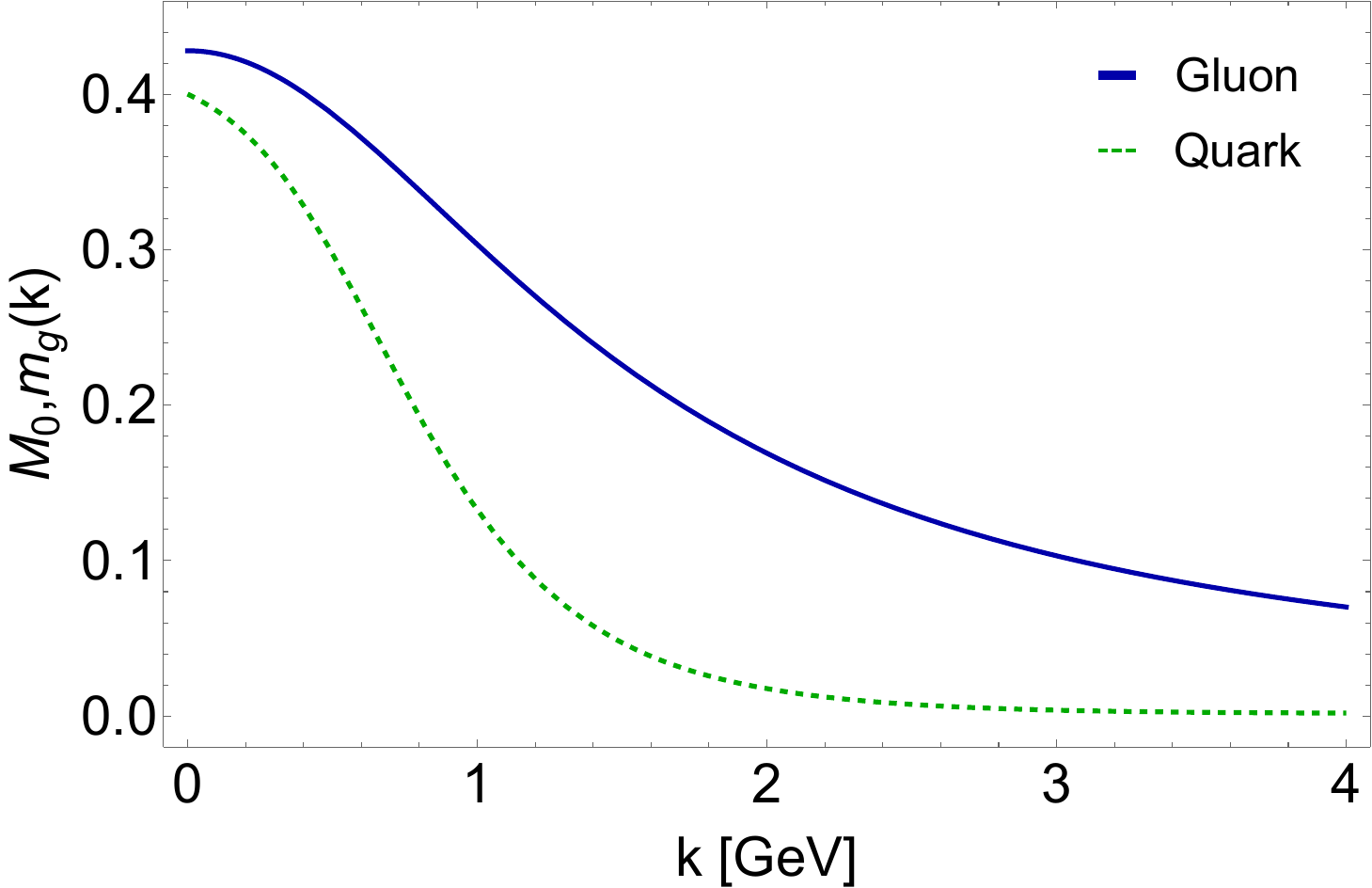} &
\includegraphics[width=0.48\textwidth]{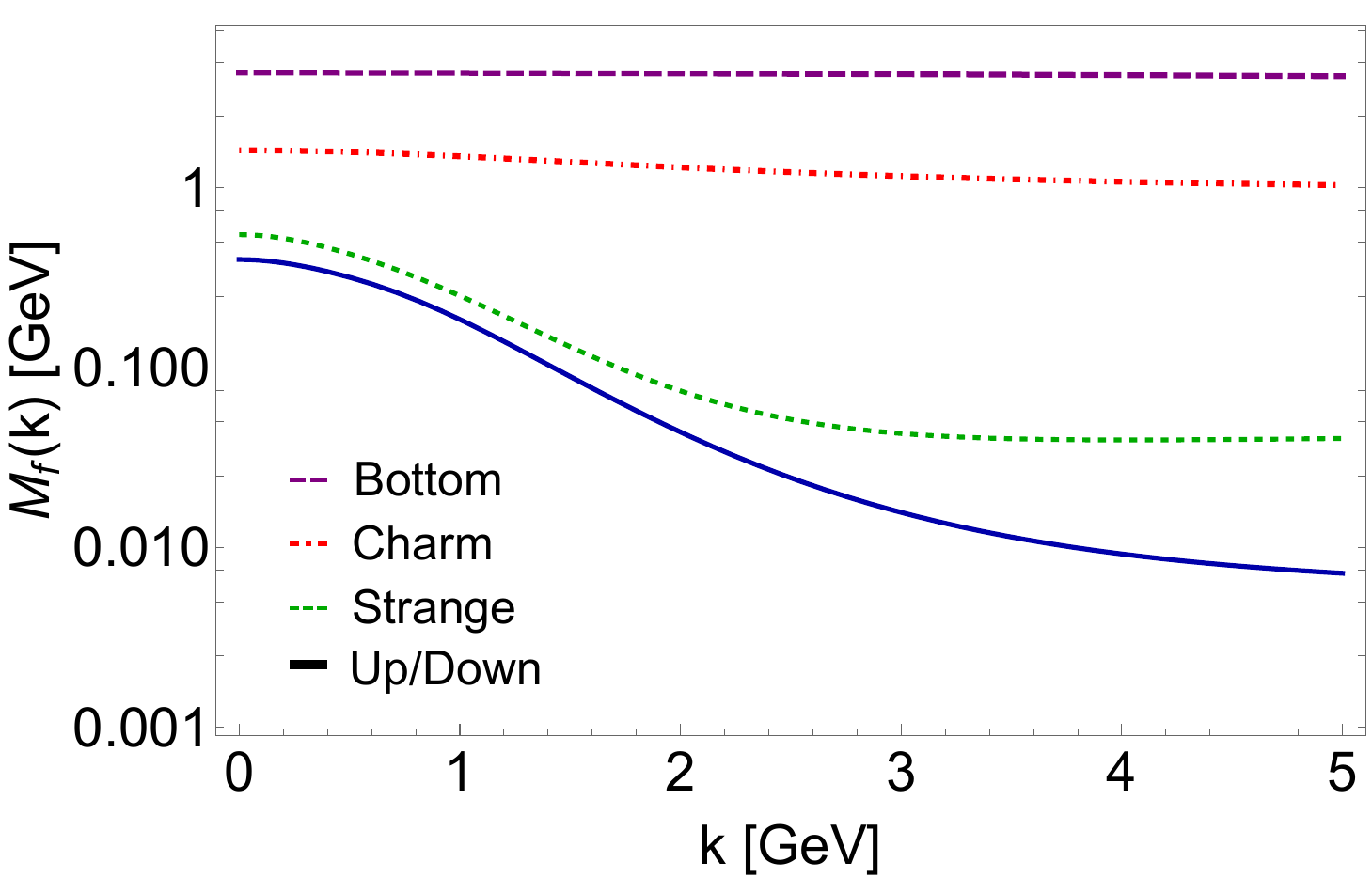}
\end{tabular}
\caption{Running masses. [left] Gluon running mass $m_g(k)$ (solid blue) and chiral limit  quark mass function $M_0(k)$ (dashed green). [right] Mass functions for different quark flavors. \label{fig:MassFuncs}}
\end{figure}

Decades of analyses have crystallized in a now widely accepted picture of dynamical gluon mass generation in QCD, which owes to a Schwinger mechanism \cite{Binosi:2022djx, Ferreira:2023fva, Deur:2023dzc}.
Amongst other important implications, this translates into the physical image of a gluon developing a running mass, $m_g(k)$, which is large in the infrared and decreases monotonically toward its perturbative (massless) limit as $k^2$ increases \cite{Cui:2019dwv}.
Likewise, in the matter sector, the nonperturbative phenomenon of DCSB generates a dressed light-quark mass that is equally large in the infrared, whether there is a Higgs mechanism or not \cite{Binosi:2016wcx, Sultan:2018tet}.
Thus, as illustrated in Fig.\,\ref{fig:MassFuncs}, dynamical mass generation is manifest in both the gauge and matter sectors.
In Fig.\,\ref{fig:MassFuncs}, the left panel highlights similarities and differences between $m_g(k)$ and the chiral limit ($m_f\equiv 0$) quark mass function; notably, both saturate in the infrared at a value $\sim 400$ MeV.
The right panel displays the influence of DCSB on the different quark flavors: although its effects are visually more conspicuous in the light sector, the magnitude of the vertical displacement above the current-mass at infrared momenta is roughly the same in each case.

\begin{figure}[t]%
\centering
\includegraphics[width=0.6\textwidth]{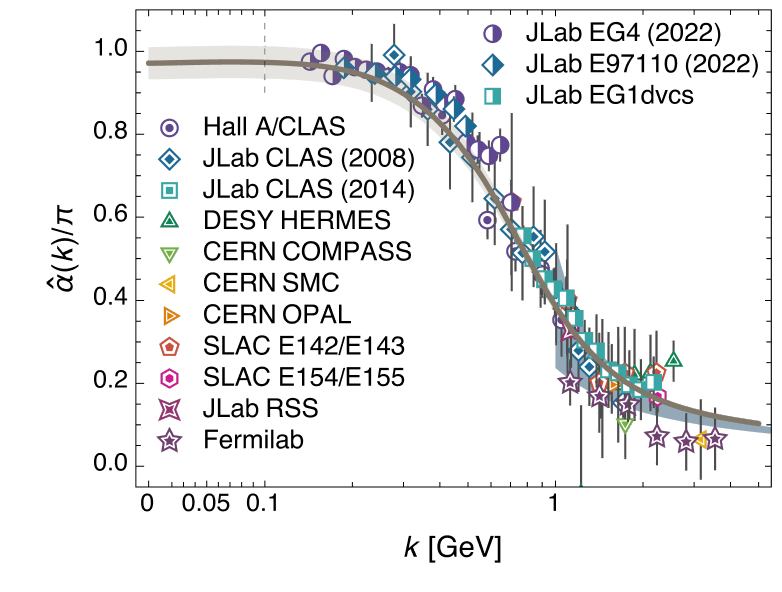}
\caption{Process-independent charge \cite{Cui:2019dwv} compared with empirical extractions \cite{Deur:2022msf, Deur:2023dzc} of the process-dependent charge defined via the Bjorken sum rule \cite{Bjorken:1969mm}.
\label{fig:alphaPI}}
\end{figure}

The generation of mass in the gauge and matter sectors is inextricably linked with the behavior of the QCD running coupling \cite{Deur:2023dzc} and recent advances in experiment and theory are leading to a solid understanding of this fundamental quantity.
Indeed, exploiting continuum advances in analyses of QCD's gauge sector, a unique QCD analogue of the Gell-Mann--Low charge \cite{GellMann:1954fq}, used widely in QED, is now available \cite{Binosi:2016nme}.
Moreover, using the best existing continuum and lattice results for low-order gauge-sector Schwinger functions, Ref.\,\cite{Cui:2019dwv} delivered a parameter-free prediction for this process-independent (PI) charge, $\hat{\alpha}(k^2)$ -- see Fig.\,\ref{fig:alphaPI}.
Even at far-infrared momenta, the result possesses an uncertainty that is just 4\%:
$\hat{\alpha}(k^2=0)=0.97(4)\pi$.
Such accuracy is quite remarkable in an \textit{ab initio} calculation of an essentially nonperturbative quantity in QCD.

The characteristics and virtues of $\hat{\alpha}(k^2)$ are canvassed elsewhere -- see, \textit{e.g}., Refs.\,\cite{Ding:2022ows, Deur:2023dzc}, and they include: the absence of a Landau pole; a match with the process-dependent charge built using the Bjorken sum rule \cite{Bjorken:1969mm, Deur:2022msf, Deur:2023dzc}; and an unambiguous infrared extension (completion) of the perturbative running coupling.

The pointwise form of $\hat{\alpha}(k^2)$, drawn in Fig.\,\ref{fig:alphaPI}, is accurately parametrized by the following expression \cite{Cui:2020dlm}:
\begin{equation}
    \label{eq:PIcharge}
    \hat{\alpha}(k) = \frac{\gamma_m \pi}{\ln\left[\mathcal{K}^2(k^2)/\Lambda_{\text{QCD}}^2 \right]}\,,\quad \mathcal{K}^2(y)=\frac{a_0^2+a_1 y+y^2}{b_0 + y}\,,
\end{equation}
where (in GeV$^2$): $a_0=0.104(1)$, $a_1=0.0975$, $b_0=0.121(1)$; and $\gamma_m=12/(33-2n_f)$, $n_f=4$, $\Lambda_{\text{QCD}}=0.234$ GeV.
Clearly and deliberately, Eq.\,\eqref{eq:PIcharge} has the appearance of the perturbative QCD running coupling at one-loop order in the $\overline{\text{MS}}$ scheme; but, in this case, $\mathcal{K}^2(k^2)$ prevents the appearance of the so-called Landau pole at $k^2=\Lambda_{\text{QCD}}^2$.
Instead, the momentum scale $\zeta_{\cal H}^2:=\mathcal{K}^2(k^2=\Lambda_{\text{QCD}}^2)\approx (0.331\,\text{GeV})^2$ defines a boundary between soft and hard physics \cite{Cui:2019dwv}, below which the gluon mass function is large enough to force these degrees-of-freedom to decouple from interactions.
As a consequence, the running practically ceases in the far-infrared, so that QCD is once again, effectively, a conformal theory.
An appreciation of the character and role of $\zeta_{\cal H}$ has led to it being identified as the ``hadron scale'', \textit{i.e}., the resolving scale at which valence dressed-parton (quasiparticle) degrees-of-freedom should be used to state and solve hadron bound state problems -- see, \textit{e.g}., Refs.\,\cite{Cui:2020dlm, Cui:2020tdf, Han:2020vjp, Xie:2021ypc, Raya:2021zrz, Cui:2021mom, Cui:2022bxn, Chang:2022jri, Lu:2022cjx, dePaula:2022pcb, Xu:2023bwv, Yin:2023dbw, Xing:2023wuk, Xing:2023pms, Lu:2023yna, Yu:2024qsd}.

Each of the unique features described above contributes to the formation of observable color-neutral bound states and determination of their properties.
In this connection, as already noted, charts of pseudoscalar meson properties provide clear windows onto such emergent phenomena in QCD.
We now proceed to illustrate this fact using CSMs, implemented via QCD's Dyson-Schwinger equations (DSEs)
\cite{Roberts:2015lja, Eichmann:2016yit, Burkert:2017djo, Fischer:2018sdj, Qin:2020rad, Roberts:2020hiw, Roberts:2021nhw, Binosi:2022djx, Roberts:2022rxm, Ding:2022ows, Ferreira:2023fva, Deur:2023dzc}.
The DSEs may be described as QCD's quantum equations of motion and have been instrumental in exposing emergent phenomena in the strong interaction.

Regarding mesons, all structural information is encoded in the Poincar\'e-covariant Bethe-Salpeter wave function (BSWF):
\begin{equation}
    \chi_{\textbf{P}}(k;P) = S(k+P/2)\Gamma_{\textbf{P}}(k;P)S(k-P/2)\,.
\end{equation}
Here, $S(p)$ refers to the dressed quark propagator and $\Gamma_{\textbf{P}}$ to the meson Bethe-Salpeter amplitude (flavor indices omitted); $P$ is the total momentum of the meson $\textbf{P}$, and $k$ is the relative momentum between the valence-quarks.
(In a Poincar\'e-invariant treatment, the choice of $k$ is practitioner dependent and no observable can depend on its definition.)

The quark propagator can be expressed as follows:
\begin{equation}
 S(p) = Z(p^2)/[i\gamma \cdot p+M(p^2)]\,,
\end{equation}
where $Z(p^2)$ is the quark wave function renormalization function and $M(p^2)$ is the quark (running) mass function.
(Herein, we largely omit a discussion of renormalization.
It is nevertheless worth noting that all renormalization point dependence of the quark propagator is contained in that of $Z(p^2)$.
$M(p^2)$ is renormalization-point independent.)
The mass functions associated with a physical range of quark current-masses are drawn in Fig.\,\ref{fig:MassFuncs}.

The structure of the Bethe-Salpeter amplitude (BSA) depends on the quantum numbers of the meson. For a pseudoscalar meson:
\begin{eqnarray}
\label{eq:BSA}
    \Gamma_{\textbf{P}}(q;P) &=& \gamma_5[i \mathbb{E}_{\textbf{P}}(q;P)+\gamma \cdot P \mathbb{F}_{\textbf{P}}(q;P)+\gamma \cdot q \mathbb{G}_{\textbf{P}}(q;P) + q_\mu \sigma_{\mu \nu} P_\nu \mathbb{H}_{\textbf{P}}(q;P)]\;.
\end{eqnarray}
The BSWF has an analogous decomposition, in terms of functions we will write as $\mathbb{E}_\chi$, $\mathbb{F}_\chi$, $\mathbb{G}_\chi$, $\mathbb{H}_\chi$.
In the meson rest-frame, $\mathbb{G}_\chi$, $\mathbb{H}_\chi$ correspond to $L=1$ components. At pion and kaon masses, one measure of their strength indicates that $L=1$ components provide $\approx 20$\% of the canonical normalization \cite{Bhagwat:2006xi}. 
Under Poinar\'e transformations, there is mixing between the terms in the wave function analogue of Eq.\,\eqref{eq:BSA}.  Consequently, pseudoscalar mesons contain nonzero quark orbital angular momentum in any reference frame.

The quark propagator and meson BSA are obtained from their corresponding Dyson-Schwinger and Bethe-Salpeter equations, namely:
\begin{eqnarray}\label{eq:quarkPropQCD}
    S^{-1}(p)&=&Z_2 (i \gamma\cdot p + m^{\rm bm})
    + \int_{dq} [\mathcal{K}^{(1)}(q,p)] S_f(q) \;, \\
    \label{eq:BSEGen}
    \Gamma_{\textbf{P}}(p;P)&=&\int_{dq} [\mathcal{K}^{(2)}(q,p;P)]\chi_{\textbf{P}}(q;P)\;,
\end{eqnarray}
where
$\int_{dq} := \int \frac{d^4q}{(2\pi)^4}$ stands  for  a  Poincar\'e  invariant  regularized  integration, $Z_2$ is a renormalization constant, and $m^{\rm bm}$ is the quark current-mass.

The 1-body kernel, $\mathcal{K}^{(1)}$, is connected to higher-order Schwinger functions (such as the quark-gluon vertex) \emph{ad infinitum}, as well as with the two-body kernel, $\mathcal{K}^{(2)}$.
Thus, these DSEs form an infinite system of coupled integral equations \cite{Roberts:1994dr}.
Consequently, a sound treatment of QCD bound states demands a systematic and symmetry-preserving truncation scheme \cite{Munczek:1994zz, Bender:1996bb, Roberts:1994hh}.

Any such truncation will ensure compliance with the Goldstone theorem \cite{Nambu:1960tm, Goldstone:1961eq}, which is most fundamentally expressed as an equivalence between the one-body quark propagator problem and the two-body meson problem in the pseudoscalar channel  \cite{Maris:1997hd, Maris:1997tm}:
\begin{equation}\label{eq:GT}
    f_\textbf{P}^0\mathbb{E}_\textbf{P}^0(q;P =0) = M_0(q^2)/Z_0(q^2)\;,
\end{equation}
where $f_\textbf{P}^0$ is the pseudoscalar meson decay constant and the super/subscript ``$0$'' indicates that the quantities are obtained in the chiral limit.
In addition to exhibiting the equivalence between the one- and two-body problems,  Eq.\,\eqref{eq:GT} states that Nambu-Goldstone bosons exist if, and only if, DCSB is realized.
The leading-order symmetry-preserving truncation, dubbed \emph{rainbow-ladder} (RL), is sufficient to guarantee the relevant symmetry principles \cite{Munczek:1994zz, Bender:1996bb}, and thus describe the properties of ground-state pseudoscalar mesons.
Ways to systematically improve upon this leading-order kernel are known; so, there are increasingly more sophisticated truncations -- see, \textit{e.g}., \cite{Chang:2009zb, Chang:2011ei, Qin:2014vya, Williams:2015cvx, Binosi:2016rxz, Qin:2020jig, Xu:2022kng}.

To close this section, it is worth stressing that the development of CSMs has reached a point from which not only a significant part of the hadron spectrum can be reproduced \cite{Eichmann:2016yit, Qin:2018dqp, Xu:2018cor, Chang:2019eob, Qin:2019hgk, Xu:2022kng}, but also, as we will see, numerous quantities that characterize hadron structural properties.
For pseudoscalar mesons, this includes:
distribution amplitudes and distribution functions \cite{Chang:2013pq, Shi:2014uwa, Binosi:2018rht, Chang:2014lva, Ding:2019lwe, Ding:2019qlr, Cui:2020tdf, Cui:2020dlm, Chang:2022jri, Lu:2022cjx, Ding:2018xwy, Chen:2018rwz}; electromagnetic and gravitational form factors \cite{Chang:2013nia, Gao:2017mmp, Raya:2015gva, Raya:2016yuj, Ding:2018xwy, Raya:2019dnh, Eichmann:2019bqf, Chen:2018rwz, Miramontes:2021exi, Xu:2023izo};
as well as light-front wave functions and generalized parton distributions \cite{Raya:2021zrz, Shi:2021nvg, Kou:2023ady, Mezrag:2023nkp}.

\section{Distribution amplitudes and functions}
\label{Sec:DAsDFs}

\subsection{Parton distribution amplitudes}
\label{sec:PDAs}

Parton distribution amplitudes (DAs) characterize the probability that a nominated parton within a hadron carries a light-front fraction $x$ of that hadron's total momentum, $P$.
Each is a particular one-dimensional projection of the hadron's light-front wave function (LFWF), which itself is the closest analogue in a quantum field theory to the wave function familiar from  quantum mechanics \cite{Brodsky:1989pv}.  Such DAs play a crucial role in the description of hard exclusive processes \cite{Lepage:1979zb, Efremov:1979qk, Lepage:1980fj}.

For a pseudoscalar meson $\textbf{P}=q\bar{h}$, where $q$, $h$ specify valence-quark flavor, a nonperturbative extension of the leading-twist DA, $\varphi^q_{\textbf{P}}(x;\zeta)$, may be obtained by projecting the meson's Poincar\'e covariant BSWF onto the light-front \cite{Chang:2013pq}:
\begin{eqnarray}
\label{eq:PDA}
f_{\textbf{P}}\varphi_{\textbf{P}}^q(x;\zeta)= Z_2(\zeta,\Lambda) \textrm{tr}_{\textrm{CD}} \int_{dk}^\Lambda \delta_{n,P}^x(k) \gamma_5 \gamma \cdot n (k;P)
S_q(k) \Gamma_{\textbf{P}}(k;P)S_h(k-P)\,,
\end{eqnarray}
where
$\delta^x_{n,P}(k) = \delta( n \cdot k  - x n \cdot P)$,
with $n$ a light-like four vector satisfying $n^2=0$ and $n \cdot P = -m_{\textbf{P}}$ in the meson's rest frame;
$\textrm{tr}_{\textrm{CD}}$ indicates a trace taken over color and spinor indices;
and we have explicitly indicated the regularization scale, $\Lambda$, and $\zeta$, the point whereat the quark propagators and meson Bethe-Salpeter amplitude are renormalized.
As written in Eq.\,\eqref{eq:PDA}, $\varphi_{\textbf{P}}^u$ is dimensionless and normalized to unity, \emph{i.e.}, $\int_0^1 dx \varphi_{\textbf{P}}^q(x;\zeta) = 1$.

Notably, in a framework that preserves the multiplicative renormalizability of QCD, both $f_{\textbf{P}}$, $\varphi_{\textbf{P}}^q(x;\zeta)$ are independent of the renormalization point.  This means that $\zeta$ need not be identified with the ERBL evolution scale \cite{Lepage:1979zb, Efremov:1979qk, Lepage:1980fj}.  That scale, $\zeta_{\rm ev}$, is a somewhat amorphous measure of the degree of collinearity of the partons in the relevant Fock components of the LFWF, \emph{i.e}., an upper bound on the $k_\perp^2$ values of partons contributing to the subject DA.  

The kinematics in Eq.\,\eqref{eq:PDA} are set such that the expression refers to the $q$ quark inside the meson $\textbf{P}$ ($q$-in-$\textbf{P}$).
The corresponding $\bar{h}$ DA is readily obtained as
\begin{equation}
\label{eq:PDAantiquark}
    \varphi_{\textbf{P}}^{\bar{h}}(x;\zeta)=\varphi_{\textbf{P}}^q(1-x;\zeta)\,.
\end{equation}
This guarantees momentum conservation and entails $\varphi_{\textbf{P}}^{\bar{h}}(x;\zeta) = \varphi_{\textbf{P}}^{q}(x;\zeta)$ in the isospin symmetric limit, \textit{viz}.\ $m_q = m_h$.
We assume the limit $m_u=m_d$ hereafter.

The Mellin moments of the DA are defined as usual:
\begin{eqnarray}
    \langle x^m \rangle_\zeta^{\varphi_{\textbf{P}}^q} &=& \int_0^1 dx \, x^m\, \varphi_{\textbf{P}}^q(x;\zeta)\;,
\end{eqnarray}
and, following from Eq.\,\eqref{eq:PDA}, these can be computed using
\begin{equation}
\label{eq:momsPDA1}
    f_{\textbf{P}}(n\cdot P)^{m+1}\phi_{\textbf{P}}^q(x;\zeta) = Z_2 \textrm{tr}_{\textrm{CD}} \int_{dk}^\Lambda (n \cdot k)^m \gamma_5 \gamma \cdot n \chi_{\textbf{P}}(k_-;P)\;.
\end{equation}

The asymptotic behavior of the leading-twist DA has been known for more than forty years  \cite{Lepage:1979zb, Efremov:1979qk, Lepage:1980fj}:
\begin{equation}
\label{eq:PDAasy}
  \varphi_{\rm asy}(x) \stackrel{m_p/\zeta_{\rm ev} \simeq 0}{=} 6x(1-x)\,.
\end{equation}
Over the years, however, it has become clear that this form is unrealistic, in the case of $\pi$ and $K$, on any domain accessible to terrestrial experiments.
It is thus imperative to compute the DA at experimentally accessible scales.

\begin{figure}[t]%
\centering
\begin{tabular}{cc}
\includegraphics[width=0.48\textwidth]{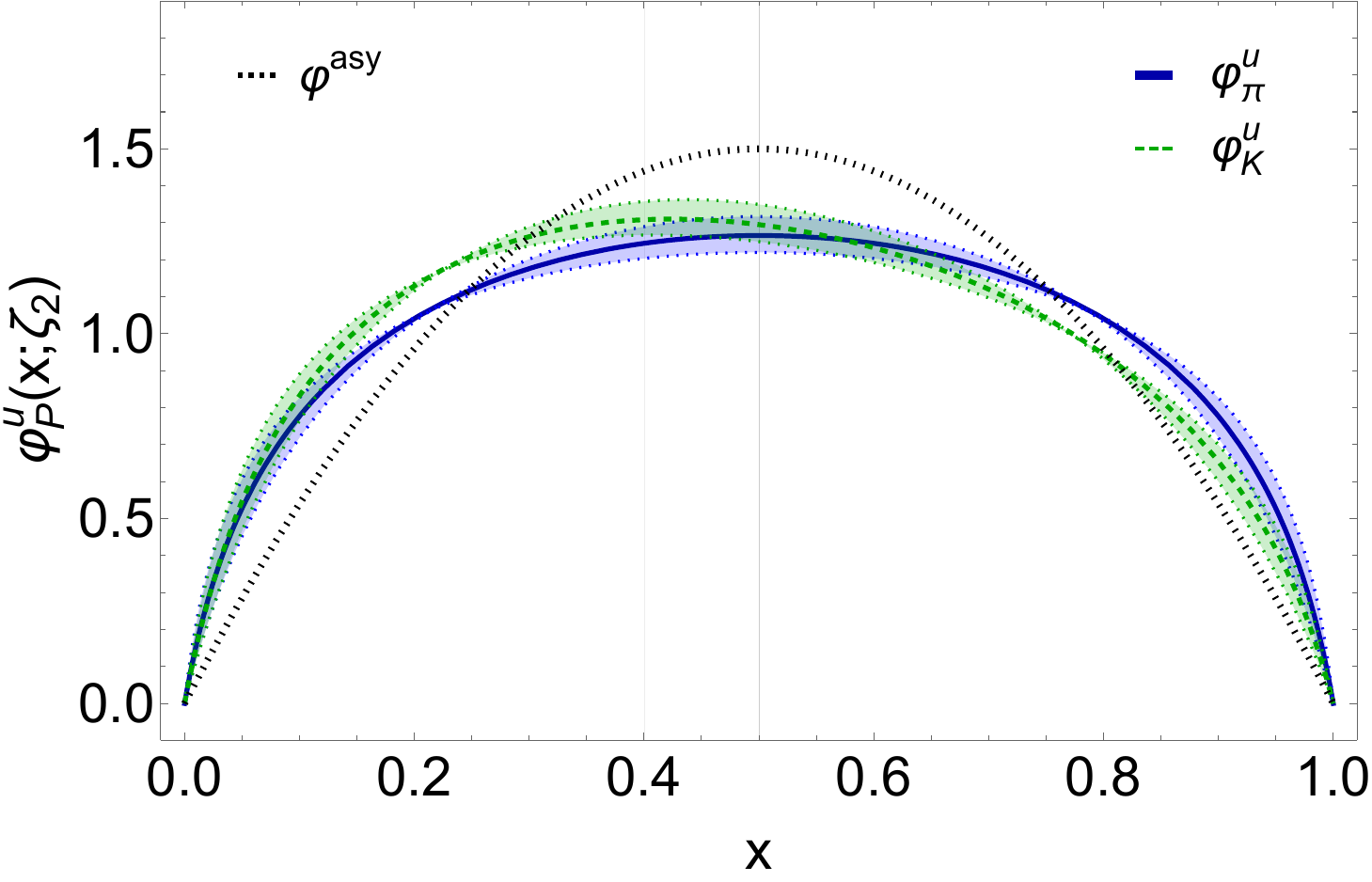} &
\includegraphics[width=0.48\textwidth]{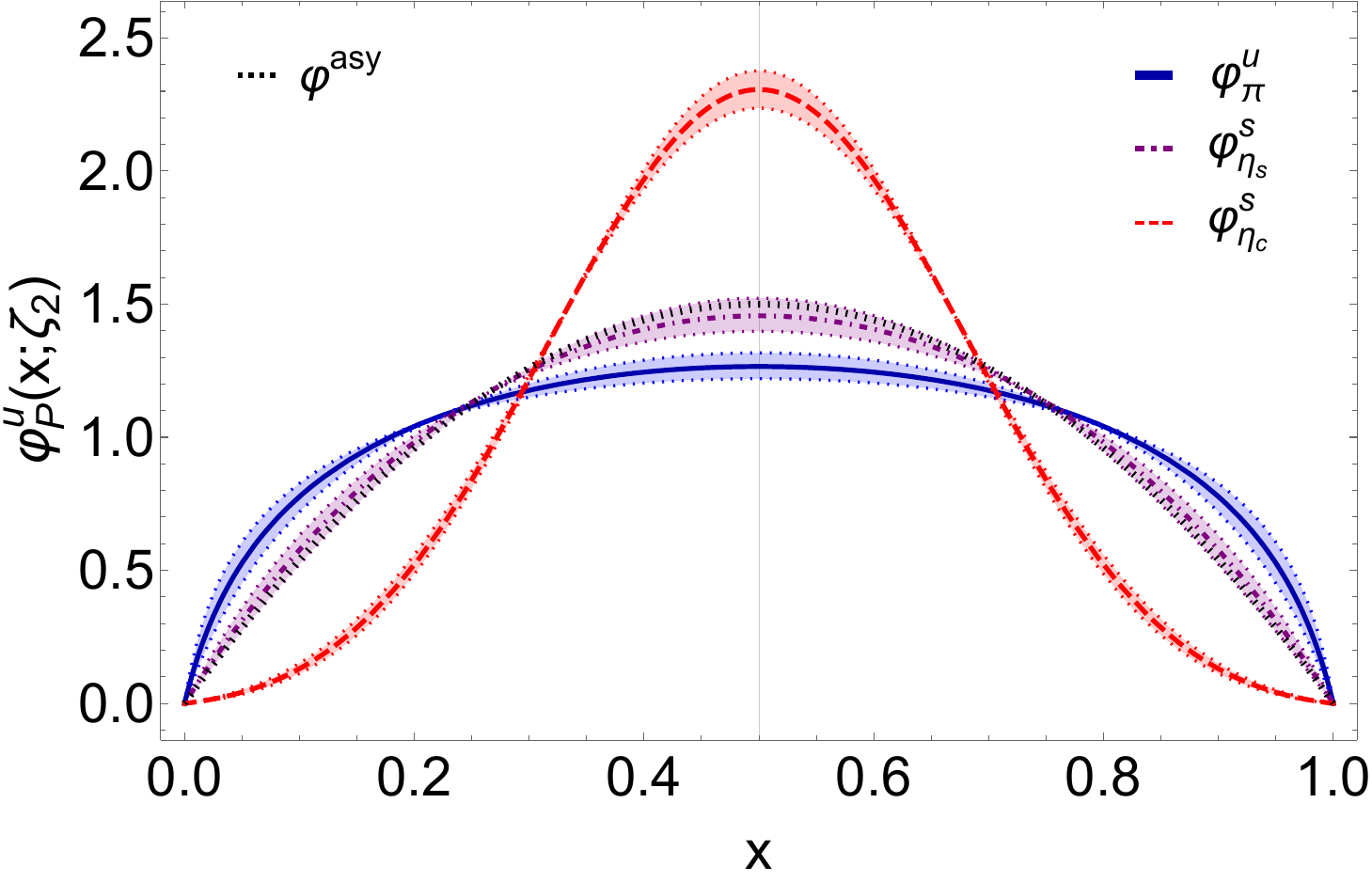}
\end{tabular}
\caption{[left] $\pi$, $K$ DAs in comparison with the asymptotic profile $\varphi_{\rm asy}$.
[right] Analogous predictions for $\pi$, $\eta_s$, $\eta_c$.
In both panels
the uncertainty bands reflect a $\pm 5$\% variation in the value of the low-order moments $\langle \xi^1 \rangle_\textbf{P}$ and $\langle \xi^2 \rangle_\textbf{P}$, $\xi = 1-2x$;
and each thin vertical line identifies the location of the maximum of a given DA -- only the $K$ DA maximum is displaced from $x=1/2$.
The predictions are taken from Refs.\,\cite{Cui:2020tdf, Ding:2015rkn}, which used a renormalisation scale $\zeta_2=2\,$GeV for all Schwinger functions.
\label{fig:PDA1}}
\end{figure}

Such efforts, using continuum and lattice methods, are canvassed elsewhere \cite{Roberts:2021nhw}.  They have concluded that, at all scales accessible to existing and foreseeable experiments, $\varphi_{\pi,K}(x;\zeta)$ is a broad concave function, \textit{viz}.\ strongly dilated and flattened in comparison with $\varphi_{\rm asy}(x)$.
This is illustrated in the left-panel of Fig.\,\ref{fig:PDA1}, which depicts CSM predictions for the $\pi$, $K$ DAs \cite{Cui:2020tdf} and contrasts them with $\varphi_{\rm asy}(x)$.

The dilation of $\varphi_{\pi,K}^u(x;\zeta)$ is a manifestation of EHM; and the DA of the pion -- Nature's lightest hadron -- is the most dilated of all.
Owing to flavor symmetry breaking, expressed in the QCD Lagrangian by a large difference between the $s$ and $u$ quark current-masses, the maximum of $\varphi_{K}^u(x;\zeta)$ is shifted slightly to $x\approx 0.4$.
This 20\% relocation is a statement that EHM dominates, but HB modulations are beginning to be felt, as may be seen by considering that $\tfrac{1}{2} \times M_u(0)/M_{s}(0) \approx \tfrac{1}{2} f_\pi/f_K \approx 0.4$.

Turning attention to heavy $q \bar q$ pseudoscalar mesons, the pattern is reversed.
Owing to the large HB-induced quark current masses, the DAs of such systems are compressed/contracted \cite{Ding:2015rkn}, something seen in the right-panel of Fig.\,\ref{fig:PDA1}.
These DAs become increasingly narrow, with greater $x=1/2$ peak height, as the current masses become larger.
Indeed, in the limit of infinitely heavy quarks:
\begin{equation}
\label{eq:PDANR}
    \varphi_{q\bar q}(x) \to \varphi_{\infty}(x) =\delta(x-1/2)\,.
\end{equation}
Considering, just for illustration, a pure $s\bar s =: \eta_s$ pseudoscalar meson, with $m_{\eta_s}\approx 0.69$\,GeV, one finds $\varphi_{\eta_s}^s(x,\zeta_2) \approx \varphi_{\rm asy}(x)$.  Thus, $s\bar s$ systems define a boundary, whereat EHM and HB mass generating effects are of roughly equal importance.
Numerical results from the simulations of lattice-regularized QCD (lQCD) confirm this CSM prediction \cite{Zhang:2020gaj}.
Moreover, CSM studies of the $\eta$-$\eta^\prime$ complex \cite{Ding:2018xwy} show that the light- and $s$-quark component DAs of the $\eta^\prime$-meson, especially, match this expectation.

\begin{figure}[t]%
\centering
\includegraphics[width=0.6\textwidth]{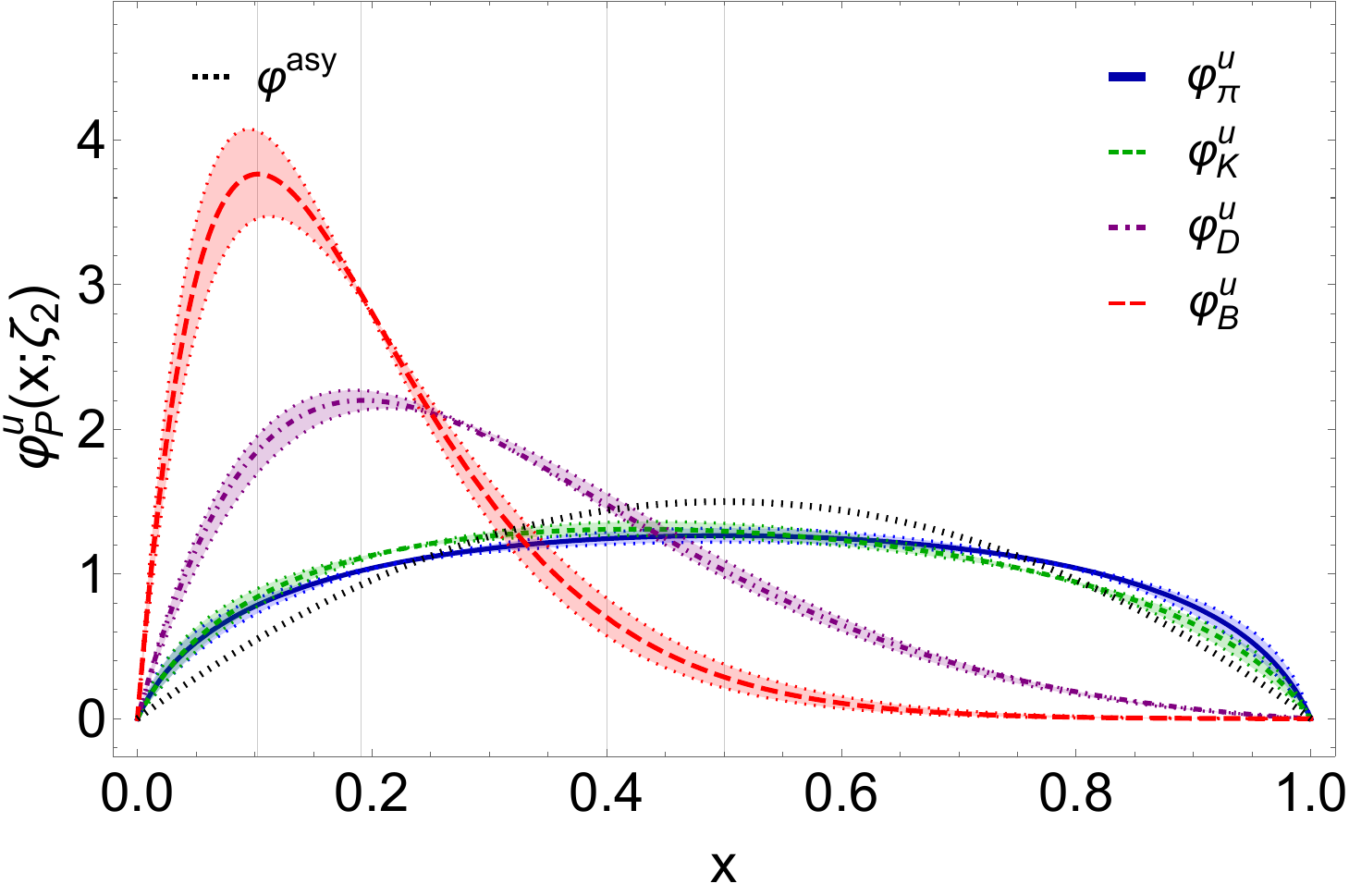}
\caption{
$\pi$, $K$, $D$, $B$ DAs.
The uncertainty bands reflect a $\pm 5$\% variation in the value of the low-order moments $\langle \xi^1 \rangle_\textbf{P}$ and $\langle \xi^2 \rangle_\textbf{P}$, $\xi = 1-2x$;
and each thin vertical line identifies the location of the maximum of a given DA.
The predictions are taken from Refs.\,\cite{Cui:2020tdf, Binosi:2018rht}.
\label{fig:PDAheavyL}}
\end{figure}

The analysis of heavy-light mesons provides additional information about the interplay between strong and weak mass generation \cite{Binosi:2018rht}.
The DAs of $D$ and $B$ mesons, compared with the $\pi$, $K$ cases, are shown in Fig.\,\ref{fig:PDAheavyL}.
Clearly, as the size of flavor symmetry breaking increases and, thus, the HB impact on the heavier quark -- see Ref.\,\cite[Fig.\,18]{Ding:2022ows}, the DA distortion becomes more pronounced.
At fixed light-quark mass, then with increasing heavy-quark mass, the location of the DA peak moves toward a minimum value $x_{\rm min}\in (0,0.5)$.
That $x_{\rm min}\neq 0$ is a special feature of heavy + light systems \cite{Neubert:1993mb}.
One may quantify this peak relocation by computing values of $\int_0^1 dx \,x \varphi(x;\zeta_2)$, which is a DA-weighted momentum fraction: $\pi$, $0.5$; $K$, $0.48$; $D$, $0.32$; $B$, $0.19$; and $0.12(1)$ in the case of one infinitely heavy quark.

For ground-state pseudoscalar mesons with masses $m_{\textbf{P}} \leq m_{\rm asy}\approx m_{\eta_s}$, the following parametrization is efficacious:
\begin{equation}
\label{eq:PDAlight}
    \varphi_\textbf{P}^{\rm light}(x;\zeta) := n_\textbf{P}\ln \left[ 1 + \frac{x(1-x)}{\rho^2_\textbf{P}}\right](1+\gamma_\textbf{P} (1-2x))\,,
\end{equation}
whereas this one works well on $m_{\textbf{P}} \geq m_{\rm asy}$:
\begin{equation}
\label{eq:PDAheavy}
    \varphi_\textbf{P}^{\rm heavy}(x;\zeta) := n_\textbf{P} x (1 - x)\,\text{exp}\left[\frac{x(1-x)}{\rho^2_\textbf{P}}+\gamma_{\textbf{P}}(1-2x)\right]\,.
\end{equation}
In both cases, $n_\textbf{P}$ ensures unit normalization and $\rho_\textbf{P}$, $\gamma_\textbf{P}$ are interpolation parameters.
These simple forms enable one to express the
endpoint behavior predicted by QCD;
dilation/compression of the DAs, via $\rho_\textbf{P}$;
and skewing in flavor asymmetric systems, via $\gamma_\textbf{P}$.

Notably, both forms reproduce $\varphi_{\rm asy}(x)$ in appropriate circumstances, \textit{viz}.\ $\rho_\textbf{P}\to \infty$, $\gamma_\textbf{P}\to 0$, and that is why they have a common boundary of applicability.
%
At the other extreme, \textit{i.e}., $\rho_\textbf{P}\to 0$, $\gamma_\textbf{P}\to 0$, one finds $\varphi_{\textbf{P}}^{\rm light}(x;\zeta) \to \varphi_{\rm SCI}(x)=1$ and $\varphi_{\textbf{P}}^{\rm heavy}(x;\zeta) \to \varphi_\infty(x)$.
The former is the broadest possible distribution and corresponds to that produced using a symmetry-preserving treatment of a vector$\,\otimes\,$vector contact interaction \cite{Roberts:2010rn}, whereas the latter is the narrowest.
Finally, the combination of numerous analyses, Refs.\,\cite{Binosi:2018rht, Ding:2018xwy, Chen:2018rwz, Cui:2020tdf}
enables us to determine the interpolation parameters shown in the Table\,\ref{tab:paramsPDA}. Associated low-order moments are also listed.

\begin{table}[t]
\centering
\captionsetup{width=.75\textwidth}
\caption{Interpolation parameters for the pseudoscalar meson DAs, to be used in Eqs.\,\eqref{eq:PDAlight}, \eqref{eq:PDAheavy}, as appropriate.
Entries above the horizontal line correspond to mesons whose masses are less than $m_{\rm asy}$.
The complement lies below this line.
In the $\eta-\eta'$ case, the superscript refers to the light $l = u/d$ and strange $s$ components of its wave function \cite{Ding:2018xwy}.
\label{tab:paramsPDA}}
\begin{tabular}[t]{c||l|l||l|l}
\hline
$\textbf{P}$ \,\,\,& $\rho_\textbf{P}$ & $\gamma_\textbf{P}$ & $\langle \xi \rangle_\textbf{P}$ & $\langle \xi^2
 \rangle_\textbf{P}$\\
\hline
$\pi$ &  0.180 & 0.0 & 0.0 & 0.247\\
$K$ &  0.224 & 0.149 & 0.036 & 0.239 \\
$\eta^{(l)}$ & 0.329 & 0.0 & 0.0 & 0.227 \\
$\eta^{(s)}$ & 0.421 & 0.0 & 0.0 & 0.220 \\
$\eta_s$ & 0.836 & 0.0 & 0.0 & 0.207 \\
\hline
$\eta'^{(l)}$ & 1.700 & 0.0 & 0.0 & 0.196 \\
$\eta'^{(s)}$ & 1.221 & 0.0 & 0.0 & 0.192 \\
$D$ & 1.887 & 2.059 & 0.365 & 0.277 \\
$D_s$ & 0.984 & 1.935 & 0.335 & 0.258 \\
$\eta_c$ & 0.294 & 0.0 & 0.0 & 0.110 \\
$B$ & 1.006 & 4.692 & 0.616 & 0.445 \\
$B_s$ & 0.747 & 4.739 & 0.607 & 0.435 \\
$B_c$ & 0.264 & 4.679 & 0.415 & 0.245 \\
$\eta_b$ & 0.200 & 0.0 & 0.0 & 0.066 \\
\hline
\end{tabular}
\end{table}%

\subsection{Parton distribution functions}
\label{sec:PDFs}
Complementing DAs, parton distribution functions (DFs) play a key role in the description of hard inclusive processes.  A given DF, ${\mathpzc p}_{\textbf{P}}(x;\zeta)$, is a number density, so that ${\mathpzc p}_{\textbf{P}}(x;\zeta) dx$ is the number of ${\mathpzc p}$ partons carrying a light-front fraction between $x$ and $x+dx$ of the total momentum of hadron $h$ at a resolving scale $\zeta$ \cite{Holt:2010vj}.

The DFs of $\textbf{P}$ are accessible via the associated forward Compton amplitude, $\gamma \textbf{P} \to \gamma \textbf{P}$ \cite{Holt:2010vj}.
Detailed considerations of that amplitude have led to the following expression for the $q$-in-\textbf{P} valence-quark DF\,\cite{Ding:2019lwe, Ding:2019qlr, Chang:2014lva}:
\begin{equation}
\label{eq:PDFdiag}
    {\mathpzc q}_\textbf{P}(x;\zeta_{\cal H}) = \text{tr}_{\text{CD}} \int_q \delta_{n,P}^x(k_\eta)  \{n\cdot \partial_{k_\eta}[\Gamma_{\textbf{P}}(k_\eta;-P)S(k_\eta)]\} \Gamma_{\textbf{P}}(k_{\bar{\eta}};P)S(k_{\bar{\eta}})\,,
\end{equation}
where $k_{\eta} = k + \eta P$, $k_{\bar{\eta}}=k-(1-\eta)P$, and the DF is independent of $\eta\in [0,1]$.
The result from Eq.\,\eqref{eq:PDFdiag} does not depend on the scale at which the quark propagator and meson Bethe-Salpeter amplitude are renormalized \cite{Cui:2020tdf}; so, what is the meaning of $\zeta_{\cal H}$?

In developing the answer to this question, it is important to note that the following identities are readily verified:
\begin{subequations}
\begin{align}
\langle x^0 \rangle_{\zeta_{\cal H}}^{{\mathpzc q}_\textbf{P}}
& =1
= \langle x^0 \rangle_{\zeta_{\cal H}}^{\bar{\mathpzc h}_\textbf{P}}\,, \label{BaryonNumber}\\
\bar{\mathpzc h}_\textbf{P}(x;\zeta_{\cal H}) & =
{\mathpzc q}_\textbf{P}(1-x;\zeta_{\cal H}) \Rightarrow
\langle x \rangle_{\zeta_{\cal H}}^{{\mathpzc q}_\textbf{P}} +
\langle x \rangle_{\zeta_{\cal H}}^{\bar{\mathpzc h}_\textbf{P}} = 1\,.
\label{MomentumVq}
\end{align}
\end{subequations}
The statements in Eq.\,\eqref{BaryonNumber} express baryon number conservation.  They must be valid, independent of the value and meaning of $\zeta_{\cal H}$.
On the other hand, those in Eq.\,\eqref{MomentumVq} mean that valence-quark degrees-of-freedom carry all the hadron's light-front momentum at the scale $\zeta_{\cal H}$.  This is a principal reason behind the identification of $\zeta_{\cal H}$ as the hadron scale \cite{Ding:2019lwe}.  We will see that $\Lambda_{\rm QCD} \lesssim \zeta_{\cal H}<m_p$.

A hadron's LFWF provides a bridge between its valence-quark DAs and DFs.  This connection is best introduced by illustration; so, suppose one has a two-body system described by the following LFWF:
\begin{equation}
\label{HLFWF}
\psi_{\textbf{P}}(x,k_\perp;\zeta_{\cal H}) = \frac{{\mathpzc n}_\psi\, M^{2\delta} x(1-x)}{[M^2 [1+x(1-x)] + k_\perp^2]^{1+\delta}}\,,
\quad
\int \frac{dx d^2 k_\perp}{16\pi^3} | \psi_{\textbf{P}}(x,k_\perp;\zeta_{\cal H}) |^2 = 1\,,
\end{equation}
where $M$ is a mass whose size is assumed to be set by EHM and ${\mathpzc n}_\psi$ is the normalisation constant.  For $\delta=0$, this LFWF exhibits the large-$k_\perp^2$ scaling behaviour of a leading-twist two-body wave function in QCD \cite[Eq.\,(2.15)]{Lepage:1980fj}.  (True QCD wave functions also include $\ln k_\perp^2$ scaling violations.  This is mimicked by $\delta \gtrsim 0$.)

\begin{figure}[t]%
\centering
\includegraphics[width=0.6\textwidth]{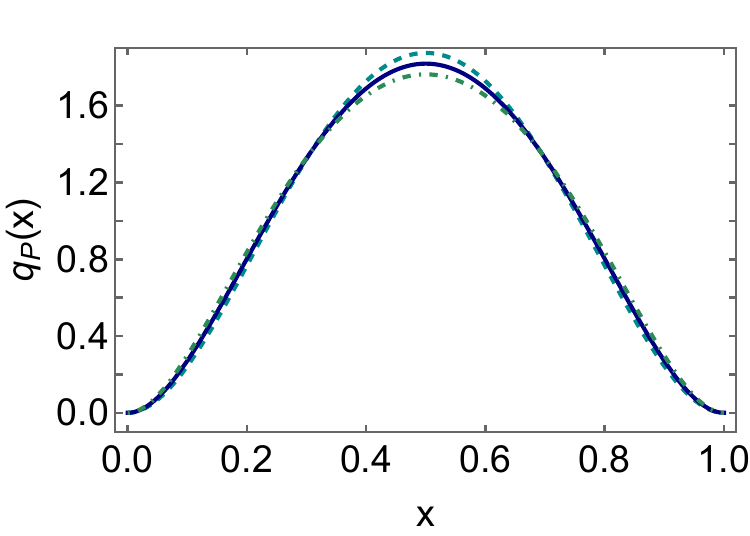}
\caption{
DF in Eq.\,\eqref{DFfromLFWF} (solid blue curve) compared with $[\tilde \varphi_{\textbf{P}}^q(x,\zeta_{\cal H})]^2$ from Eq.\,\eqref{rescaleDA2} (dot-dashed green curve).
Dashed cyan curve: scale-free function in Eq.\,\eqref{DFscalefree}.
\label{DFcfDA}}
\end{figure}

The valence-quark DF is obtained as
\begin{subequations}
\label{DFfromLFWF}
\begin{align}
{\mathpzc q}_\textbf{P}(x;\zeta_{\cal H}) & =
\int \frac{d^2 k_\perp}{16\pi^3} | \psi_{\textbf{P}}(x,k_\perp;\zeta_{\cal H}) |^2 
\label{DFfromLFWFa}\\
& \stackrel{\delta \to 0}{=}
\frac{30\, x^2 (1-x)^2}{\left[1+x(1-x)\right] \left[24 \sqrt{5} \tanh
   ^{-1}\left(\frac{1}{\sqrt{5}}\right)-25\right]}\,.
\end{align}
\end{subequations}
This is a scale-free function, \textit{i.e.}, it is not explicitly dependent on the mass-scale in the LFWF.  However, the presence of that scale and, hence, EHM is manifested in the denominator structure $1+x(1-x)$, which, as evident Fig.\,\ref{DFcfDA}, introduces a dilation with-respect-to the numerator function alone:
\begin{equation}
    {\mathpzc q}_{\rm sf}(x) = 30 x^2 (1-x)^2\,.  \label{DFscalefree}
\end{equation}
It is worth noting that one may recover this EHM-insensitive function by changing $M^2[1+x(1-x)]\to M^2[1 + \epsilon x(1-x)]$ in Eq.\,\eqref{HLFWF} and then taking $\epsilon \to 0$.

The meson's DA is obtained via
    \begin{align}\label{DAfromLFWF}
    f_{\textbf{P}} \varphi_{\textbf{P}}^q(x,\zeta_{\cal H}) & =
    \int \frac{d^2 k_\perp}{16\pi^3} \psi_{\textbf{P}}(x,k_\perp;\zeta_{\cal H}) \,,
    \end{align}
where extraction of the leptonic decay constant, $f_{\textbf{P}}$, means $ \int_0^1 dx\, \varphi_{\textbf{P}}^q(x,\zeta_{\cal H})=1$.
With $\delta = 0$, this integral is $\ln$-divergent and that explains the renormalization constant in Eq.\,\eqref{eq:PDA}.  For the purpose of our illustration, we define the result by expressing the integral as a series in $\delta$ on $\delta \simeq 0$, discarding the $1/\delta$ piece that characterizes the $\ln$-divergence, then taking the limit $\delta\to 0$.  This procedure yields
\begin{subequations}
    \begin{align}
    \varphi_{\textbf{P}}^q(x,\zeta_{\cal H}) & =
    2.399 x(1-x) (2.949 - 2.474 \ln[1+x(1-x)])\,,\\
    f_{\textbf{P}} & = 0.0809 M \,.
    \end{align}
\end{subequations}
Here, the only explicit dependence on $M$ is contained in the decay constant; so, the DA is seemingly independent of this mass-scale.  However, as with the DF, EHM is expressed in the $1+x(1-x)$ term.  (To make these things readily apparent, we have replaced special-functions at fixed arguments  by their numerical values.)
Once again, akin to the DF, one may recover the asymptotic DA in Eq.\,\eqref{eq:PDAasy} by making the replacement $M^2[1+x(1-x)]\to M^2[1 + \epsilon x(1-x)]$ in Eq.\,\eqref{HLFWF} and then taking $\epsilon \to 0$.

Consider now the following rescaled DA:
\begin{equation}
    \tilde \varphi_{\textbf{P}}^q(x,\zeta_{\cal H}) =  {\mathpzc r}_{\varphi^2} \varphi_{\textbf{P}}^q(x,\zeta_{\cal H})
    \ni \int_0^1 dx\, [\tilde \varphi_{\textbf{P}}^q(x,\zeta_{\cal H})]^2 = 1\,.
    \label{rescaleDA2}
\end{equation}
Figure~\ref{DFcfDA} compares $[\tilde \varphi_{\textbf{P}}^q(x,\zeta_{\cal H})]^2$ with the DF derived from the same LFWF.  The two functions are practically indistinguishable: mathematically, the ${\mathpzc L}_1$-difference between these curves is just 2.7\%.

Following the same procedure with the asymptotic DA in Eq.\,\eqref{eq:PDAasy} yields exactly the scale-free DF in Eq.\,\eqref{DFscalefree}.
This highlights a simple fact. Namely, $M^2[1+x(1-x)]\to M^2$ in Eq.\,\eqref{HLFWF} produces a factorized LFWF: $\psi_{\textbf{P}}(x;\zeta_{\cal H}) = \varphi_{\textbf{P}}(x;\zeta_{\cal H}) \psi_{\textbf{P}}(k_\perp^2;\zeta)$; and whenever this is a good approximation for quantities obtained by integration -- it need not be pointwise precise, then
\begin{equation}
\label{DFeqDA2}
    {\mathpzc q}_\textbf{P}(x;\zeta_{\cal H}) =  [\tilde \varphi_{\textbf{P}}^q(x,\zeta_{\cal H})]^2\,.
\end{equation}
Such factorized representations are known \cite{Raya:2021zrz} to be a good approximation for ground-state mesons in which either the valence quarks are mass degenerate or EHM leads to significant suppression of HB-induced flavor symmetry violation, \emph{e.g}., kaons.  Its value in treating other mesons, such as excited and heavy + light states, is currently being explored.

\subsection{Hadron scale distributions}
The reliability of Eq.\,\eqref{DFeqDA2} for pions and kaons has been exploited to good effect \cite{Cui:2020tdf, Cui:2020dlm, Chang:2021utv, Chang:2022jri, Lu:2022cjx, Xing:2023pms}.  The first step is to associate the DA obtained using Eq.\,\eqref{eq:PDA}  with the hadron scale, $\zeta_{\cal H}$. This places it on the same level as the DF calculated using Eq.\,\eqref{eq:PDFdiag}.  (Recall that both expressions produce distributions that are independent of the propagator and Bethe-Salpeter amplitude renormalization scale.)


\begin{figure}[t]%
\centering
\begin{tabular}{lr}
\includegraphics[width=0.48\textwidth]{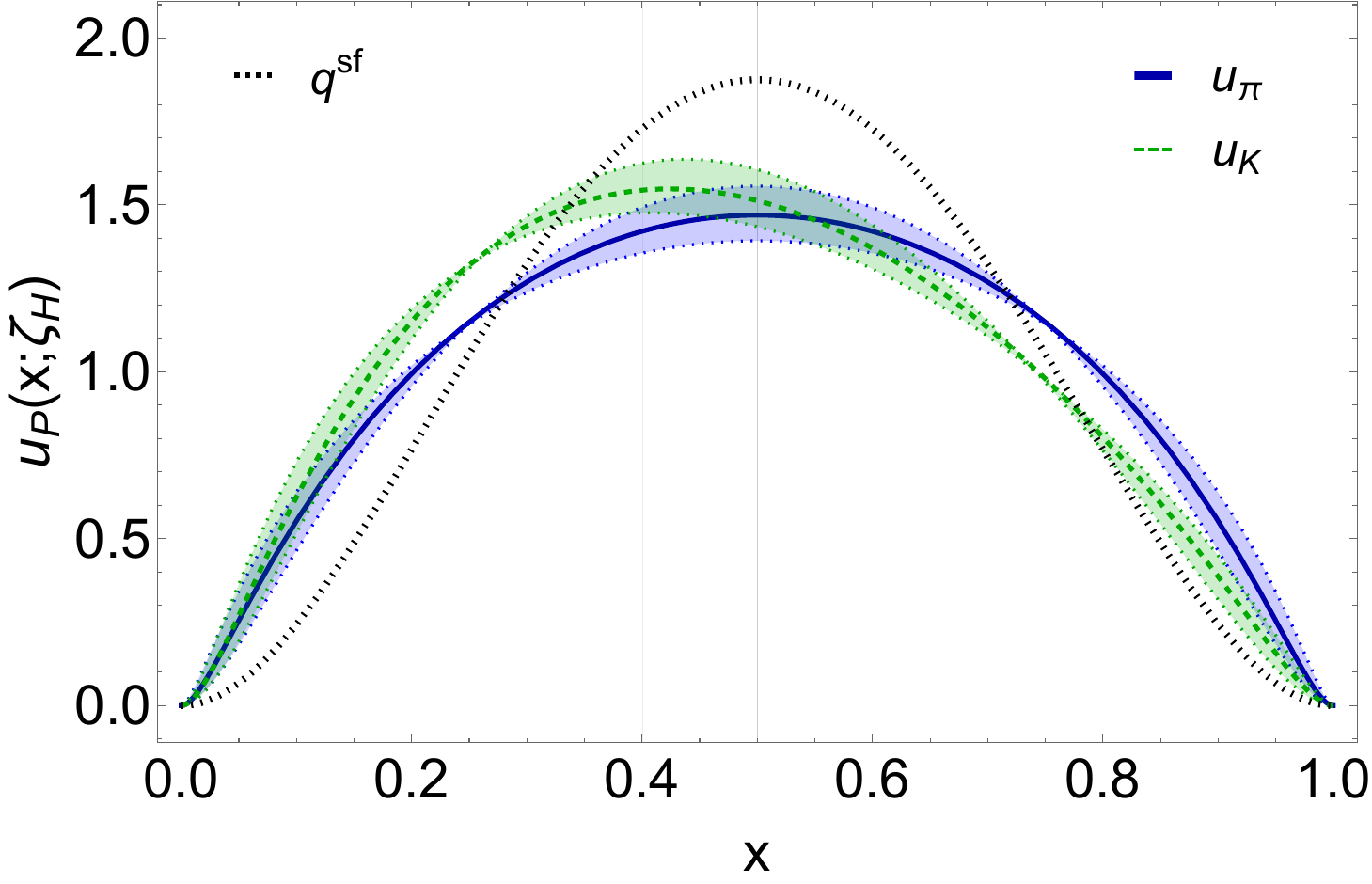}
\includegraphics[width=0.48\textwidth]{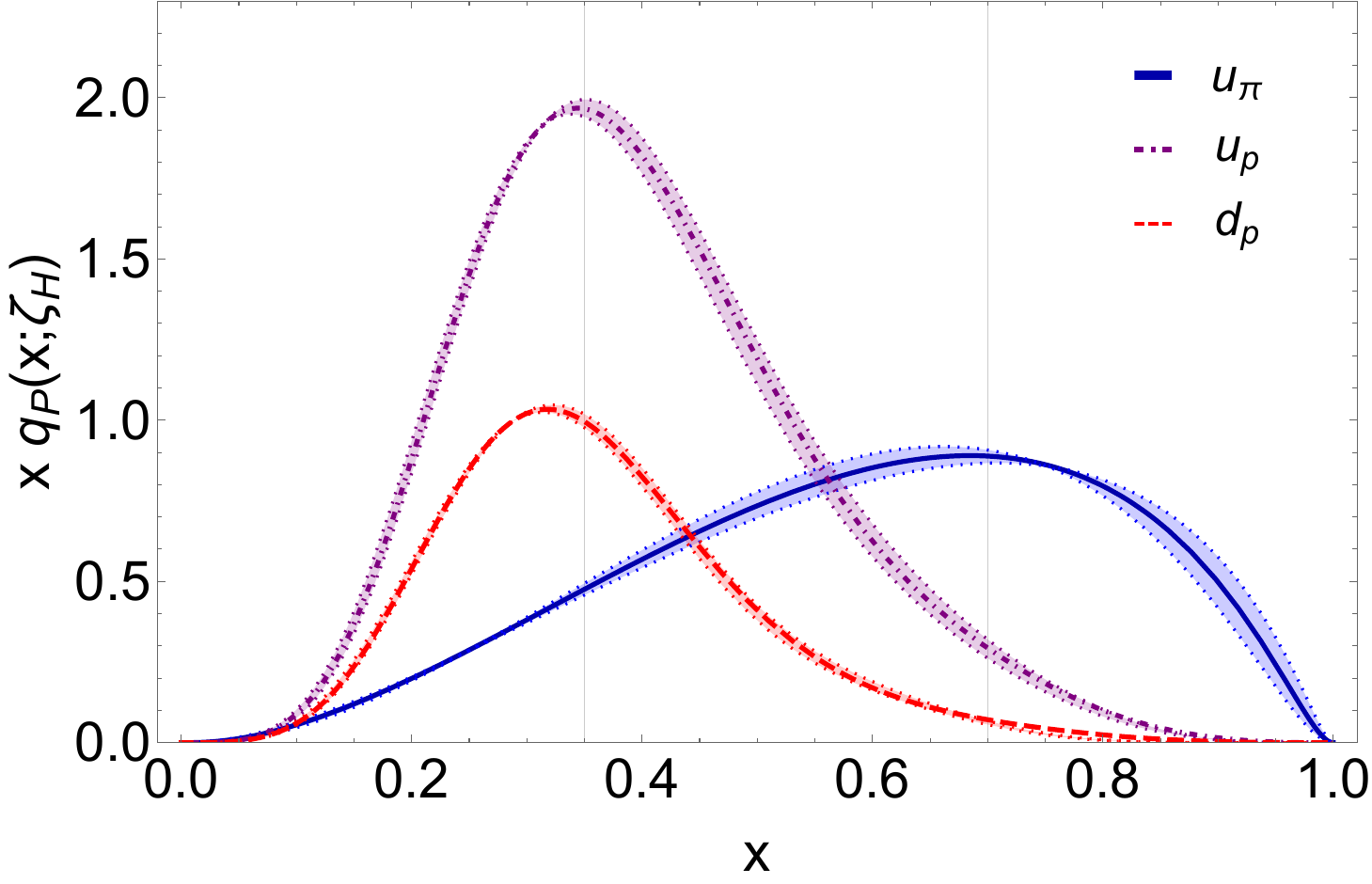}
\end{tabular}
\caption{Hadron scale DFs. [left] $\pi$, $K$ distributions at $\zeta_{\cal H}$, in constrast with the scale-free profile $q_{\rm sf}$ \cite{Cui:2020tdf}.
[right] Analogous comparison for ${\mathpzc u}_\pi$ ${\mathpzc u}_p$, ${\mathpzc d}_p$ \cite{Lu:2022cjx}.
The $\pi$, $K$ uncertainty bands stem from that associated with the corresponding DAs, whereas those of the proton express a $10$\% variation in the values of the low-order moments
$\langle x-x_0^{\mathpzc q}
 \rangle_{{\mathpzc q}_P}$ and $\langle (x-x_0^{\mathpzc q})^2
 \rangle_{{\mathpzc q}_P}$,
where $x_0^{{\mathpzc u},{\mathpzc d}}\approx 0.35,\,0.32$ corresponds to the maximum of the $x {\mathpzc u}_p$ and $x{\mathpzc d}_p$ distributions, respectively.
In each panel, the vertical lines indicate the maximum of a given $u$-quark DF. \label{fig:PDF1}}
\end{figure}

Following this procedure and working with the DAs drawn in Fig.\,\ref{fig:PDA1}, one obtains the $\pi$ and $K$ DFs drawn in Fig.\,\ref{fig:PDF1}\,--\,left panel.  They are noticeably dilated in comparison with ${\mathpzc q}_{\rm sf}(x)$.  The DFs produce the following low-order Mellin moments:
\begin{subequations}
\begin{align}
    \langle x \rangle_{\zeta_{\cal H}}^{{\mathpzc u}_\pi} & = \tfrac{1}{2}\,,\;
    \langle x \rangle_{\zeta_{\cal H}}^{{\mathpzc u}_K} = 0.473(3)\,,\;
    \langle x \rangle_{\zeta_{\cal H}}^{\bar{\mathpzc s}_K} = 0.527(3)\,,\\
 \langle x^2 \rangle_{\zeta_{\cal H}}^{{\mathpzc u}_\pi} & = 0.300(3)\,,\;
\langle x^2 \rangle_{\zeta_{\cal H}}^{({\mathpzc u}_K+\bar{\mathpzc s}_K)/2}
= \langle x^2 \rangle_{\zeta_{\cal H}}^{{\mathpzc u}_K} + \tfrac{1}{2}
- \langle x \rangle_{\zeta_{\cal H}}^{{\mathpzc u}_K}
= 0.295(2)
\end{align}
\end{subequations}
\textit{N.B}.\ At $\zeta_{\cal H}$, $({\mathpzc u}_K+\bar{\mathpzc s}_K)/2$ is a symmetric distribution, so its moments can be compared directly with those of the pion DF: evidently, HB modulation does not significantly affect the dilation of the kaon DF.
Using the scale-free DF, these low-order moments are, respectively, $\tfrac{1}{2}$, $\tfrac{2}{7} = 0.286$.
It is worth stressing that, despite the EHM-induced dilation, each DF is compatible with QCD constraints \cite{Brodsky:1994kg, Yuan:2003fs, Holt:2010vj}.

An interpolating DF parametrization that simultaneously expresses the EHM-induced dilation, soft endpoint behavior, and skewing, when present, is provided by the following function:
\begin{equation}
\label{eq:PDFlight}
    {\mathpzc u}_\textbf{P}(x;\zeta_{\cal H}) := n_\textbf{P}\ln \left[ 1 + \frac{x^2(1-x)^2}{\rho^2_\textbf{P}}\right](1+\gamma_\textbf{P} (1-2x))\,.
\end{equation}
The DFs in Fig.\,\ref{fig:PDF1}\,--\,left panel are reproduced with
\begin{equation}
\rho_\pi=0.069\,, \gamma_\pi=0\,,\quad
\rho_K=0.087\,, \gamma_K=0.295\,.
\end{equation}
For some purposes, such as the calculation of kaon fragmentation functions \cite{Xing:2023pms}, one may use an alternative, practically equivalent form:
\begin{equation}
        {\mathpzc u}_K(x;\zeta_{\cal H}) =
        \tilde {\mathpzc n}_K 
        \ln \left[ 1 + \frac{x^2(1-x)^2}{\tilde\rho^2_K}
        (1+\tilde \gamma_K^2 x^2(1-x)^4)\right]\,,
\end{equation}
$\tilde\rho_K=0.062$, $\tilde\gamma_K=13.83$.

Calculation of proton DFs is described in Ref.\,\cite{Chang:2022jri, Lu:2022cjx, Yu:2024qsd}. Such analyses yield the following light-front momentum fractions:
\begin{equation}
    \langle x \rangle_{\zeta_{\cal H}}^{u_p} = 0.69 \neq 2/3\,,\;\langle x^1 \rangle_{\zeta_{\cal H}}^{d_p} = 0.31\neq 1/3\,.
\end{equation}
Contrary to the pion, and despite also being composed of light quarks, the proton momentum distributions associated with the different flavors are not the same, even accounting for the $2\times u\,$:$\,1\times d$ ratio.
This is a manifestation of $SU(4)$ spin-flavor symmetry breaking in the proton wave function, which may be attributed to the emergence of strong, nonpointlike diquark correlations \cite{Barabanov:2020jvn}.
Furthermore, as highlighted by Fig.\,\ref{fig:PDF1}\,--\,right panel, the valence parton DFs in the pion and proton have markedly different profiles.
This owes partly to the difference in the number of valence degrees-of-freedom; but it is also an expression of EHM, with the pion valence-quark DF being the most dilated amongst all hadrons. For instance, the $\eta^\prime$ meson DF profile is much less dilated.

\subsection{Evolved distributions}
In order for data to be connected with DFs, the experiments should involve energy and/or momentum transfers (far) in excess of $m_p$, \textit{i.e}., be conducted on a kinematic domain for which QCD factorization is valid.  Consequently, before comparisons with such data can be made, the hadron scale DFs must be evolved to the energy scale appropriate to a given experiment.  That can be accomplished using the all-orders (AO) evolution scheme explained elsewhere \cite{Yin:2023dbw}, which is a particular realization of DGLAP evolution \cite{Dokshitzer:1977sg, Gribov:1972ri, Lipatov:1974qm, Altarelli:1977zs} that has proved efficacious in numerous applications -- see, \emph{e.g}., Refs.\,\cite{Cui:2020tdf, Han:2020vjp, Xie:2021ypc, Raya:2021zrz, Chang:2022jri, Lu:2022cjx, dePaula:2022pcb, Cheng:2023kmt, Xing:2023wuk, Xing:2023pms, Lu:2023yna, Yu:2024qsd}.

Regarding DF Mellin moments, the AO scheme provides closed algebraic relations between them.
 For instance, the moments of any given valence-parton DF are related as follows:
\begin{equation}\label{eq:intval}
\langle x^n \rangle^{\mathpzc p}_{\zeta} 
=  \langle x^n \rangle^{\mathpzc p}_{\zeta_{\cal H}} \left[ \frac{\langle x \rangle^{\mathpzc p}_{\zeta} }{\langle x \rangle^q_{\zeta_{\cal H}} } \right]^{\gamma_{qq}^{n}/\gamma_{qq}^1}\,,
\end{equation}
where $\gamma_{qq}^n$ are the appropriate 1-loop anomalous dimensions \cite{Dokshitzer:1977sg, Gribov:1972ri, Lipatov:1974qm, Altarelli:1977zs}.
This identity states that all Mellin moments of the DF at any $\zeta>\zeta_{\cal H}$ are completely determined by the valence-quark momentum fraction at this scale, so long as all moments are known at the hadron scale.

Recall that, by definition, $\zeta_{\cal H}$ is the scale at which valence degrees-of-freedom carry all the hadron's properties.  This entails that glue and sea DFs are identically zero at the hadron scale: 
\begin{equation}
{\mathpzc g}_{\textbf{P}}(x;\zeta_{\cal H})\equiv 0 \equiv {\mathpzc S}_\textbf{P}(x;\zeta_{\cal H})\,.  
\end{equation}
It has thus far been found that the same value of $\zeta_{\cal H}$ serves well for all hadrons.  Its actual value is immaterial and need not be specified.  Notwithstanding that, practical analyses of lQCD results indicate that $\zeta_{\cal H}\approx 0.35\,$GeV \cite{Lu:2023yna}; and using the PI charge described above, one predicts \cite[Eq.\,(15)]{Cui:2020tdf}:
\begin{equation}
    \zeta_{\cal H}= 0.331(2)\,{\rm GeV }\,.
\end{equation}
It is worth stressing that, under evolution, glue and sea DFs are nonzero $\forall \zeta>\zeta_{\cal H}$; moreover, even on $\zeta \simeq m_p$, a significant fraction of a given hadron's light-front momentum is lodged with glue and sea.  This entails that, even without recourse to ``intrinsic charm'' \cite{Brodsky:1980pb}, roughly $1.5$\% of the hadron's momentum is lodged with the $c$-quark sea at $\zeta \simeq 1.5\,m_p$.

\begin{figure}[t]%
\centering
\begin{tabular}{cc}
\includegraphics[width=0.49\textwidth]{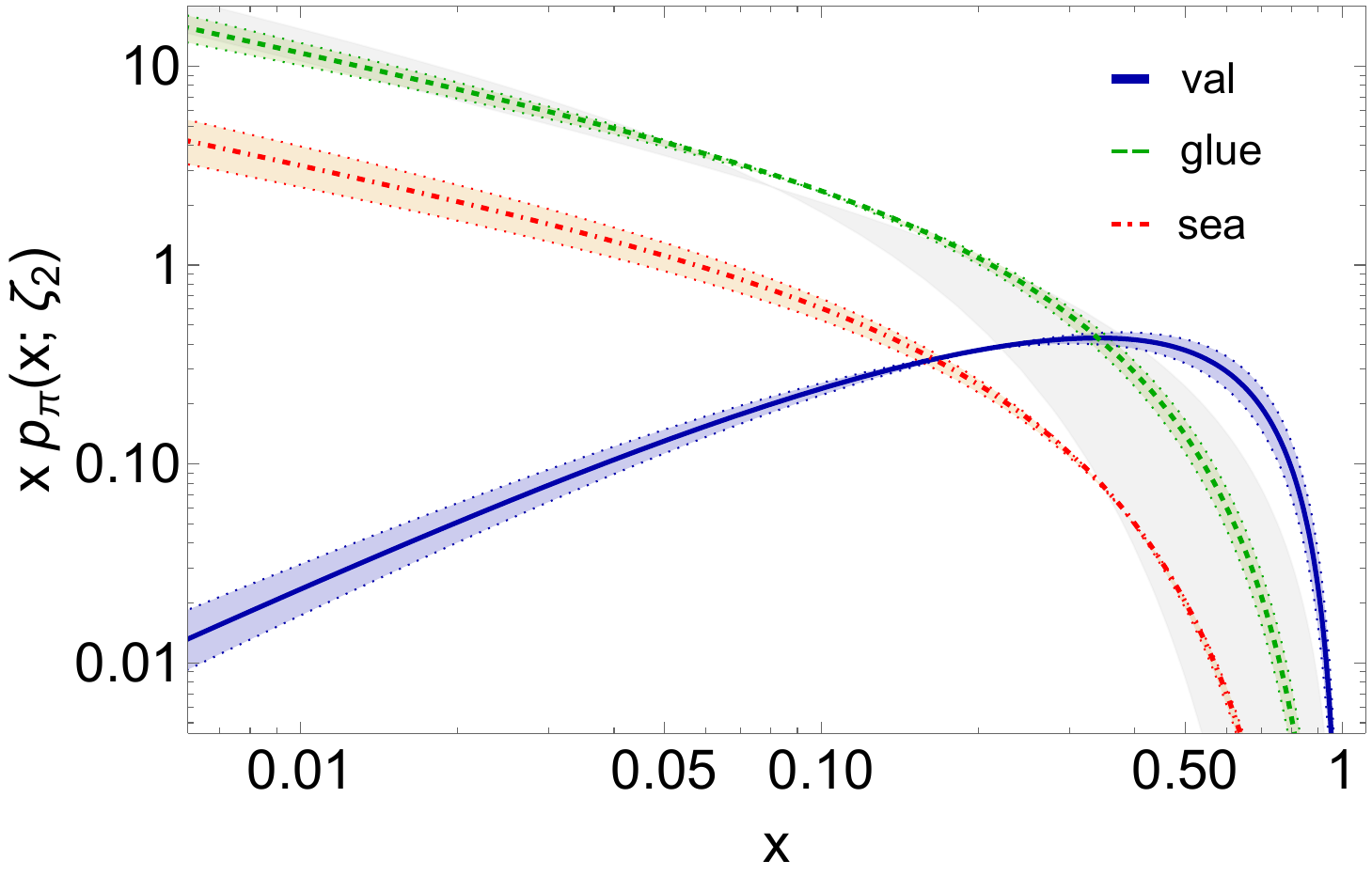}
\includegraphics[width=0.49\textwidth]{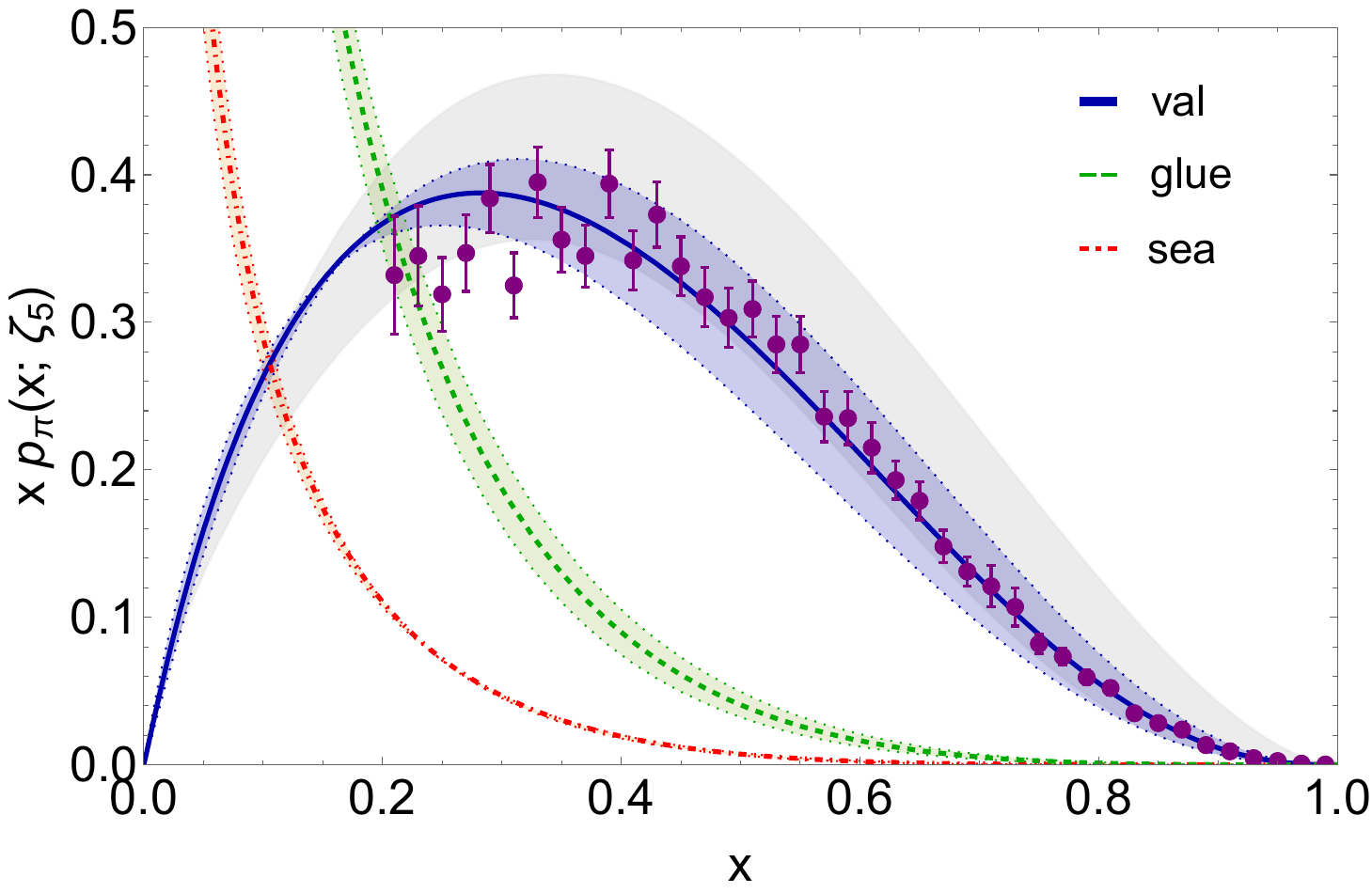}
\end{tabular}
\caption{Pion DFs:
valence (solid blue),
glue (dashed green),
and sea (dot-dashed orange).
[left] Results obtained after evolution of hadron scale predictions to $\zeta_2=2\,$GeV \cite{Cui:2020tdf}.
(Gluon and sea distributions rescaled by a factor $1/\langle x g_\pi(x;\zeta_2) \rangle = 1/0.41 = 2.46$.)
lQCD result for the gluon (grey band) \cite{Fan:2021bcr, Chang:2021utv}.
[right] CSM predictions at $\zeta_5$.
lQCD extracton of $u_\pi(x;\zeta_5)$ \cite{Sufian:2020vzb} (grey band).
Data from the reanalysis of Ref.\,\cite[E615]{Conway:1989fs} described in Ref.\,\cite{Aicher:2010cb}.
\label{fig:PDFpievol}}
\end{figure}

These features are illustrated for the pion in Fig.\,\ref{fig:PDFpievol}.
The left panel displays the valence, glue, and sea DFs calculated in Ref.\,\cite{Cui:2020tdf}.
Notably, the parameter-free CSM prediction for the glue DF agrees well with a recent lQCD computation \cite{Fan:2021bcr, Chang:2021utv}.
At this scale, $\zeta=\zeta_2=2\,$GeV, referred to the light-front \cite{Cui:2020tdf}:  valence degrees-of-freedom carry $48(4)$\% of the pion's momentum; 
glue, $41(2)$\%, and four-flavor sea, $11(2)$\%.
(In Ref.\,\cite{Cui:2020tdf}, quark current-mass effects were not included in the evolution equations.  More recent analyses have introduced mass thresholds \cite{Lu:2022cjx, Yin:2023dbw}.)

Figure~\ref{fig:PDFpievol}\,--\,right depicts the CSM predictions at $\zeta_5=5.2\,$GeV, \textit{i.e}., the scale of Ref.\,\cite[E615]{Conway:1989fs}.
There is excellent agreement with the analysis of that data described in Ref.\,\cite{Aicher:2010cb}.
At this scale, referred to the light-front \cite{Cui:2020tdf}:  valence degrees-of-freedom carry $41(4)$\% of the pion's momentum; glue, $45(2)$\%, and four-flavor sea, $14(2)$\%.
The lQCD computation from Ref.\,\cite{Sufian:2020vzb} is also displayed in Fig.\,\ref{fig:PDFpievol}\,--\,right.
That study used a novel techique for extracting DF pointwise behavior from a Euclidean lattice.
Further discussion of these and related points can be found in Refs.\,\cite{Cui:2021mom, Cui:2022bxn, Lu:2023yna}.

\begin{figure}[t]%
\centering
\begin{tabular}{cc}
\includegraphics[width=0.48\textwidth]{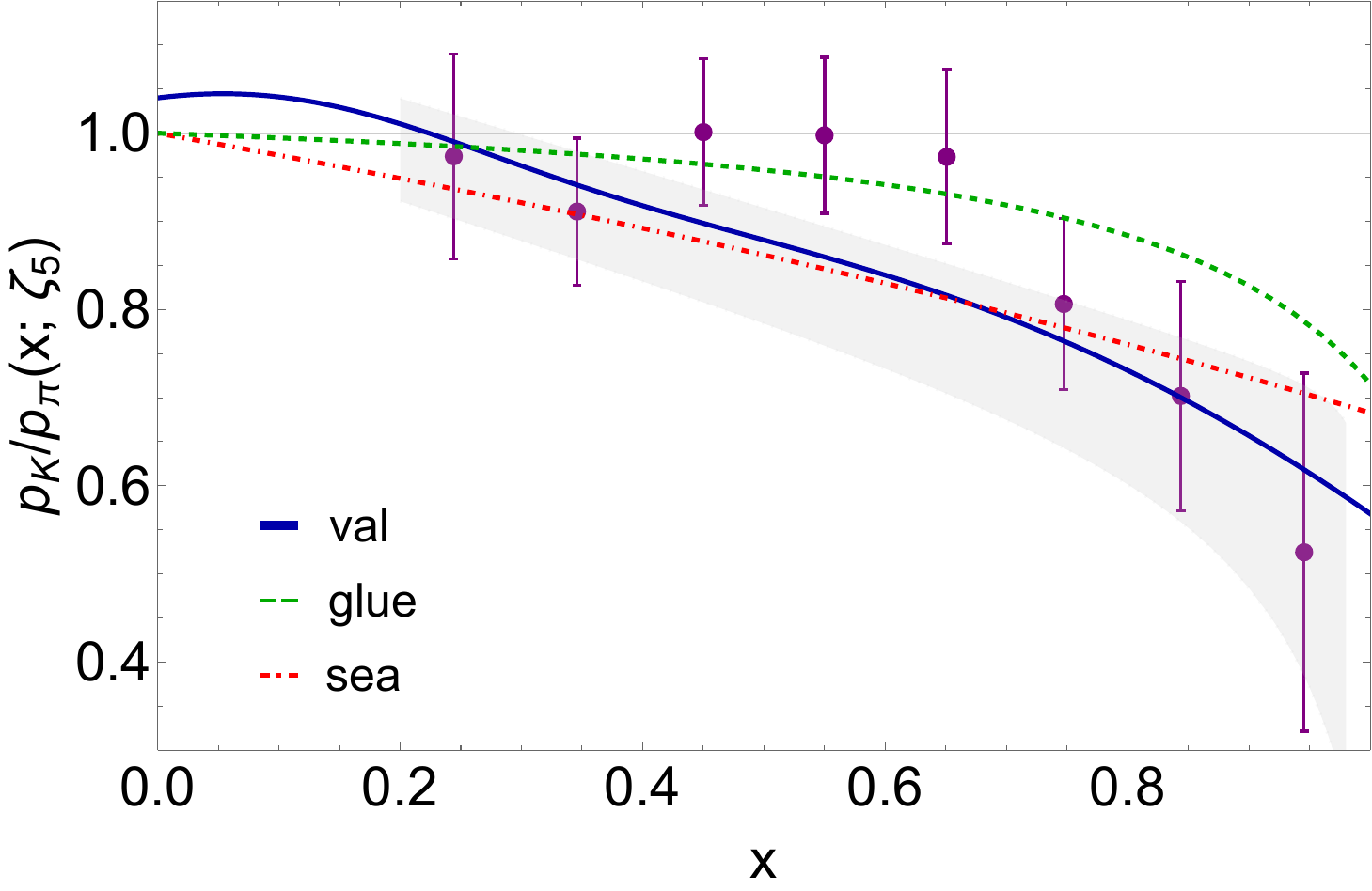} &
\includegraphics[width=0.48\textwidth]{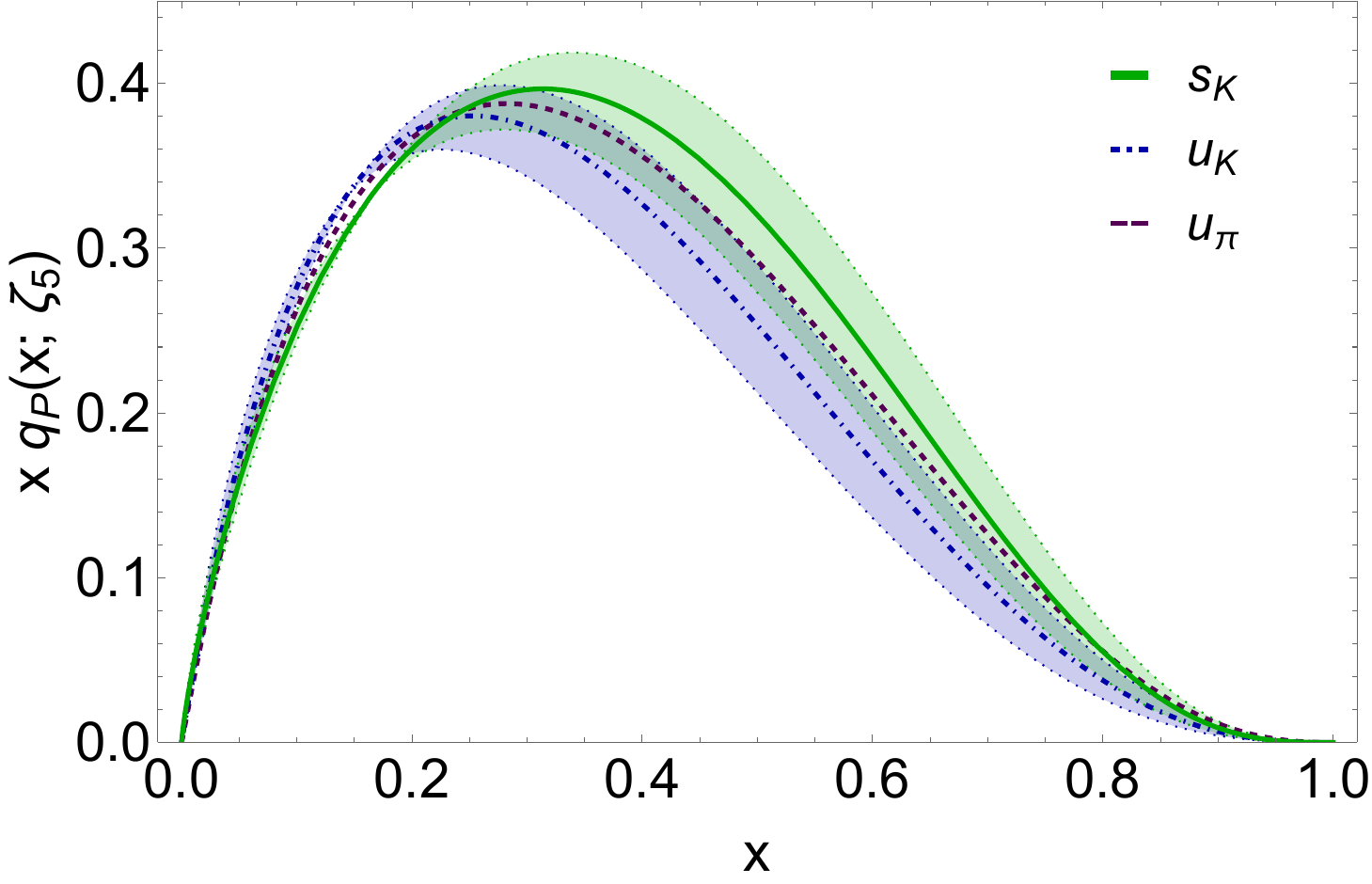}
\end{tabular}
\caption{Kaon parton DFs.
[left] Ratio ${\mathpzc p}_K/{\mathpzc p}_\pi$ for different parton species: $u$ valence-quark (solid blue), gluon (dashed green), and sea (dot-dashed red).
Data \cite{Badier:1980fhh} and lQCD (grey band) \cite{Lin:2020ssv} correspond to the ratio valence ratio $u_K/u_\pi$.
[right] CSM predictions for $\bar{s}_K(x),\, u_K(x)$ and $u_\pi(x)$. The error bands associated with the kaon correspond to a $\pm 10$\% variation of $\zeta_{\cal H}$.
\label{fig:PDFKevol}}
\end{figure}

Regarding the kaon, empirical information is scarce.
Only eight points are available and those relate solely to the valence-quark ratio ${\mathpzc u}_K(x;\zeta_5)/{\mathpzc u}_\pi(x;\zeta_5)$ \cite{Badier:1980fhh}.
Figure~\ref{fig:PDFKevol}\,--\,left displays the CSM prediction for this ratio compared with experimental data and a lQCD result \cite{Lin:2020ssv}: plainly, all results are compatible.
Actually, compared individually, the CSM and lQCD results for  ${\mathpzc u}_K(x ;\zeta_5)$, ${\mathpzc u}_K(x ;\zeta_5)$ are quite different.
Evidently, therefore, the ratio is a forgiving measure and data on the individual DFs would provide a far keener tool for discriminating between pictures of kaon structure.
This panel also displays CSM predictions for analogous glue and sea ratios.

The right panel of Fig.\,\ref{fig:PDFKevol} contrasts the valence-quark DFs within the kaon and pion.  In this calculation \cite{Cui:2020tdf}, recognizing that gluon splitting must produce less heavy $s+\bar s$ pairs than light $u+\bar u$ pairs and heavy $\bar s$ quarks should produce less gluons via bremsstrahlung, quark current-mass threshold factors were included in the evolution kernel.  As a consequence, when compared with mass-independent evolution results, low-order Mellin moments of the $\bar s$-in-$K$ valence DF are increased by $4.8(8)$\% and the glue moments are commensurately smaller.

Before closing this section, we list CSM predictions for low-order Mellin moments of valence parton DFs in the pion and kaon:
\begin{equation}
    \begin{array}{l|ll|ll}
         &  \pi_u & & K_{\mathpzc u} & K_{\bar{\mathpzc s}} \\\hline
         & \zeta_2 & \zeta_5 & \zeta_5 & \zeta_5 \\ \hline
      \langle x\rangle  & 0.24(2) & 0.21(2) & 0.19(2) & 0.23(2) \\
      \langle x^2\rangle & 0.094(13)  & 0.074(10) & 0.067(09) & 0.085(20) \\
      \langle x^3\rangle & 0.047(08) & 0.035(06) & 0.030(08) & 0.070(12)\\ \hline
    \end{array}
\end{equation}
These predictions may be viewed as benchmarks for phenomenology.  Existing approaches to fitting relevant data typically see such phenomenology place too much momentum in the sea with the cost paid by the valence fraction \cite{Cui:2021mom}.

Finally, comparisons between pion and proton DFs are drawn in Ref.\,\cite{Lu:2022cjx}.  Notably, in all cases, QCD-connected CSM predictions for evolved DFs comply with SM constraints on the large-$x$ behavior.

\section{Electromagnetic and gravitational form factors}
\label{Sec:FFs}
\subsection{Hard-scattering formulae and scaling violations}
Electromagnetic form factors (EFFs) also provide opportunities for the examination of diverse aspects of hadron internal structure.
Obvious examples are the charge and magnetization distributions, but it goes much further than that.
For instance, such EFFs of pseudoscalar mesons present an ideal platform for testing fundamental QCD predictions, since rigorous connections have been drawn between them and the DAs discussed above \cite{Lepage:1979zb, Efremov:1979qk, Lepage:1980fj}.

Focusing first on pseudoscalar meson elastic and transition electromagnetic form factors (EFFs and TFFs), QCD predicts the following behavior.
\smallskip

\noindent \underline{Elastic}: $\gamma^\ast(Q) \textbf{P} \to \textbf{P}$,
\begin{align}
Q^2 F_{\textbf{P}}(Q^2) \stackrel{Q^2 \gg m_p^2}{\approx} 16 \pi \alpha_s(Q^2) f_\textbf{P}^2 \mathpzc{w}_\textbf{P}^2(Q^2)\,,
\label{FPUV}
\end{align}
where $\alpha_s$ is the one-loop strong running coupling, which agrees with $\hat\alpha$ in Fig.\,\ref{fig:alphaPI} on the applicable domain, 
$f_\pi = 0.092\,$GeV, $f_K=0.110\,$GeV,
and $\mathpzc{w}_\textbf{P}^2 = e_{\bar q} \mathpzc{w}_{\bar q}^2(Q^2) + e_{u}\mathpzc{w}_u^2(Q^2)$,
\begin{equation}
\label{weightings}
\mathpzc{w}_{f} = \tfrac{1}{3}\int_0^1 dx\, \mathpzc{g}_f(x) \,\varphi_M(x;Q^2) \,,
\end{equation}
$\mathpzc{g}_u(x) = 1/x$, $\mathpzc{g}_{\bar q}(x) = 1/(1-x)$,  $e_{u}=2  e_{\bar q} = (2/3)$,  $\bar q = \bar s$ ($K^+$) or $\bar d$ ($\pi^+$).
The $\pi^0$ elastic form factor is identically zero owing to charge conjugation invariance; and a prediction for the neutral kaon is obtained via $e_u \to e_d = (-1/3)$.
\smallskip

\noindent \underline{Transition}: $\gamma^\ast(Q) \gamma \to \textbf{P}^0$, considering any $q\bar q$ component of $\textbf{P}^0$,
\begin{equation}
\label{EqHardScattering}
Q^2 G_\textbf{P}^q(Q^2) \stackrel{Q^2 \gg m_p^2}{\approx} 12 \pi^2 \, f_{\textbf{P}}^q\, {\mathpzc e}_q^2\, \mathpzc{w}_q(Q^2),
\end{equation}
where:
$f_\textbf{P}^q$ is the $q\bar q$-component contribution to the pseudovector projection of the meson's wave function onto the origin in configuration space, \emph{i.e}., a leptonic decay constant; and ${\mathpzc e}_q$ is the electric charge of quark $q$.
The complete transition form factor is obtained as a sum over the various $q\bar q$ subcomponent contributions:
\begin{equation}
G_\textbf{P} = \sum_{q\in \textbf{P}} \psi_\textbf{P}^q G_\textbf{P}^q,
\end{equation}
where $\psi_\textbf{P}^q$ is a flavour weighting factor originating in the meson's wave function.

It is made plain by Eq.\,\eqref{FPUV} that QCD is not seen in EFF $Q^2$-scaling, but in the violations of scaling that reveal the character of the running coupling and evolution of the DA.  Regarding TFFs, scaling violations are also evident in the leading order result, Eq.\,\eqref{EqHardScattering}, through the evolution of the DA.
Importantly, the absolute magnitude of either the EFF or TFF on the ultraviolet domain is set by the leptonic decay constant of the meson involved.  This quantity is an order parameter for DCSB; hence, a measure of EHM.
In addition, when considering the neutral pion TFF in the neighbourhood of the chiral limit, one has \cite{Adler:1969gk, Bell:1969ts, Adler:2004ih, Holl:2005vu}
\begin{equation}
    2 f_\pi^0 G_{\pi^0}^0(Q^2=0) = 1\,.
\end{equation}
Thus EHM sets the infrared scale as well; and deviations from this result for other (heavier) mesons are a measure of EHM + HB interference.

The longstanding question is: \smallskip

\hspace*{0.05\textwidth}\parbox[c]{0.85\textwidth}{At what value of $Q^2 \gg m_p^2$ do Eqs.\,\eqref{FPUV}, \eqref{EqHardScattering} begin to serve as good approximations, \textit{i.e}., how hard is hard for exclusive processes?}

\smallskip

Forty years of experiment and theory have shown that if $\varphi_{\rm asy}(x)$ is used in these equations, then that domain is beyond the reach of terrestrial experiments.
This opens up the possibility that existing and foreseeable EFF and TFF measurements might be interpreted as placing constraints on the $1/x$ moment of meson DAs.
Such potential is the subject of much debate and analysis -- see, \textit{e.g}., Refs.\,\cite{Raya:2015gva, Raya:2016yuj, Gao:2017mmp, Stefanis:2020rnd, Chen:2020ijn, Arrington:2021biu}.  The $\eta$ and $\eta^\prime$ TFFs are of additional interest because they can expose observable consequences of the non-Abelian anomaly and topological effects within hadrons \cite{Christos:1984tu, Bhagwat:2007ha, Ding:2018xwy}.

\subsection{Electromagnetic elastic form factors}
Consider the elastic process $\gamma^*(Q)\textbf{P}\to \textbf{P}$. At leading order in the systematic, symmetry-preserving DSE truncation scheme \cite{Munczek:1994zz, Bender:1996bb}, the amplitude for this process is expressed as follows \cite{Roberts:1994hh, Maris:2003vk, Chang:2013nia, Chen:2018rwz, Xu:2023izo}:
\begin{subequations}
\label{RLFKq}
\begin{align}
F_\textbf{P}(Q^2) & = e_q F_\textbf{P}^q(Q^2) + e_{\bar h} F_\textbf{P}^{\bar h}(Q^2)\,, \label{FKFuFs}\\
P_\mu F_\textbf{P}^q(Q^2) & = {\rm tr}_{\rm CD} \int_{dk}^\Lambda
\chi_\mu^q(k+p_o,k+p_i) \Gamma_\textbf{P}(k_i;p_i)\,S_h(k)\,\Gamma_\textbf{P}(k_o;-p_o)\,,  \label{RLAmp1}
\end{align}
\end{subequations}
with a similar expression for $F_\textbf{P}^{\bar h}(Q^2)$, where $Q$ is the in\-coming photon momentum, $p_{o,i} = P\pm Q/2$, $k_{o,i}=k+p_{o,i}/2$, $p_{o,i}^2 = -m_\textbf{P}^2$, $m_\textbf{P}$ is the meson mass.  The calculation also requires quark propagators, $S_f$, $f=q$, $h$,
which, consistent with Eq.\,\eqref{RLAmp1}, should be obtained from the rainbow-truncation gap equation; the meson Bethe-Salpeter amplitude, $\Gamma_\textbf{P}$, computed in rainbow-ladder truncation; and consistent unamputated dressed-quark-photon vertices, $\chi_\mu^f$.

\begin{figure}[t]%
\centering
\begin{tabular}{lr}
\includegraphics[width=0.48\textwidth]{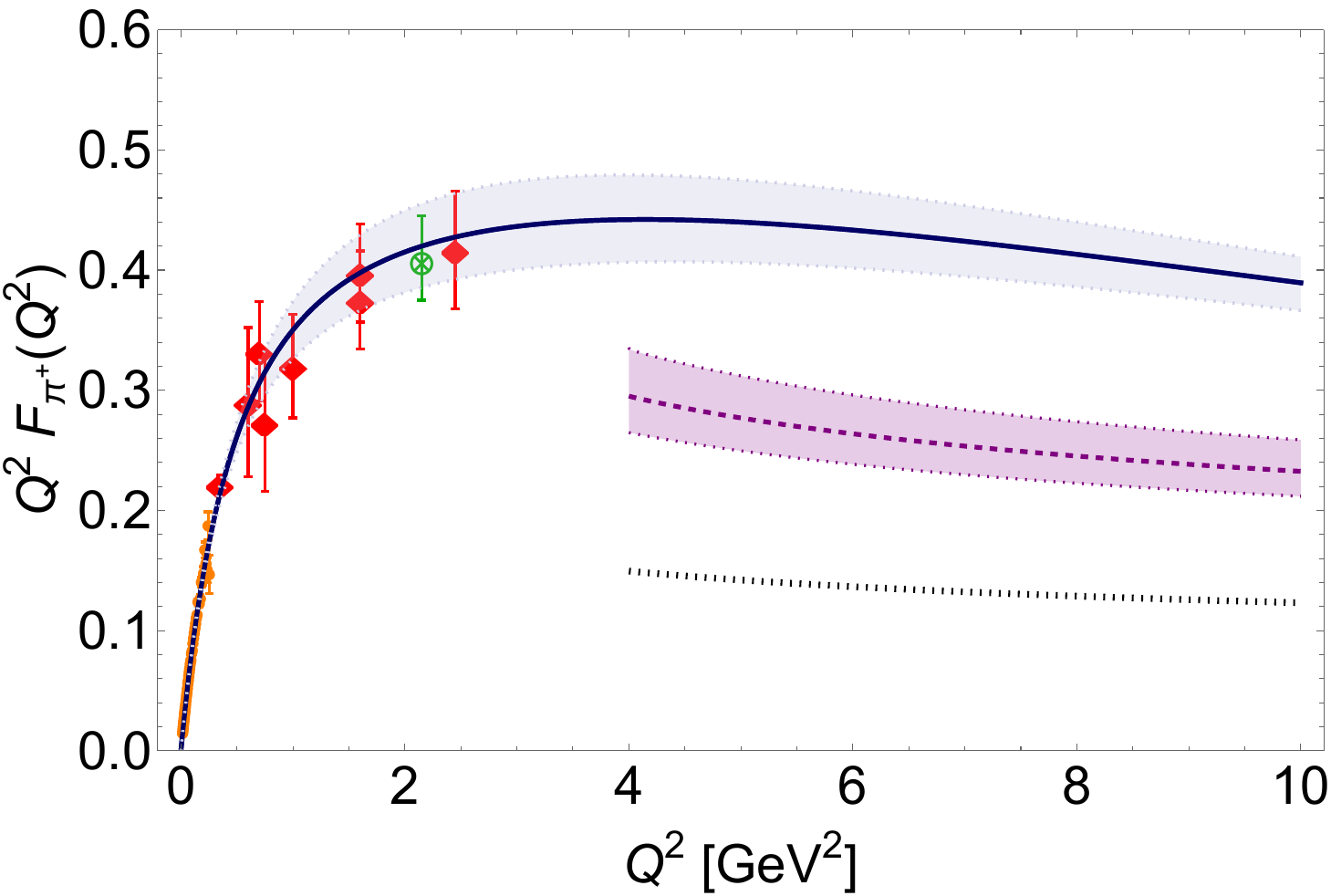} &
\includegraphics[width=0.48\textwidth]{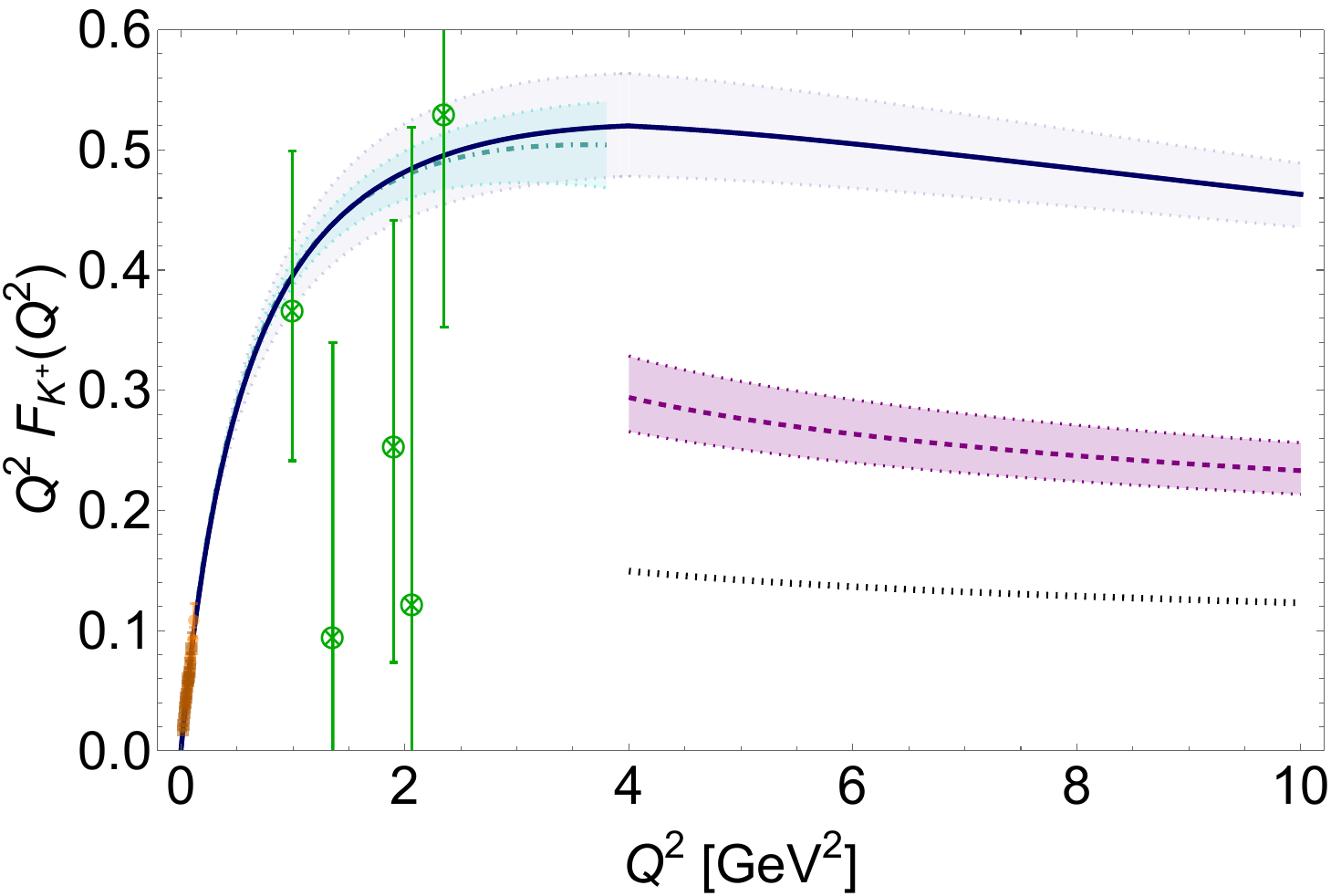}
\end{tabular}
\caption{[left] $\pi^+$ and [right] $K^+$ EFFs \cite{Xu:2023izo}.
Results obtained using Eq.\,\eqref{FPUV}: broadened pion DA, dashed purple curve and like-colored band; and $\varphi_{\rm asy}(x)$, Eq.\,\eqref{eq:PDAasy}, dotted black curve.
Pion data from Refs.\,\cite{NA7:1986vav, Horn:2007ug, JeffersonLab:2008jve}; and kaon data from Refs.\,\cite{Dally:1980dj, Amendolia:1986ui, Carmignotto:2018uqj}.
[right] Dot-dashed cyan curve within like-colored band: $K^+$ lQCD result \cite{Davies:2018zav}.
\label{fig:EFF}}
\end{figure}

Parameter-free CSM predictions for the $\pi^+$ and $K^+$ EFFs \cite{Chang:2013nia, Xu:2023izo}, calculated from Eq.\,\eqref{RLFKq}, are displayed in Fig.\,\ref{fig:EFF}. Plainly, where precise data are available \cite{Horn:2007ug, JeffersonLab:2008jve}, they agree with the CSM predictions.  Importantly, unlike many other approaches, CSMs deliver QCD-connected predictions on the entire domain of spacelike momentum transfer.  Timelike momenta are also accessible -- see, \textit{e.g}., Ref.\,\cite{Cloet:2008fw}, but this requires careful treatment of resonance contributions to the photon-quark vertex.

Considering the low-$Q^2$ domain, the CSM results deliver predictions for charged pion and kaon radii via the usual definition:
\begin{equation}
    r_\textbf{P}^2 = \left. - 6 \frac{d F_\textbf{P}(Q^2)}{dQ^2}\right|_{Q^2=0} .
\end{equation}
Using the curves in Fig.\,\ref{fig:EFF}, one obtains $r_\pi = 0.64(2)\,$fm, $r_K \approx 0.9\,r_\pi$.  These values are consistent with modern determinations \cite{Cui:2021aee}.

Given the range accessible to CSM analyses, it is possible to develop an answer to the question ``how hard is hard?''\  Focusing first on the pion, consider Fig.\,\ref{fig:EFF}\,--\,left.
Precise higher-$Q^2$ data are available \cite{Horn:2007ug, JeffersonLab:2008jve}: in comparison with the CSM prediction, one finds $\chi^2/{\rm datum} = 1.0$.
Extending beyond the range of extant data, the CSM prediction follows a monopole function, whose scale is fixed by the pion charge radius, until $Q^2 \approx 6\,$GeV$^2$.
Thereafter, the CSM prediction breaks away from the simple scaling result, trending below with a separation that grows as $Q^2$ increases.
This onset of scaling violation is the signal for QCD in hard exclusive scattering; and based on the anticipated precision of forthcoming data, experiments that probe above $Q^2 \approx 9\,$GeV$^2$ will be sensitive to this signal \cite{Chen:2020ijn, Roberts:2021nhw, Arrington:2021biu}.

Figure~\ref{fig:EFF}\,--\,left also includes the result obtained from Eq.\,\eqref{FPUV} by using a dilated pion DA of the type displayed in Fig.\,\ref{fig:PDA1} (dashed purple curve) and that produced by $\varphi_{\rm asy}(x)$ in Eq.\,\eqref{eq:PDAasy} (dotted black curve).  Evidently, the broadened DA provides semi-quantitative agreement with the CSM prediction.  Indeed, the quantitative difference between these curves may be explained by a combination of higher-order and -twist corrections to Eq.\,\eqref{FPUV} in perturbative QCD on the one hand, and shortcomings in the rainbow-ladder truncation, which predicts the correct power-law behaviour for the form factor but not precisely the right anomalous dimension in the strong coupling calculation, on the other hand.
Empirical support for the broadened pion DA is also found in analyses of pion + proton Drell-Yan data \cite{Xing:2023wuk}.

The CSM prediction for the charged kaon EFF is drawn in Fig.\,\ref{fig:EFF}\,--\,right.  
At this time, on the entire domain, precise data are lacking \cite{Cui:2021aee, Cui:2022fyr}, but that is expected to change in the foreseeable future \cite{Chen:2020ijn, Roberts:2021nhw, Arrington:2021biu}.  
Nevertheless, the figure reveals that 
scaling violations should be visible in the charged kaon EFF on $Q^2 \gtrsim 6\,$GeV$^2$.

\subsection{Two-photon transition form factors}
The TFFs of interest are obtained from the following amplitude:
\begin{eqnarray}
\hspace{-0.26cm} \mathcal{T}_{\mu\nu}(k_1,k_2) &=& \mathcal{T}_{\mu\nu}(k_1,k_2) + \mathcal{T}_{\nu\mu}(k_2,k_1) = \frac{e^2}{4\pi^2}\epsilon_{\mu \nu \alpha \beta} k_{1\alpha}k_{2\beta}G_{\textbf{P}}(k_1^2,k_2^2,k_1 \cdot k_2) \,,
\end{eqnarray}
where the momentum of the meson is $P=k_1+k_2$, with $k_1$, $k_2$ the incoming photon momenta. At leading-order in the most commonly used CSM truncation (rainbow-ladder) \cite{Raya:2015gva}:
\begin{eqnarray}
    \mathcal{T}_{\mu\nu}(k_1,k_2)
    =e^2\mathcal{Q}^2_{\textbf{P}}\mbox{tr} \int_{l} i\chi_{\mu}(l,l+k_1)\Gamma_{\textbf{P}}(l+k_1,l-k_2)  S(l-k_2)i \Gamma_{\nu}^q(l-k_2,l) \,,\label{eq:Tmunu}
\end{eqnarray}
where 
$\mathcal{Q}^2_{\pi,\,\eta_c,\,\eta_b}=\{1/3 ,\,4/9,\,1/9\}$.
(The $\eta$, $\eta^\prime$ cases require some adjustments owing to the non-Abelian anomaly \cite{Ding:2018xwy}.)
Placing one of the photons on-shell, then $k_1^2=Q^2$, $k_2^2=0$, $2k_1\cdot k_2=-(m_{\textbf{P}}^2+Q^2)$.

\begin{figure}[t]%
\centering
\begin{tabular}{lr}
\includegraphics[width=0.48\textwidth]{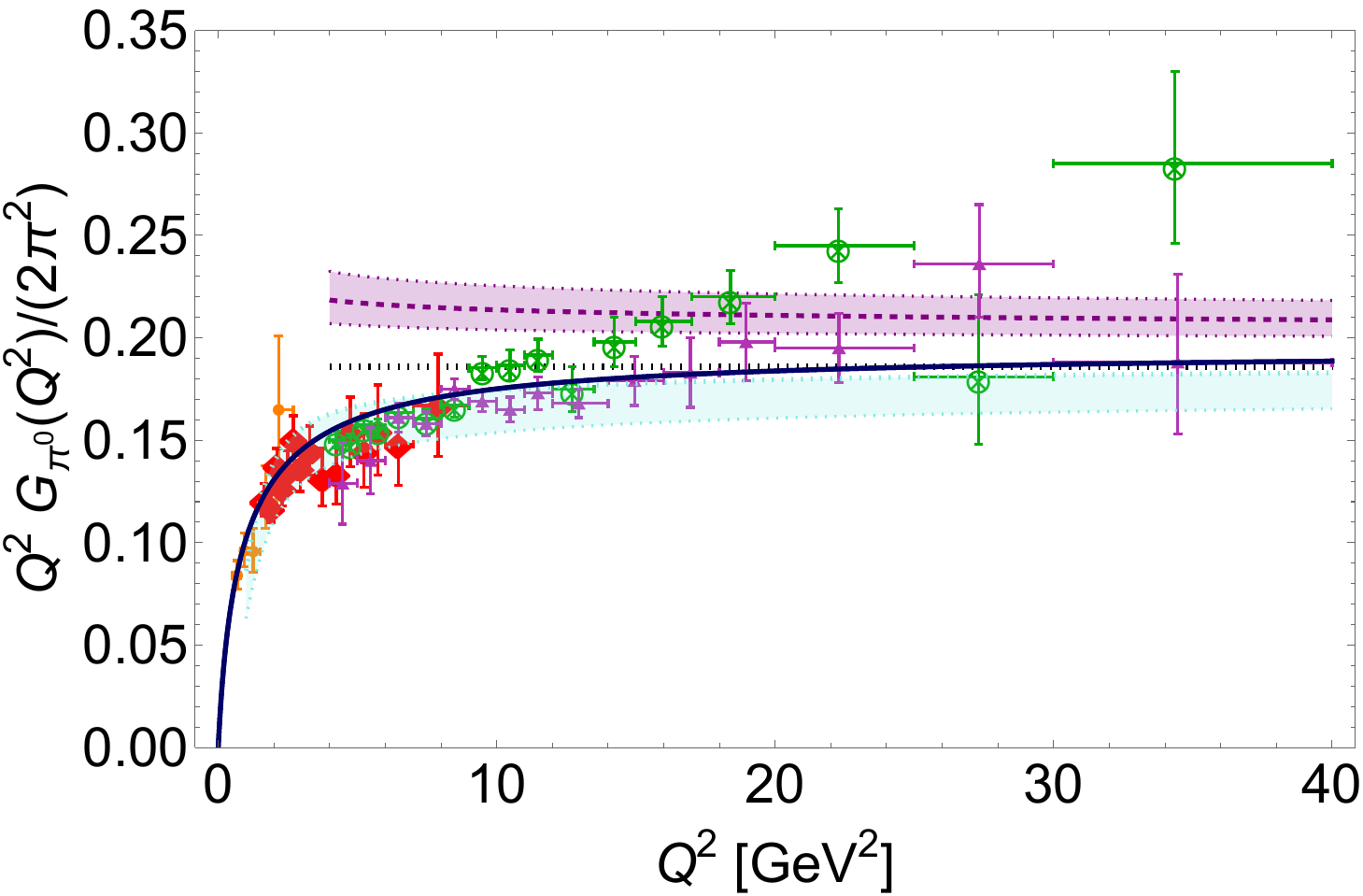} &
\includegraphics[width=0.48\textwidth]{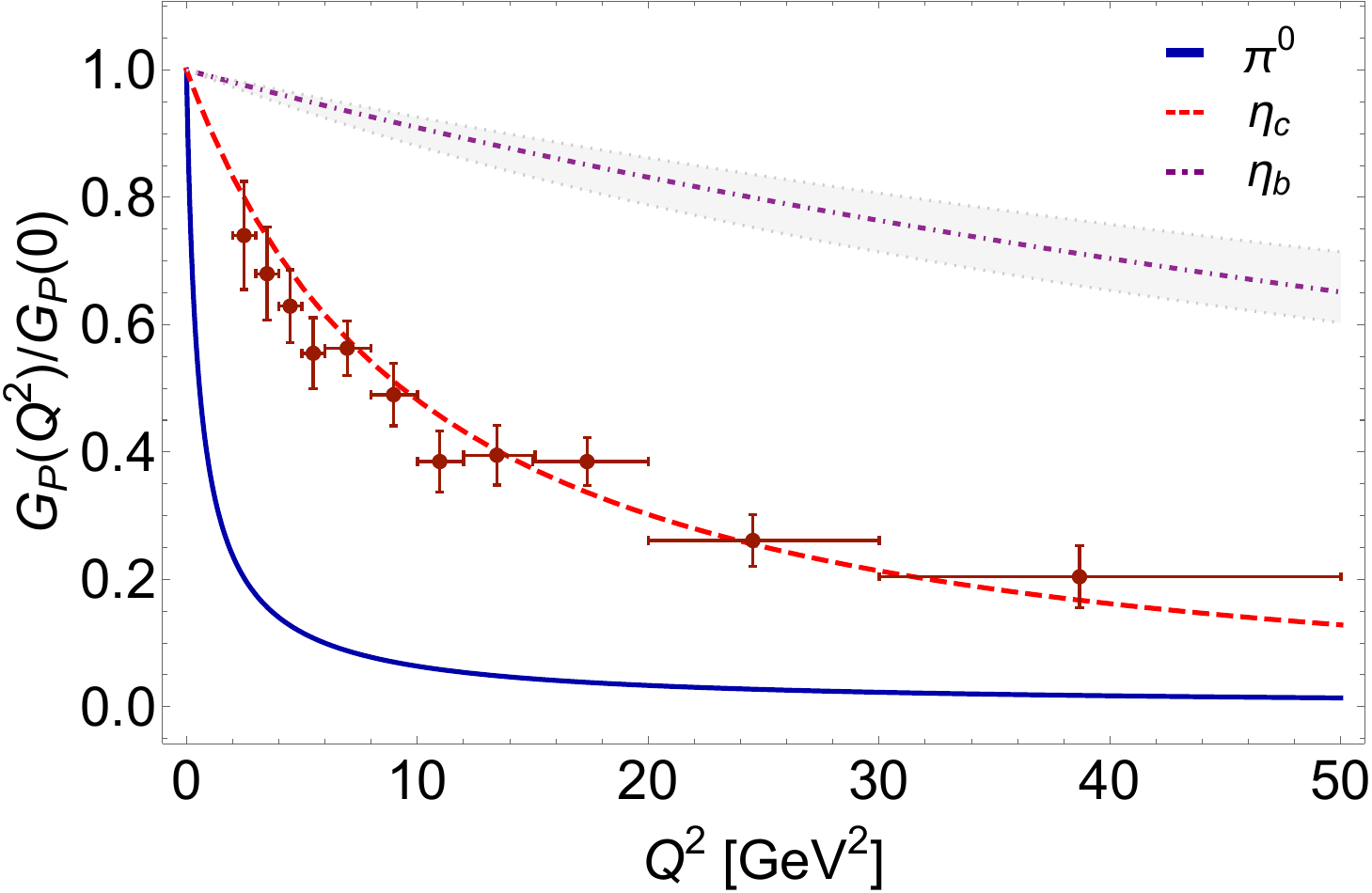}
\end{tabular}
\caption{$\gamma^*(Q)\gamma\to \textbf{P}$ TFFs.
[left] $\gamma^*\gamma\to \pi^0$.
Solid blue curve, CSM prediction \cite{Raya:2015gva, Ding:2018xwy, Raya:2019dnh};
dashed purple curve within like-colored band, Eq.\,\eqref{EqHardScattering} evaluated using the broad, concave pion DA drawn in Fig.\,\ref{fig:PDA1};
dotted black line, Eq.\,\eqref{EqHardScattering} evaluated using $\varphi_{\rm asy}(x)$, Eq.\,\eqref{eq:PDAasy}, which produces $Q^2 G_{\pi^0}(Q^2)/[2\pi^2] = 2 f_\pi$.
The light-blue band represents the analysis in Ref.\,\cite{Bakulev:2012nh}.
Data \cite{CELLO:1990klc, PhysRevD.57.33, PhysRevD.80.052002, PhysRevD.86.092007}.
[right] CSM predictions for $\gamma^*(Q)\gamma\to \{\pi^0, \eta_c, \eta_b\}$ TFFs \cite{Raya:2015gva, Raya:2016yuj}.
The shaded band bracketing the $\eta_b$ result derives from the non-relativistic QCD analysis in Ref.\,\cite{Feng:2015uha}.
Data \cite[Babar]{BaBar:2010siw}.
\label{fig:TFF}}
\end{figure}

CSM predictions for such TFFs are drawn in Fig.\,\ref{fig:TFF}.
Once again, we focus first on the $\pi^0$ case.
This is interesting because available data extend to $Q^2 \gg m_p^2$ \cite{CELLO:1990klc, PhysRevD.57.33, PhysRevD.80.052002, PhysRevD.86.092007};
all data agree on $Q^2\lesssim 10\,$GeV$^2$ and are compatible with Eq.\,\eqref{EqHardScattering};
but, thereafter, the two available sets, Ref.\,\cite[BaBar]{PhysRevD.80.052002} and Ref.\,\cite[Belle]{PhysRevD.86.092007}, display conflicting trends in their evolution with photon virtuality.
This conflict has attracted much attention -- see, e.g., Refs.\,\cite{Stefanis:2012yw, Raya:2015gva, Nedelko:2016vpj, Eichmann:2017wil, Choi:2020xsr, Stefanis:2020rnd, Zhou:2023ivj}.

It is worth reiterating that the CSM analysis in Ref.\,\cite{Raya:2015gva} generates a broad, concave pion DA of the type drawn in Fig.\,\ref{fig:PDA1}; expresses the hard-QCD limit, Eq.\,\eqref{EqHardScattering}; and, as we shall discuss briefly below, also delivers a unification of $\gamma^\ast \gamma \to \eta, \eta^\prime, \eta_c, \eta_b$ transition form factors \cite{Raya:2016yuj, Ding:2018xwy}.  Hence, the following comparisons have weight:
\begin{equation}
\begin{array}{l|c|c|c}
{\rm sources} & \mbox{Refs.\,\cite{CELLO:1990klc, PhysRevD.57.33, PhysRevD.80.052002}}
& \mbox{Refs.\,\cite{CELLO:1990klc, PhysRevD.57.33, PhysRevD.86.092007}}
& \mbox{Refs.\,\cite{CELLO:1990klc, PhysRevD.57.33, PhysRevD.80.052002, PhysRevD.86.092007}} \\\hline
\chi^2/{\rm datum} & 2.97 & 1.78 & 2.34
\end{array}\,.
\end{equation}
Plainly, the BaBar Collaboration data \cite{PhysRevD.80.052002} are not compatible with the CSM prediction, whereas the Belle Collaboration data \cite{PhysRevD.86.092007} match well.
Going further and focusing on data at $Q^2>10\,$GeV$^2$, one finds $\chi^2/{\rm datum} = 4.14$ \cite[BaBar]{PhysRevD.80.052002} and $\chi^2/{\rm datum} = 0.64$ \cite[Belle]{PhysRevD.86.092007}.
These comparisons suggest to us that Eq.\,\eqref{EqHardScattering} is confirmed by the bulk of existing data and, hence, such data support a picture of the pion DA as a broad, concave function at experimentally accessible probe momenta.  This perspective may be tested by new data \cite{Belle-II:2018jsg}.

The $\gamma^*(Q)\gamma\to \{\eta_c,\eta_b\}$ TFFs are drawn in Fig.\,\ref{fig:TFF}\,--\,right and compared with their $\pi^0$ analogue, all normalized to unity at $Q^2=0$.
The parameter-free CSM prediction agrees well with the Ref.\,\cite[BaBar]{BaBar:2010siw} data.
There is no data on $\gamma^*(Q)\gamma\to \eta_b$, but a result is available from an analysis made using a non-relativistic QCD effective field theory \cite{Feng:2015uha}.  This result and the CSM prediction are in excellent agreement.

The impact of HB couplings into QCD on these TFFs is readily apparent in Fig.\,\ref{fig:TFF}\,--\,right.
As the HB generated quark current-mass increases, the associated TFF falls more slowly.
This effect is manifest in the TFF radii: $r_{\pi^0}\approx 0.66(2)\,$fm, $r_{\eta_c}/r_{\pi^0}\approx 0.25$, $r_{\eta_b}/r_{\pi^0} \approx 0.06$.
Consequently, with increasing current-mass, the boundary of the domain upon which the hard-QCD limit, Eq.\,\eqref{EqHardScattering}, delivers a good approximation is pushed ever deeper into the spacelike region.

Notably, the successes of CSMs in describing all neutral pseudoscalar meson TFFs \cite{Raya:2015gva, Raya:2016yuj, Ding:2018xwy, Raya:2019dnh}, have made the approach a credible contributor in matters related to the anomalous magnetic moment of the muon\,\cite{Aoyama:2020ynm}.

\subsection{Gravitational form factors}
The interaction of a pseudoscalar meson with a $J=2$ probe is characterized by the current:
\begin{equation}
    \label{eq:matElemGFF}
    \mathbf{\Lambda}_{\mu\nu}(Q,P) = 2P_\mu P_\nu \theta_2^{\textbf{P}}(Q^2)+ \frac{1}{2}[Q^2 \delta_{\mu\nu}-Q_\mu Q_\nu]\theta_1^{\textbf{P}}(Q^2) + 2 m_{\textbf{P}}^2 \delta_{\mu\nu} \bar{c}^{\textbf{P}}(Q^2)\,,
\end{equation}
which corresponds to the in-meson expectation value of the QCD energy-momentum tensor \cite{Polyakov:2018zvc}.
Here $\theta_{2,1}^{\textbf{P}}$ are the meson gravitational form factors (GFFs) associated with the mass and pressure distributions.
Symmetries impose the following relations:
\begin{equation}
\label{eq:consGFF}
    \theta_2^{\textbf{P}}(0) = 1\,,
    \quad \theta_1^{\textbf{P}}(0) \overset{m_{\textbf{P}}^2=0}{=} 1\,,
    \quad \bar{c}^{\textbf{P}}(Q^2) = 0\,.
\end{equation}
The first identity is a statement of mass normalization;
the second is connected with a soft-pion theorem \cite{Polyakov:2018zvc, Mezrag:2014jka}, a corollary of EHM;
and the third is a direct consequence of energy-momentum conservation, \emph{i.e.}, $Q_{\mu} \mathbf{\Lambda}_{\mu\nu}(Q,P) \equiv 0 \equiv Q_\nu \mathbf{\Lambda}_{\mu\nu}(Q,P)$.

The GFFs  defined in Eq.\,\eqref{eq:matElemGFF} are the result of adding up the individual contributions of each type of parton.
Working at $\zeta_{\cal H}$, however, only fully dressed valence quarks play a role and $\mathcal{F}_\textbf{P}=\mathcal{F}_\textbf{P}^q + \mathcal{F}_\textbf{P}^{\bar{h}}$ ($\mathcal{F}_\textbf{P}=\theta_{2,1}^\textbf{P},\,\bar{c}^\textbf{P}$).
Consequently, the GFFs maybe be extracted from an analogue of Eq.\,\eqref{RLFKq}, obtained by replacing the photon-quark vertex by the ``graviton''-quark vertex \cite{Xu:2023izo}.

\begin{figure}[t]%
\centering
\begin{tabular}{lr}
\includegraphics[width=0.48\textwidth]{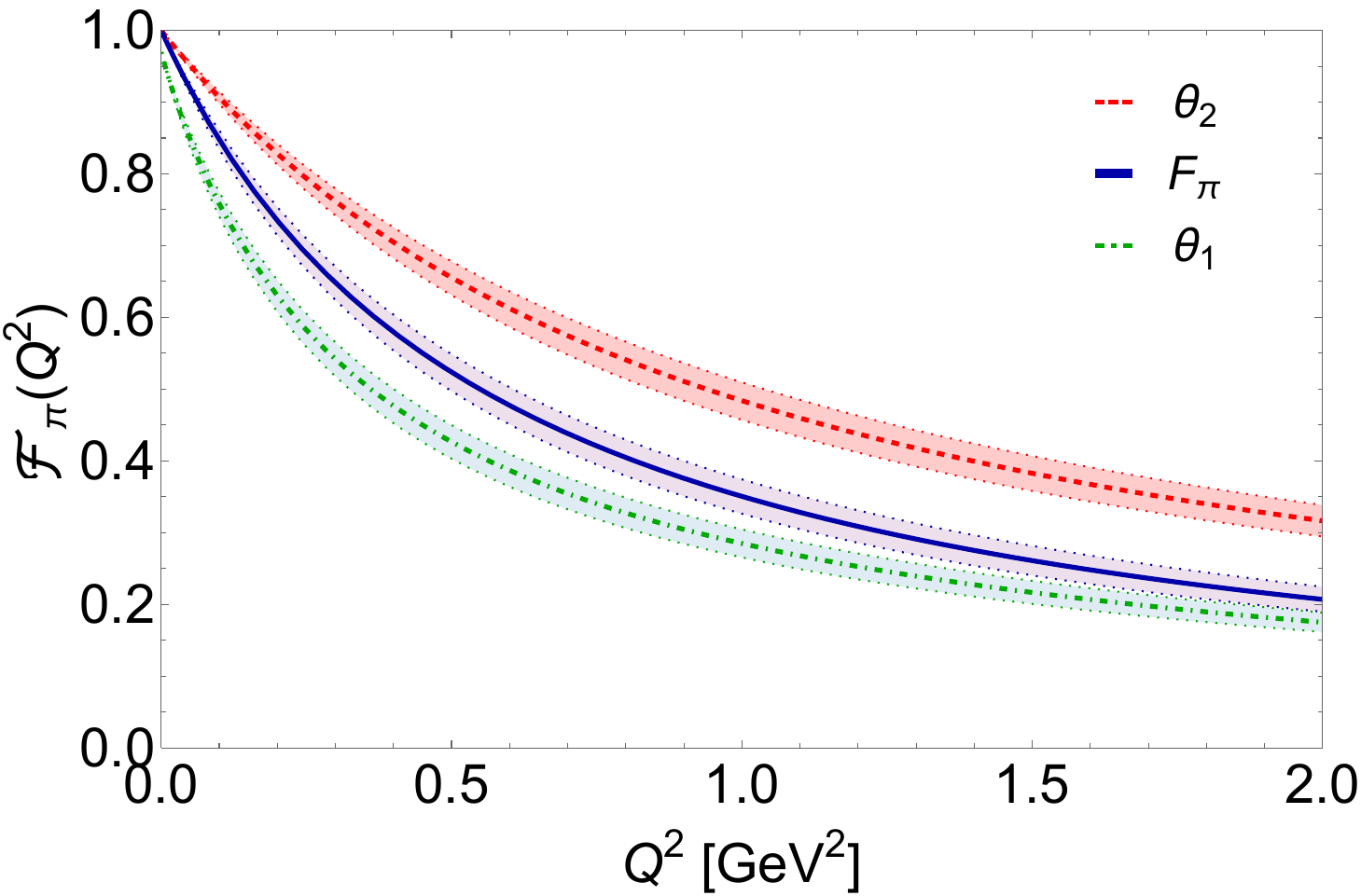} &
\includegraphics[width=0.48\textwidth]{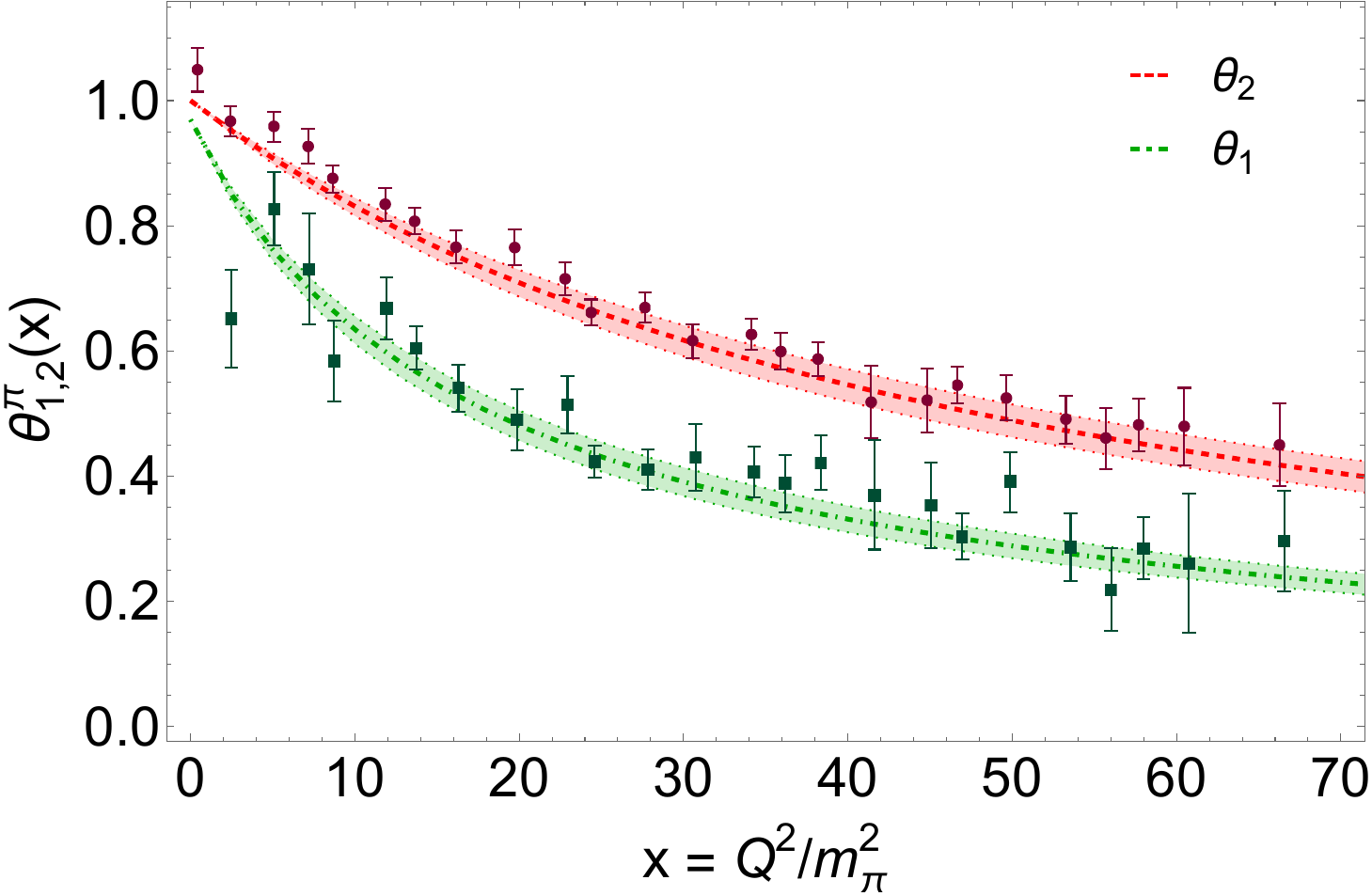}
\end{tabular}
\caption{Charged pion GFFs \cite{Xu:2023izo}.
[left] Electromagnetic ($F_\pi$) and gravitational form factors ($\theta_{2,1}^\pi$). Experimental data on $F_\pi$ included for reference \cite{JeffersonLab:2008jve, NA7:1986vav, Horn:2007ug}.
[right] CSM predictions for $\theta_{2,1}$  compared with recent lQCD results \cite{Hackett:2023nkr}.
As the latter correspond to simulations with $m_\pi \approx 0.170$ GeV, both CSM and lQCD results are displayed in terms of $x=Q^2/m_\pi^2$, constructed using the appropriate value of the pion mass. \label{fig:GFF}}
\end{figure}

CSM predictions for pion GFFs are drawn in Fig.\,\ref{fig:GFF}.
The left panel compares $\theta_{2,1}^\pi$ with $F_\pi$:
$\theta_2$ shows the least rapid decay with $Q^2$,
$F_\pi$ falls faster,
but $\theta_1$ decays most rapidly.
Naturally, therefore, the corresponding radii are ordered as follows:
\begin{equation}
\label{eq:OrdRad}
    r^\pi_{\theta_1}\,\text{(pressure)} \,> \,r^\pi_{F}\,\text{(charge)} \,> \,r^\pi_{\theta_2} \,\text{(mass)}\,;
\end{equation}
in other words, the mass distribution is more compact than the charge distribution ($r^\pi_{\theta_2}/r^\pi_{F} \approx 0.74$), which is, in turn, tighter than the pressure distribution ($r^\pi_{F}/r^\pi_{\theta_1} =0.79$). 
These patterns are also consistent with available lQCD results \cite{Hackett:2023nkr} -- see Fig.\,\ref{fig:GFF}\,--\,right, and have been confirmed empirically \cite{Kumano:2017lhr, Xu:2023bwv}.

It is worth noting here that by exploiting general physical constraints on the form of the pion DF \cite{Cui:2022bxn, Cui:2021mom, Lu:2023yna}, the following bounds can be established: 
\begin{equation}\label{eq:constraints}
\frac 1 {\sqrt{2}} \approx 0.71 \leq \frac{r^\pi_{\theta_2}}{r^\pi_{F}} \leq 1 \;.
\end{equation}
The results above are consistent with these limits.

Kaon FFs exhibit similar profiles \cite{Xu:2023izo}.
Naturally, being heavier than the pion, the kaon is more compact when judged by the measures discussed herein.
Quantitatively, averaging all relevant radii, one finds $r_K/r_\pi = 0.85(6)$.

Gravitational form factors are also accessible via generalized parton distributions (GPDs)  \cite{Belitsky:2005qn, Mezrag:2022pqk, Mezrag:2023nkp}, which themselves will be discussed below.  However, following that route, the pressure distribution, $\theta_1$, suffers from the so-called $D$-term ambiguity \cite{Chouika:2017rzs}.  This is avoided when one calculates the form factors directly from the analogue of Eq.\,\eqref{RLFKq} and this makes $\theta_1$ especially interesting.
The CSM analysis in Ref.\,\cite{Xu:2023izo} predicts 
\begin{equation}
\theta_1^\pi(0) = 0.97\,, \quad \theta_1^K(0) = 0.77(10)\,,     
\end{equation}
both in agreement with estimates made using chiral effective field theory \cite{Polyakov:2018zvc}.
The deviations from Eq.\,\eqref{eq:consGFF} are a measure of HB modulation of EHM in the systems considered: the magnitudes match expectations.

\begin{figure}[t]%
\centering
\begin{tabular}{lr}
\includegraphics[width=0.48\textwidth]{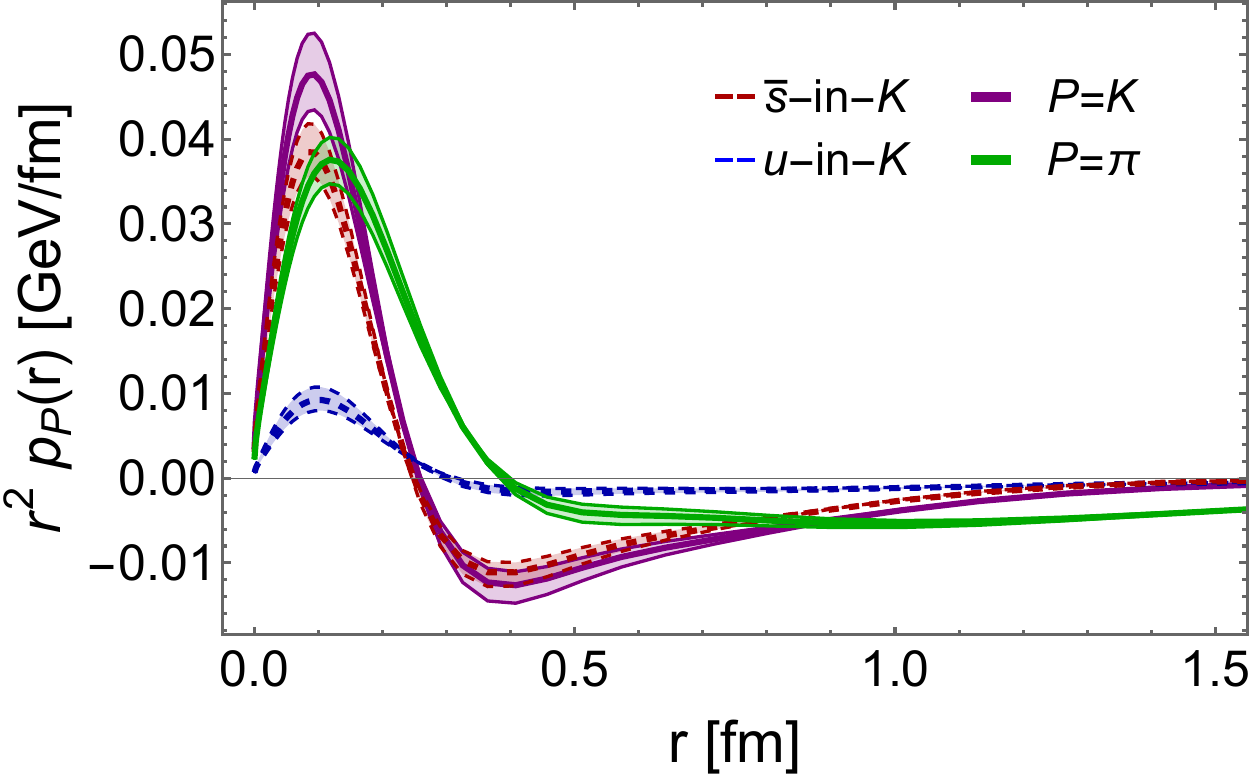} &
\includegraphics[width=0.48\textwidth]{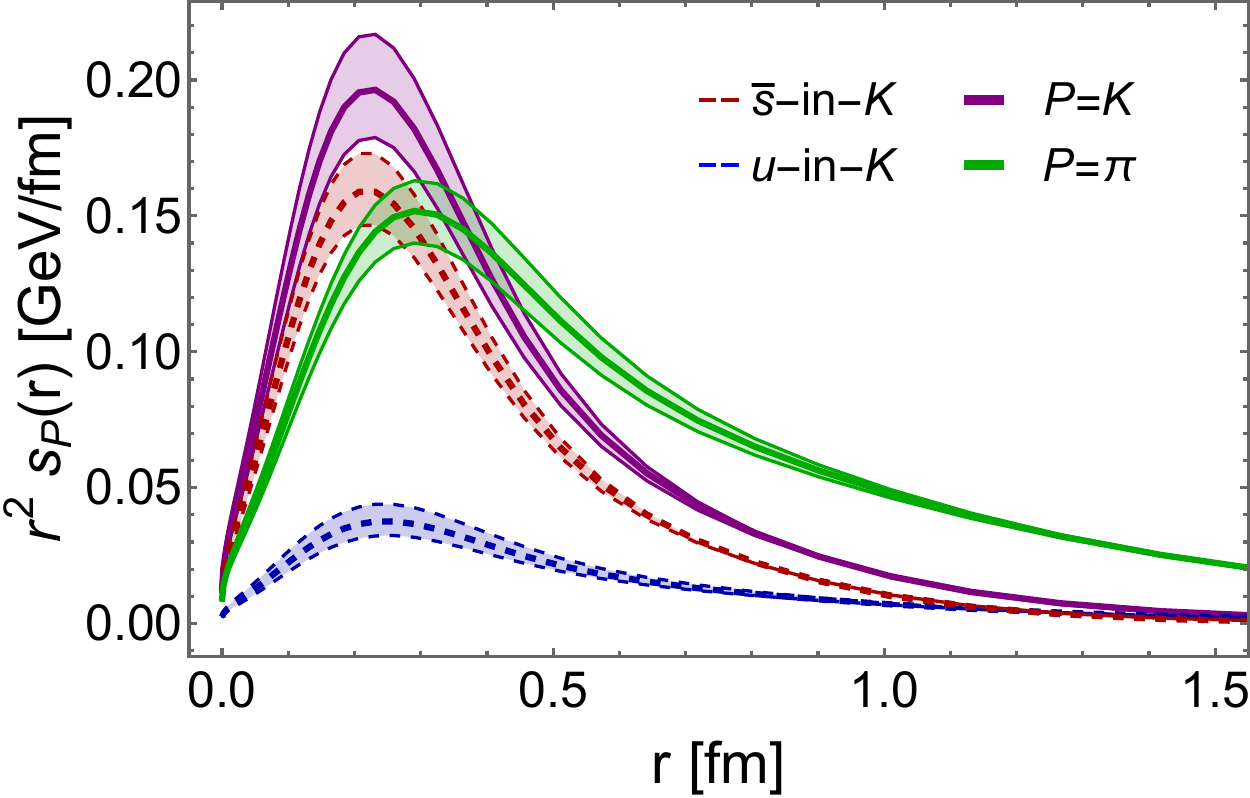}
\end{tabular}
\caption{[left] Pressure and [right] shear force distributions, calculated using Eqs.\,\eqref{eq:PressureAB} -- see Ref.\,\cite{Xu:2023izo}. \label{fig:Press}}
\end{figure}

Importantly, too, pressure and shear profiles -- $p_\textbf{P}(r)$ and $s_\textbf{P}(r)$, respectively, can be obtained from $\theta_1$ as follows \cite{Polyakov:2018zvc}:
\begin{subequations}
\label{eq:PressureAB}
\begin{align}
p_\textbf{P}(r)  & =
 \frac{1}{6\pi^2 r} \int_0^\infty d\Delta \,\frac{\Delta}{2 E(\Delta)} \, \sin(\Delta r) [\Delta^2\theta_1^{\textbf{P}}(\Delta^2)] \,, \label{eq:PressureA} \\
  s_\textbf{P} (r)  & =\frac{3}{8 \pi^2} \int_0^\infty d\Delta \,\frac{\Delta^2}{2 E(\Delta)} \, {\mathpzc j}_2(\Delta r) \, [\Delta^2\theta_1^{\textbf{P}}(\Delta^2)] \,, \label{eq:PressureB}
\end{align}
\end{subequations}
where $2E(\Delta)=\sqrt{4 m_{\textbf{P}}^2+\Delta^2}$ and ${\mathpzc j}_2(z)$ is a spherical Bessel function. 
The pressure is shown in Fig.\,\ref{fig:Press}\,--\,left: it is large and positive at small separations; but at some critical value, $r_c$, it changes sign.
This point may be interpreted as marking the beginning of the domain on which confinement forces become dominant.
The locations are (in fm): $r_{c}^\pi = 0.39(1)$, $r_c^K = 0.26(1)$, $r_{c}^{Ku} = 0.30(1)$, $r_c^{Ks} = 0.25(1)$. 
Notably, this qualitative change occurs when the deformation forces, shown in Fig.\,\ref{fig:Press}\,--\,right, are maximal. 
Finally, it is worth highlighting that the meson core pressures are commensurate with those in neutron stars \cite{Raya:2021zrz, Ozel:2016oaf}.

\section{Toward a 3-dimensional picture}
\label{Sec:GPDs}

\subsection{Light-front wave functions}
Light-front wave functions provide probability amplitudes for different parton configurations within a hadron. 
The following light-front projection of a pseudoscalar meson BSWF connects with a leading-twist LFWF:
\begin{equation}
\label{eq:defLFWF}
\psi_{\textbf{P}}^u(x,k_\perp^2;\zeta_{\cal H})=Z_2 \text{tr}_{\text{CD}}  \int \frac{d^2k_\parallel}{\pi}   \delta^x_n(k) \gamma_5 \gamma \cdot n \chi_\textbf{P}(k-P;P_{\textbf{P}})\,.
\end{equation}
Associations with the discussion in Sec.\,\ref{Sec:DAsDFs} are readily made. 

Owing to the open dependence on $k_\perp^2$, explicit calculation of $\psi_{\textbf{P}}^u(x,k_\perp^2;\zeta_{\cal H})$ from Eq.\,\eqref{eq:defLFWF} is more complicated than that of the DA via Eq.\,\eqref{eq:PDA}.  The challenges can be overcome by developing perturbation theory integral representations (PTIRs) of the numerical solutions for $\chi_\textbf{P}(k-P;P_{\textbf{P}})$ -- see, \textit{e.g}., Ref.\,\cite{Shi:2021nvg}. 

Another, simpler approach, which is nevertheless insightful and realistic, can be found in Refs.\,\cite{Xu:2018eii, Raya:2021zrz, Raya:2022eqa}.  In character, it exploits PTIRs, but the representations are less complex.  For instance, one may write 
\begin{eqnarray}
\label{eq:BSWFam}
{\mathpzc n}_\textbf{P} \chi_{\textbf{P}}\left(k_-;P\right) &=&
{\mathpzc M}(k;P)
\int_{-1}^1\,dw\,\rho_{\textbf{P}}(w) {\mathpzc D}(k;P)\,,\\
\nonumber
{\mathpzc M}(k;P)  & =& -\gamma_5[ \gamma\cdot P  M_{q}+ \gamma\cdot k (M_q-M_h) + \sigma_{\mu\nu} k_\mu P_{\nu}] \,,\\
\nonumber {\mathpzc D}(k;P) & =&  \Delta(k^2,M_q^2) \Delta((k-P)^2,M_h^2) \hat\Delta(k_{w-1}^2,\Lambda_{\textbf{P}}^2)\,,
\end{eqnarray}
where 
$M_{q,h}$ are constituent-quark mass-scales, associated with the infrared size of the dressed-quark mass function -- see Fig.\,\ref{fig:MassFuncs}, and $\Lambda_{\textbf{P}}$ is a mass-dimension parameter;
$\Delta(s,t) = 1/[s+t]$, $\hat \Delta(s,t) = t \Delta(s,t) $;
$k_w = k+ (w/2) P_{\textbf{P}}$, with $P^2_{\textbf{P}}=-m_{\textbf{P}}^2$;
$\rho_{\textbf{P}}(w)$ is a spectral density; 
and ${\mathpzc n}_{\textbf{P}}$ ensures canonical normalization.
Introducing two Feynman parameters, ($\alpha$, $v$), the BSWF can be re-expressed as
\begin{align}
\label{X2a}
\chi_{\textbf{P}}\left(k_-, P \right) & = {\mathpzc M}(k;P) \int_0^1 \,d\alpha\,2\, {\mathpzc X}_{\textbf{P}}(\alpha;\sigma^3(\alpha))\,,
\end{align}
with $\sigma(\alpha) = (k-\alpha P)^2+ \Omega_{\textbf{P}}^2$,
\begin{align}
\nonumber
\Omega_{\textbf{P}}^2 & = v M_q^2 + (1-v)\Lambda_{\textbf{P}}^2  + (M_h^2-M_q^2)\left(\alpha - \tfrac{1}{2}[1-w][1-v]\right) \\
& \quad + ( \alpha [\alpha-1] + \tfrac{1}{4} [1-v] [1-w^2]) m_{\textbf{P}}^2\,,
\label{Omega}\\
{\mathpzc X}_{\textbf{P}}(\alpha;\sigma^3) & =
\left[
\int_{-1}^{1-2\alpha} \! dw \int_{1+\frac{2\alpha}{w-1}}^1 \!dv  + \int_{1-2\alpha}^1 \! dw \int_{\frac{w-1+2\alpha}{w+1}}^1 \!dv \right]\frac{\rho_{\textbf{P}}(w) }{{\mathpzc n}_{\textbf{P}} } \frac{\Lambda_{\textbf{P}}^2}{\sigma^3}\,.\label{X2c}
\end{align}

Inserting these expressions into Eq.\,\eqref{eq:defLFWF}, then after a  series of algebraic manipulations, the Mellin moments can be expressed as:
\begin{eqnarray}
 \int_0^1 dx \, x^m\,  \psi_{\textbf{P}}^{q}(x,k_\perp^2;\zeta_{\cal H})
=  \int_0^1 d\alpha \, \alpha^m\, 12[M_q+\alpha(M_h-M_q)]{\mathpzc X}_{\textbf{P}}(\alpha;\sigma_\perp^2)\,;
\label{eq:xnLFWF}
\end{eqnarray}
with $\sigma_\perp=k_\perp^2+\Omega_\textbf{P}^2$. Capitalizing now on uniqueness properties of Mellin moments, one may immediately conclude that 
\begin{eqnarray}
\label{eq:LFWFPTIR}
    \psi_{\textbf{P}}^q(x,k_\perp^2;\zeta_{\cal H}) &=& 12[M_q+x(M_h-M_q)]{\mathpzc X}_{\textbf{P}}(x;\sigma_\perp^2)\,.
\end{eqnarray}

\begin{figure}[t]%
\centering
\begin{tabular}{lr}
\includegraphics[width=0.48\textwidth]{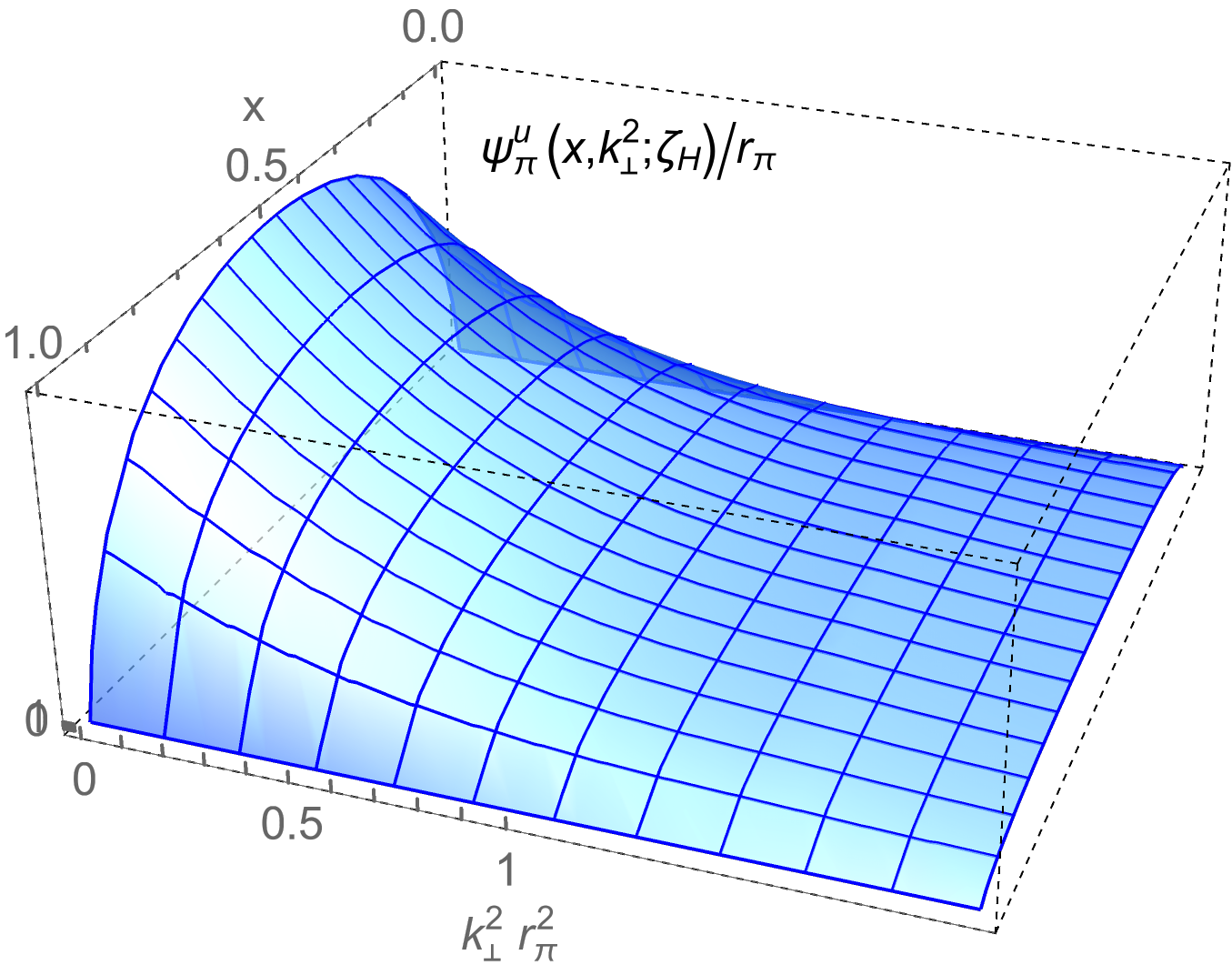} &
\includegraphics[width=0.48\textwidth]{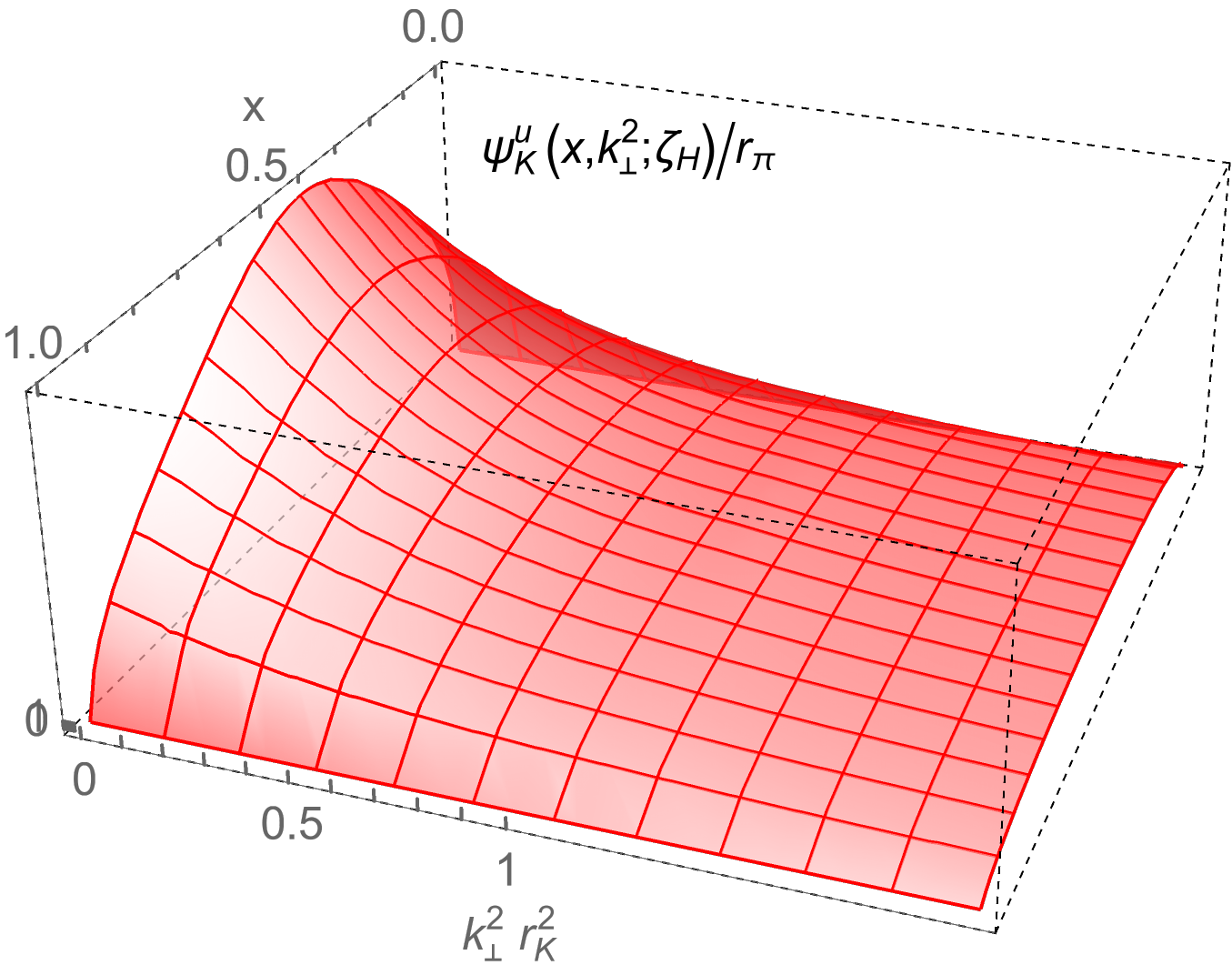}
\end{tabular}
\caption{Leading-twist projected pion [left] and kaon [right] LFWFs at $\zeta_{\cal H}$, obtained using Eqs.\,\eqref{X2a}-\eqref{eq:LFWFPTIR} and informed by the DAs and DFs discussed in Secs.\,\ref{sec:PDAs}, \ref{sec:PDFs}. \label{fig:LFWF}}
\end{figure}

Plainly, the spectral function, $\rho_{\textbf{P}}(w)$, must play a key role in determining the LFWF's profile and, hence, the related DAs and DFs.  Alternately, as highlighted by the discussion in Sec.\,\ref{Sec:DAsDFs}, if one has reliable results for a given meson's DA and/or DF, then those results can be used to determine the spectral density.  This is the 
procedure followed in Refs.\,\cite{Raya:2021zrz, Raya:2022eqa} and the resulting LFWFs are displayed in Fig.\,\ref{fig:LFWF}.

Looking deeper into Eqs.\,\eqref{X2a}-\eqref{eq:LFWFPTIR}, some interesting and useful features are revealed. 
Working in the chiral limit ($M_q = M_h$, $m_\textbf{P}^2=0$) and setting $\Lambda_{\textbf{P}} = M_q$, then one finds $\Omega_{\textbf{P}}^2 = M_q^2$. 
In this case, direct evaluation of all integrals in Eq.\,\eqref{X2c} is possible and, using Eqs.\,\eqref{DFfromLFWFa}, \eqref{DAfromLFWF}, \eqref{rescaleDA2}, one arrives at:
\begin{eqnarray}
\psi_{\textbf{P}}^{q}(x,k_\perp^2;\zeta_{\cal H}) =\tilde{\psi}_{\textbf{P}}^q(k_\perp^2) \tilde\varphi_\textbf{P}^q(x;\zeta_{\cal H}) =\tilde{\psi}_{\textbf{P}}^q (k_\perp^2)\,[{\mathpzc q}_\textbf{P}(x;\zeta_{\cal H})]^{1/2}\,, \label{eq:facLFWF}
\end{eqnarray}
\textit{viz}.\ the $x$-$k_\perp$ dependences factorize and Eq.\,\eqref{DFeqDA2} is recovered.
Focusing on Eq.\,\eqref{Omega}, it is evident that any violation of factorization is tied to $M_h-M_q\neq 0$; hence, is small for modest values of $[M_h-M_q]2/[M_h+M_q]$. 
Similar conclusions were reached in Refs.\,\cite{Albino:2022gzs, Almeida-Zamora:2023bqb}.
Consequently, one may reliably proceed with factorized representations for the LFWFs of light pseudoscalar mesons or, if desiring to express factorization violations, a simple wave function of the following form: 
\begin{subequations}
\begin{align}
    \label{eq:Adnan1}
    \psi_{\textbf{P}}^{q}(x,k_\perp^2;\zeta_{\cal H}) &\sim \frac{ \tilde \varphi_\textbf{P}^q(x;\zeta_{\cal H}) }{[k_\perp^2 + M_q^2+ x(M_h^2-M_q^2)-m_\textbf{P}^2 x(1-x)]^2}\,,
\\
    \label{eq:Adnan2}
    \Rightarrow {\mathpzc q}_\textbf{P}(x;\zeta_{\cal H}) &\sim  \frac{ [\tilde \varphi_\textbf{P}^q(x;\zeta_{\cal H})]^2 }{[M_q^2+ x(M_h^2-M_q^2)-m_\textbf{P}^2 x(1-x)]}\;.
\end{align}
\end{subequations}

\subsection{Generalized parton distributions}
\label{sec:GPDs}
Generalized parton distributions (GPDs) are useful because they connect an array of hadron properties, such as those relating to DFs, EFFs, GFFs, and transverse spatial distributions -- see, \textit{e.g}., Refs.\,\cite{Mezrag:2022pqk, Mezrag:2023nkp}. 
In the case of pseudoscalar mesons at the hadron scale, it is especially convenient to work with the valence-quark GPD obtained via the overlap representation \cite{Diehl:2000xz}:
\begin{eqnarray}
H^q_{\textbf{P}}(x,\xi,-\Delta^2;\zeta_{\cal H}) =  \int \frac{d^2{\bf k_\perp}}{16 \pi^3}
\psi^{q \ast}_{\textbf{P}}\left(\frac{x-\xi}{1-\xi},\hat{\bf k}_\perp^2;\zeta_{\cal H} \right)
\psi^{q}_{\textbf{P}}\left( \frac{x+\xi}{1+\xi},\tilde{\bf k}_\perp^2;\zeta_{\cal H} \right) \,,
\label{eq:overlap}
\end{eqnarray}
where:
$2P=p^\prime + p$, $p^\prime$, $p$ are the final, initial meson momenta in the defining scattering process;
$\Delta = p^\prime-p$, $P\cdot \Delta =0$;
the skewness $\xi =-n\cdot \Delta/n\cdot P$;
$\Delta_\perp^2=\Delta^2(1-\xi^2)-4\xi^2 m_\textbf{P}^2$; 
and 
\begin{eqnarray}
\hat{\bf k}_\perp = {\bf k}_\perp + \frac{1-x}{1-\xi} \frac{{\bf \Delta}_\perp}{2}\,,\,\tilde{\bf k}_\perp = {\bf k}_\perp - \frac{1-x}{1+\xi}
\frac{{\bf \Delta}_\perp}{2} \;.
\label{eq:hatandtilde}
\end{eqnarray}

\begin{figure}[t]%
\centering
\begin{tabular}{lr}
\includegraphics[width=0.48\textwidth]{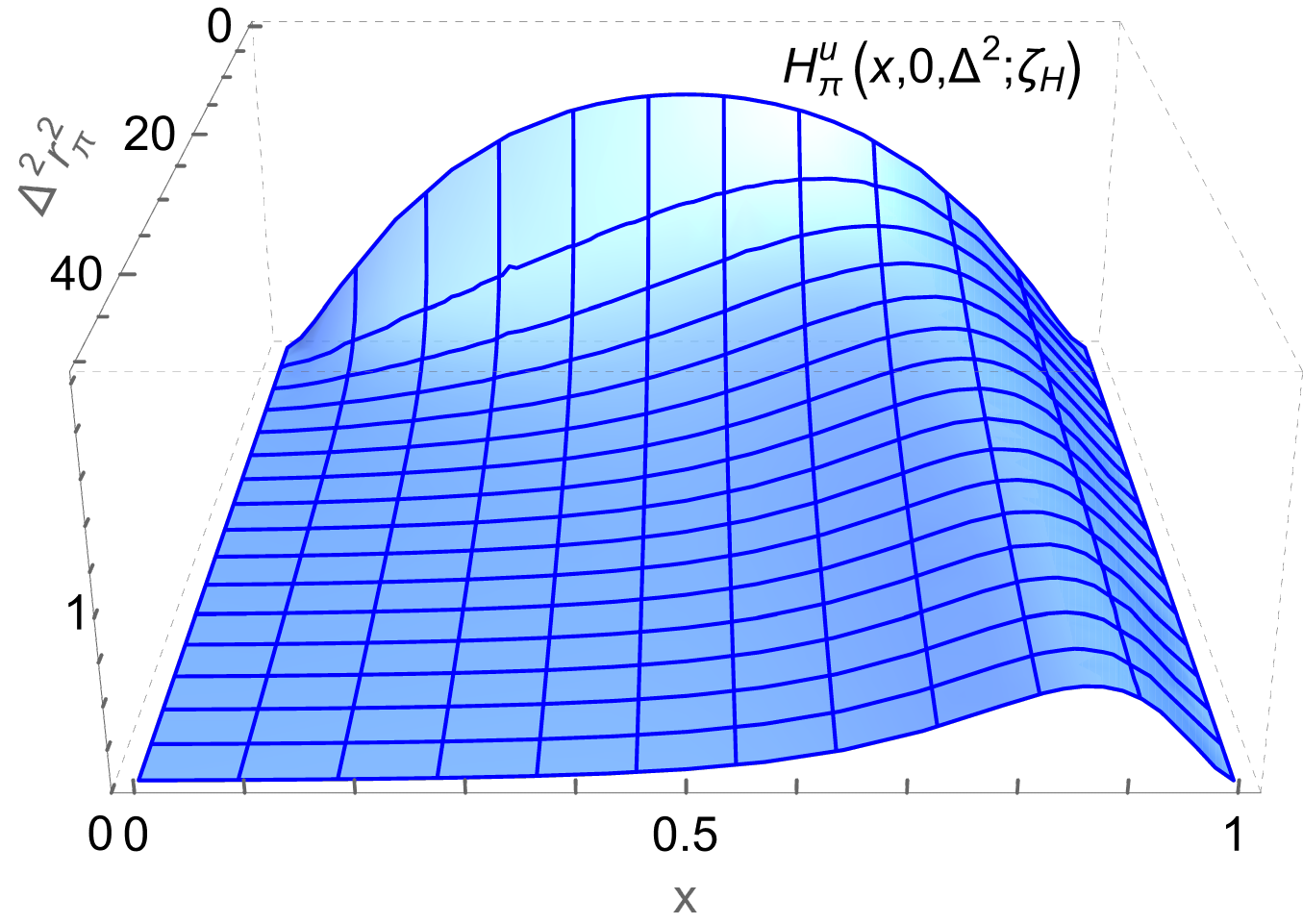} &
\includegraphics[width=0.48\textwidth]{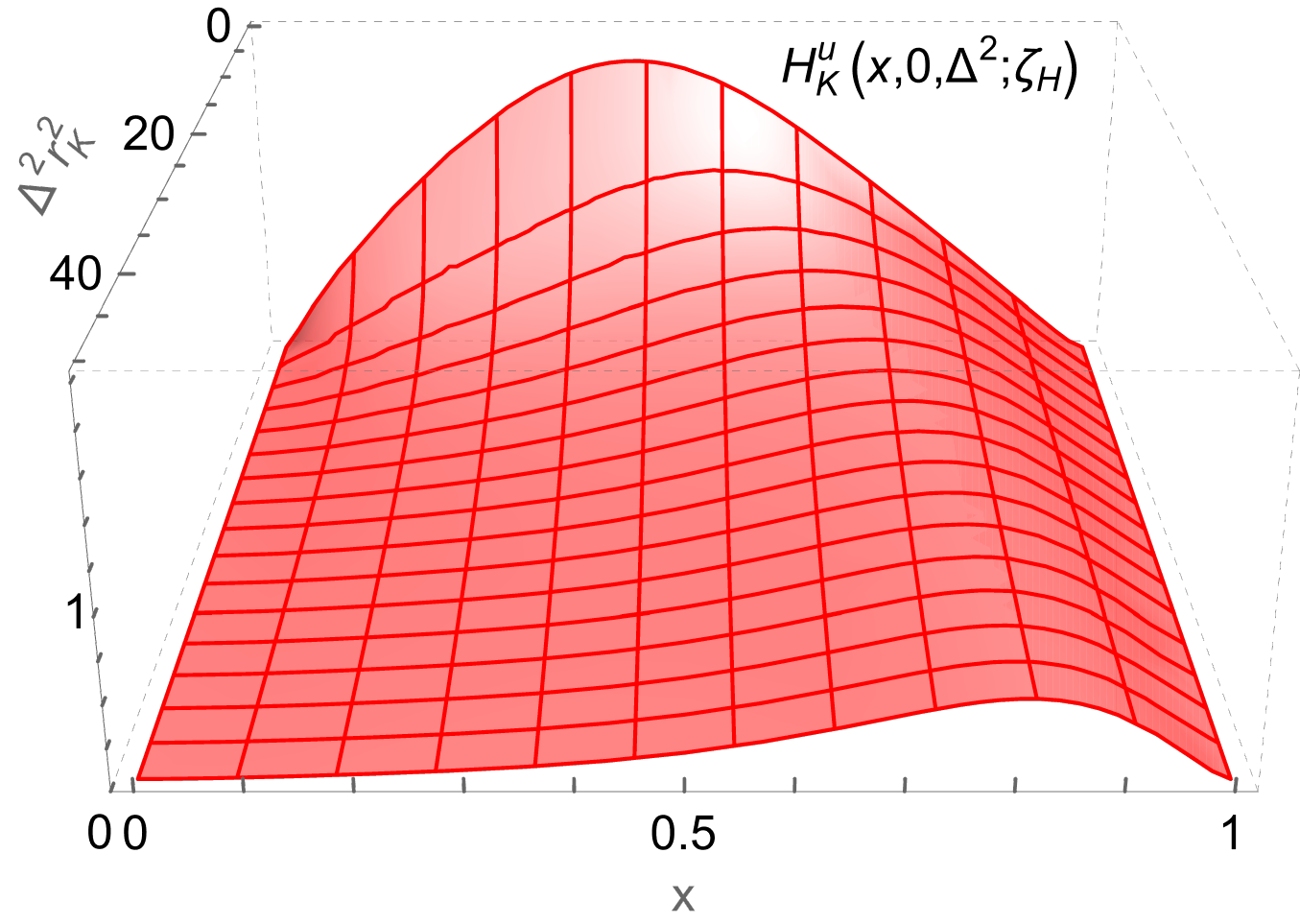}
\end{tabular}
\caption{Pion [left] and kaon [right] valence-quark GPDs at $\zeta_{\cal H}$ and zero skewness \cite{Raya:2021zrz}. These distributions were obtained using Eq.\,\eqref{eq:overlap} informed by the DAs and DFs discussed in Secs.\,\ref{sec:PDAs}, \ref{sec:PDFs}. \label{fig:GPD}}
\end{figure}

The overlap representation is only valid within the so-called DGLAP kinematic domain ($|x|\geq \xi$).  
This is sufficient for many purposes and the associated pion and kaon valence-quark GPDs are shown in Fig.\,\ref{fig:GPD}.

To obtain a complete GPD, one must also have knowledge of the defining matrix element on the complementary ERBL domain ($|x| < \xi$).  
Methods have been developed to extend a DGLAP-defined GPD onto the ERBL domain \cite{Chouika:2017dhe, Chouika:2017rzs, Chavez:2021llq, DallOlio:2024vjv}.

Considering Eq.\,\eqref{DFfromLFWFa}, it is apparent that a meson's hadron-scale DF and forward-limit GPD are identical:  
\begin{equation}
\label{eq:FL}
    {\mathpzc q}_\textbf{P}(x;\zeta_{\cal H})=H_\textbf{P}^q(x,0,0;\zeta_{\cal H})\;.
\end{equation}
Moreover, the following identities have been established:
\begin{subequations}
\label{GFFGPD:eq}
\begin{align}
\int_{-1}^1 dx\, H_\textbf{P}^q(x,\xi,-\Delta^2;\zeta_{\cal H}) &=  F_\textbf{P}^q(\Delta^2)\,,
\label{eq:EFFGPD}\\
\int_{-1}^1dx x\, H_\textbf{P}^q(x,\xi,-\Delta^2;\zeta_{\cal H}) &=  \theta_2^{\textbf{P}q}(\Delta^2)-\xi^2 \theta_1^{\textbf{P}q}(\Delta^2)\,.\label{eq:GFFGPD}
\end{align}
\end{subequations}
A meson's complete FF is obtained by properly summing the valence quark and antiquark contributions. 
In principle, both the direct amplitude approach to form factor calculations, discussed in Sec.\,\ref{Sec:FFs}, and the GPD representations, Eqs.\,\eqref{GFFGPD:eq}, are entirely equivalent. 

Novelty is provided by the $\xi=0$ impact parameter space (IPD) GPD: 
\begin{eqnarray}
\mathpzc{q}_{\textbf{P}}(x,b_\perp^2;\zeta_{\cal H}) =  \int_0^\infty \frac{d\Delta_\perp}{2\pi} \Delta_\perp J_0(b_\perp \Delta_\perp) H_{\textbf{P}}^q(x,0,-\Delta_\perp^2;\zeta_{\cal H}) \,.
\label{eq:IPDHgen}
\end{eqnarray}
This quantity is a true density, which relates to the number of partons within the light-front at a transverse distance $|b_\perp|$ from the meson's centre of transverse momentum. 
Predictions for $\mathpzc{q}_{\pi, K}(x,b_\perp^2;\zeta_{\cal H})$ can be found in Refs.\,\cite{Raya:2021zrz, Raya:2022eqa} and those for additional states in Refs.\,\cite{Albino:2022gzs, Almeida-Zamora:2023bqb}.

\begin{figure}[t]%
\centering
\begin{tabular}{lr}
\includegraphics[width=0.46\textwidth]{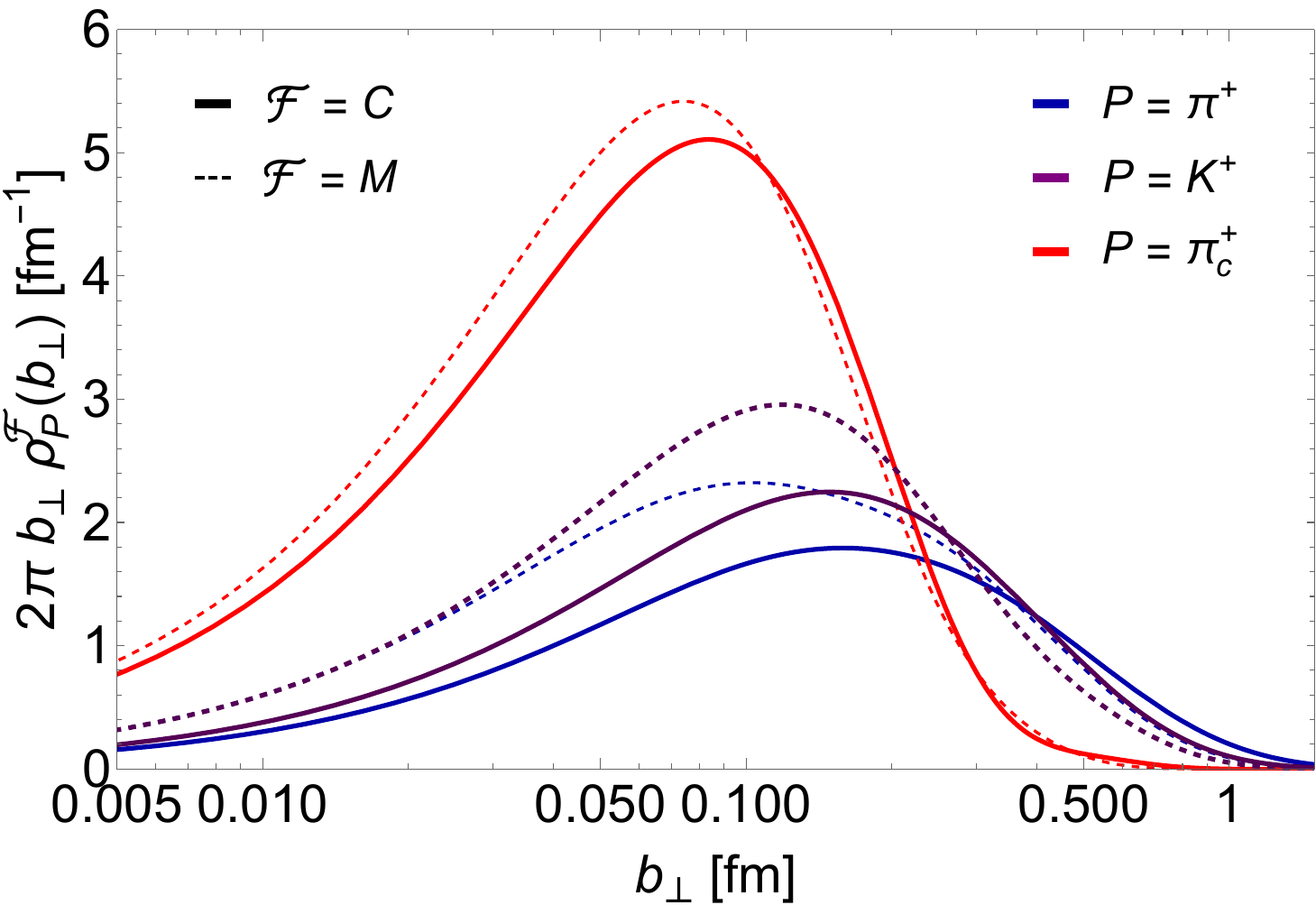} &
\includegraphics[width=0.49\textwidth]{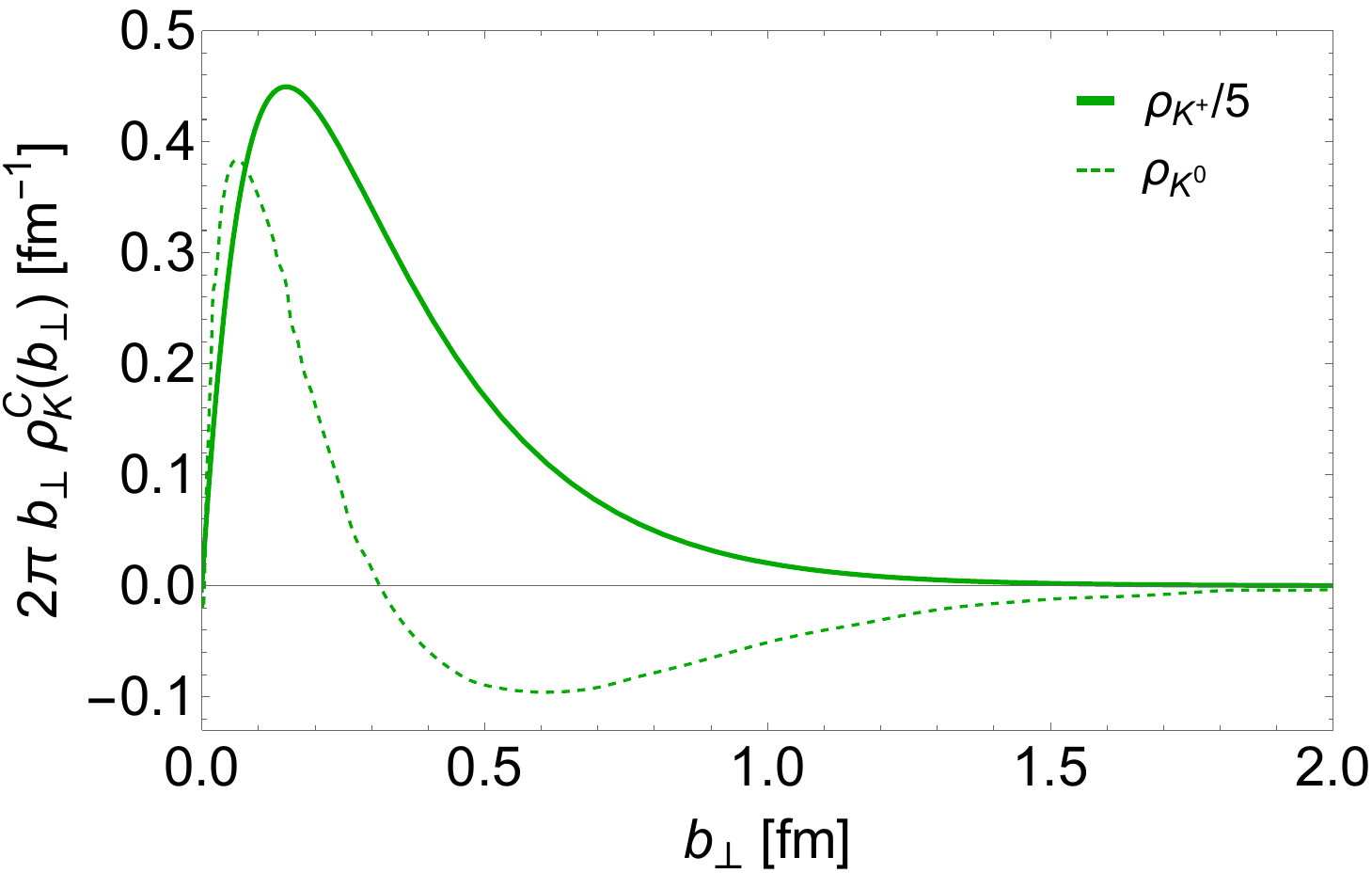}
\end{tabular}
\caption{Charge ($C$) and mass ($M$) distributions. 
[left] $\pi^+$, $K^+$ and $\pi^+_c$ cases. The last is $\pi^+$-like with $c$-massive valence quarks. 
[right] $K^+$ and $K^0$ charge distributions. The charged case has been multiplied by $0.2$ so that it shares the same range as the neutral one. \label{fig:dists}}
\end{figure}

Herein, we limit ourselves to a discussion of $\pi$, $K$ charge and mass distributions derived from the IPD GPD.  For a $q$-in-$\textbf{P}$ quark:
\begin{equation}
\label{eq:dists}
\rho_{\textbf{P}q}^{\{C,M\}}(|b_\perp|)=\int_{0}^1 dx\,\{1,\,x\}\mathpzc{q}_{\textbf{P}}(x,b_\perp^2;\zeta_{\cal H})\,;
\end{equation}
so, the total meson distributions are:
\begin{subequations}
\begin{align}
    \rho_{\textbf{P}}^{C}(|b_\perp|) &= \mathbf{e}_q  \rho_{\textbf{P}q}^{C}(|b_\perp|) + \mathbf{e}_{\bar{h}}  \rho_{\textbf{P}\bar{h}}^{C}(|b_\perp|)\,,\\
    \rho_{\textbf{P}}^{M}(|b_\perp|) &=\   \rho_{\textbf{P}q}^{M}(|b_\perp|) +  \rho_{\textbf{P}\bar{h}}^{M}(|b_\perp|)\,.
\end{align}
\end{subequations}
The results are displayed in Fig.\,\ref{fig:dists}.
The left panel contrasts $\pi^+$, $K^+$, and $\pi_c^+$, where this last system is a pion-like state with quark and antiquark possessing $c$-quark masses, all calculated using a LFWF as in Eq.\,\eqref{eq:Adnan1}. 
Consistent with Fig.\,\ref{fig:GFF}, mass distributions are more compact than charge distributions. 
Moreover, with increasing meson mass, both distributions are squeezed toward $|b_\perp|=0$ -- recall, the area under each curve is the same. 
This can be quantified by recording the associated radii ratios, $r_\text{M}^\perp/r_\text{C}^\perp$: $\pi^+=0.77(6)$; $K^+=0.80(7)$; $\pi_c^+=0.97(3)$. 
(The uncertainties were obtained as explained in Ref.\,\cite{Xu:2023bwv}.) 
Plainly, pion and kaon ratios are equal, within uncertainties.
Notwithstanding that, they do hint at a trend, which is highlighted by the $\pi_c^+$ ratio being nearly unity. 

Figure~\ref{fig:dists}\,--\,right depicts the $K^{+,0}$ transverse-plane charge distributions.
The $K^+$ profile is similar to that of the $\pi^+$.  
Regarding the $K^0$, destructive interference between $d$ and $\bar s$ distributions leads to a zero at $|b_\perp| \approx 0.3\,$fm.  
Below this value, the distribution is positive, which indicates that the positively-charged $\bar s$ valence-quark is more likely to be found nearer to the centre of transverse momentum than the lighter $d$ quark.

\subsection{Empirical determination of the pion GPD}
It will be very difficult to extract sufficient precise data from deeply virtual Compton scattering experiments to reconstruct meson GPDs \cite{Chavez:2021llq, Mezrag:2022pqk, Mezrag:2023nkp}. 
Considering Eqs.\,\eqref{eq:FL}, \eqref{eq:EFFGPD}, it is plain that a different approach is feasible, however.
Namely, exploiting the AO evolution scheme and the fact that a factorized representation of the pion LFWF is reliable for integrated quantities, a pion GPD can be recovered from independent data on the pion EFF and valence-quark DF \cite{Xu:2023bwv}.  

Given a factorized LFWF, Eq.\,\eqref{eq:facLFWF}, the pion GPD takes a simple form: \cite{Raya:2021zrz}:
\begin{equation}
    \label{GPD:fac}
    H_\pi^u(x;\xi,-\Delta^2;\zeta_{\cal H})=\theta(x_-)\sqrt{u_\pi(x_-;\zeta_{\cal H})u_\pi(x_+;\zeta_{\cal H})}\Phi_\pi(z^2;\zeta_{\cal H})\;,
\end{equation}
where $x_\pm=(x\pm\xi)/(1\pm\xi)$ and $z^2=\Delta_\perp^2(1-x)^2/(1-\xi^2)^2$.  
Using AO evolution, any sound extraction of $u_\pi(x;\zeta)$ from inclusive data can be mapped back to $u_\pi(x;\zeta_{\cal H})$ \cite{Cui:2022bxn, Lu:2023yna}; then, $\Phi_\pi(z^2;\zeta_{\cal H})$ can be determined via Eq.\,\eqref{eq:EFFGPD}.

\begin{figure}[h]%
\centering
\begin{tabular}{lr}
\includegraphics[width=0.48\textwidth]{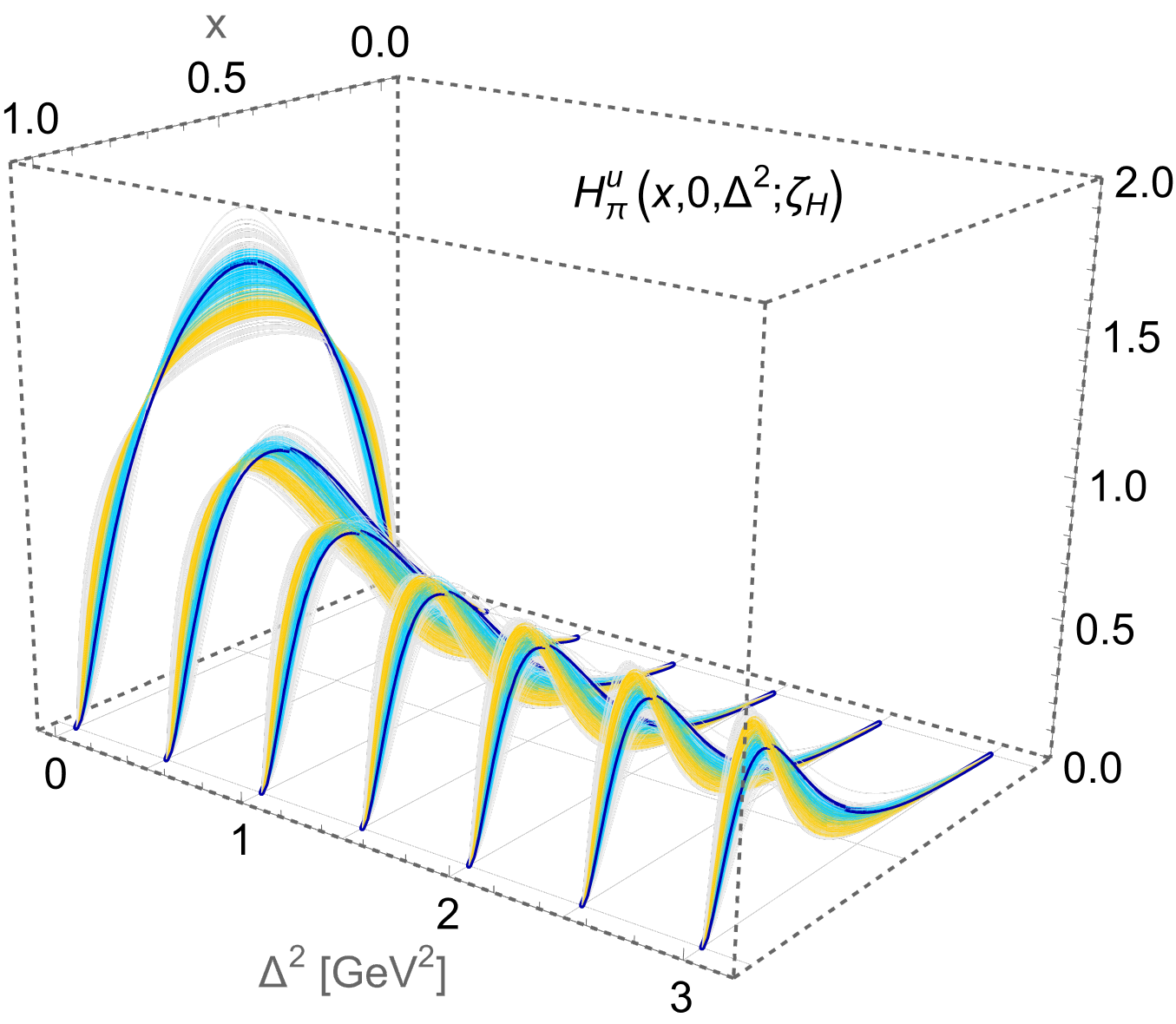} &
\includegraphics[width=0.48\textwidth]{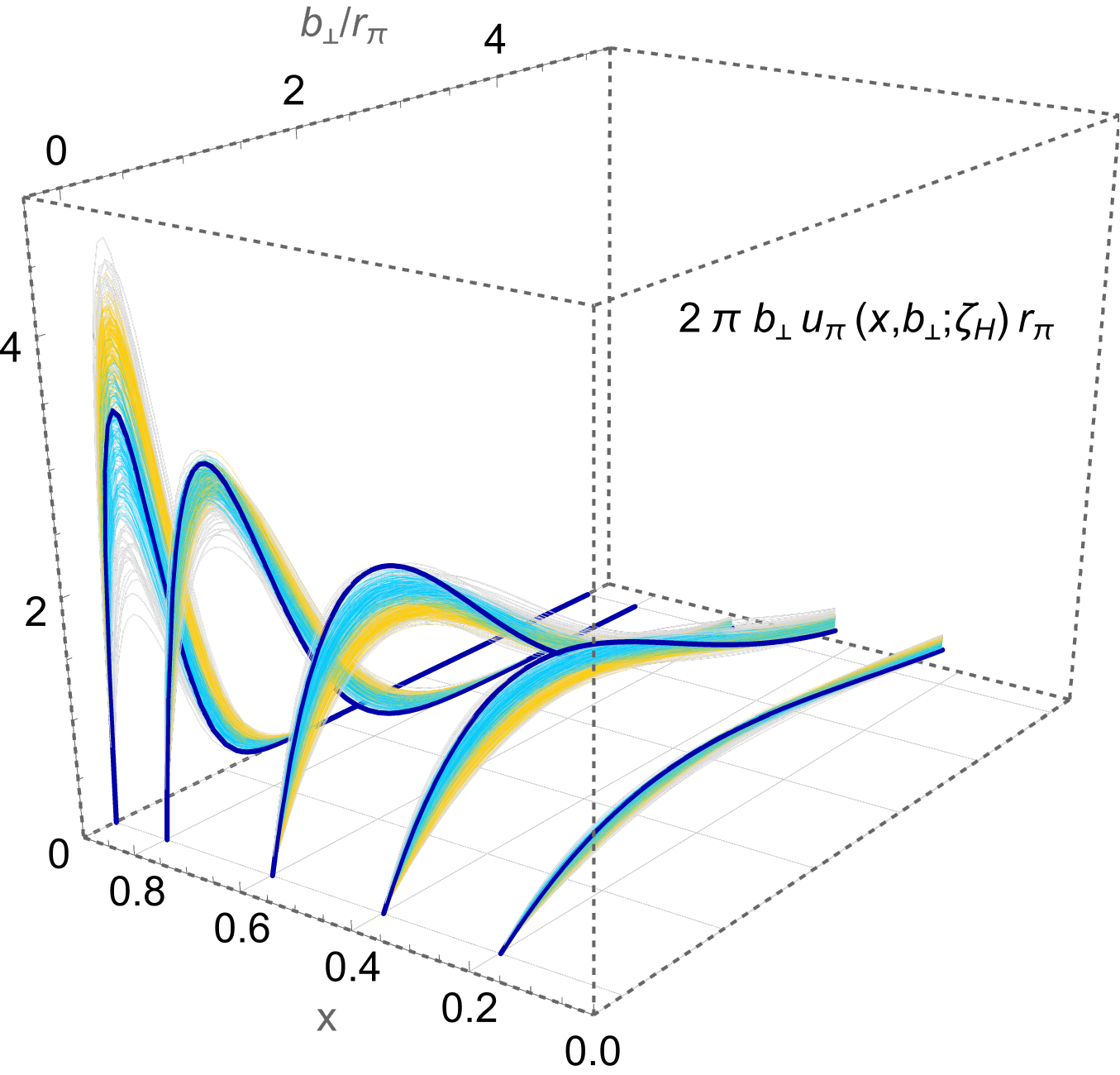}
\end{tabular}
\caption{[left] Pion GPD inferred in Ref.\,\cite{Xu:2023bwv} from existing pion + nucleus Drell-Yan and electron + pion scattering data \cite{Conway:1989fs, Aicher:2010cb, Cui:2021mom, NA7:1986vav, Horn:2007ug, JeffersonLab:2008jve}. 
[right] Associated IPS GPD. 
Both panels. 
Dark-blue curve: CSM prediction discussed in Sec.\,\ref{sec:GPDs}.
Blue band obtained from Ref.\,\cite{Aicher:2010cb} analysis of Ref.\,\cite{Conway:1989fs} Drell-Yan data;
orange bands, from Ref.\,\cite[Sec.\,8]{Cui:2021mom} pion DFs;
and grey bands, from moments computed using lQCD \cite{Joo:2019bzr, Sufian:2019bol, Alexandrou:2021mmi}. \label{fig:DD}}
\end{figure}

These ideas were implemented in Ref.\,\cite{Xu:2023bwv} as follows.  
Beginning with existing analyses of pion + nucleus Drell-Yan and electron + pion scattering data \cite{Conway:1989fs, Aicher:2010cb, Cui:2021mom, NA7:1986vav, Horn:2007ug, JeffersonLab:2008jve}, ensembles of model-independent representations of the pion GPD were developed.  
They are illustrated in Fig.\,\ref{fig:DD}.  
Using those GPD ensembles and Eq.\,\eqref{eq:GFFGPD}, data-driven predictions for the pion mass distribution form factor, $\theta_2$, were obtained.  
Compared with the pion elastic EFF obtained simultaneously from Eq.\,\eqref{eq:EFFGPD}, $\theta_2$ is harder: the ratio of the radii derived from these two form factors is $r_\pi^{\theta_2}/r_\pi = 0.79(3)$, which is in accord with Eq.\,\eqref{eq:constraints}.
The Ref.\,\cite{Xu:2023bwv} data-driven predictions for the pion GPD, related form factors and distributions should prove valuable as constraints on theories of pion structure.
Improvements to those results are possible if new, precise data relating to pion DFs are obtained and EFF data is secured at larger momentum transfers.
An extension to the kaon is currently being explored, but in this case serious impediments are presented by the lack of data and the imprecision of that which is available.

\section{Perspective}
\label{Sec:Conclusions}
Emergent phenomena in QCD are responsible for a diverse array of measurable outcomes. 
Of primary importance is the fact that the parton degrees-of-freedom used to express the QCD Lagrangian are not directly observable.  
Only color-neutral systems, seeded by valence partons, can be captured in detectors. This is an empirical definition of confinement.  
A mathematical definition remains elusive and those being discussed can be contentious.
Nevertheless, a growing body of evidence suggests that the source of confinement can be traced to the dynamical generation of running masses in the gauge and matter sectors.  
Moreover, that the associated masses explain the emergence of $\sim 99$\% of the visible mass in the Universe. 
Finally, that these masses, too, eliminate the Landau pole in QCD and the Gribov ambiguity, thereby enabling the calculation of a unique process-independent running coupling, which is everywhere finite and practically momentum-independent in the infrared, so that QCD is effectively a conformal theory at long range. 
This being the case, then QCD is potentially the first mathematically well defined quantum field theory ever formulated in four-dimensions \cite{Deur:2023dzc}. 

The properties of pseudoscalar mesons provide ideal means by which to elucidate these facets of QCD; and we exploited this in providing many explanations and illustrations. 
Indeed, as both bound-states of valence-quark and valence-antiquark and Nambu-Goldstone bosons in the chiral limit, pseudoscalar mesons provide a clean link to emergent hadron mass (EHM) and a route to exposing its interference with the other known mechanism of mass generation in the Standard model, \textit{viz}.\ the Higgs boson.  The progress achieved and standpoint of theory today, continuum and lattice, were, to many, unimaginable a decade ago.  Nevertheless, controversies remain.  

Many of the things we discussed are or will be the focus of experiments at modern and planned facilities.  Confirmation of the predictions will go far toward resolving the disputes and validating the EHM paradigm discussed herein.  Their complements are also critical, \textit{e.g}., studies of nucleon structure, of course; but importantly, too, development of an understanding of meson and nucleon excited states.  EHM is expressed in each such system; yet, the manifestations need not be everywhere identical.  With the advent of high-energy, high-luminosity facilities, it will become possible to map and link the expressions of EHM across the entire field of strong interaction phenomena.  Finally, thereby, will science test QCD and decide whether it is one part of an effective field theory of Nature or a theory in its own right, with the potential to guide extensions of the Standard Model.  

\backmatter

\bmhead{Declaration of Competing Interest}
One of the authors (C.\,D.\ Roberts) is a Field Editor for Few Body Systames in the area of Elementary Particle Physics.
Apart from this, the authors declare that they have no other known competing financial interests or personal relationships that could have appeared to influence the work reported in this paper.


\bmhead{Funding Information}
This work was supported by
Spanish MICINN grant PID2022-140440NB-C22;
regional Andalusian project P18-FR-5057;
National Natural Science Foundation of China grant no.\ 12135007;
and 
STRONG-2020 ``The strong interaction at the frontier of knowledge: fundamental research and applications” which received funding from the European Union's Horizon 2020 research and innovation programme (grant agreement no.\ 824093).

\bmhead{Acknowledgments}
This contribution is based on results obtained and insights developed through collaborations with many people, to all of whom we are greatly indebted.
%







\bibliography{main}


\begin{thebibliography}{190}
\ifx \bisbn   \undefined \def \bisbn  #1{ISBN #1}\fi
\ifx \binits  \undefined \def \binits#1{#1}\fi
\ifx \bauthor  \undefined \def \bauthor#1{#1}\fi
\ifx \batitle  \undefined \def \batitle#1{#1}\fi
\ifx \bjtitle  \undefined \def \bjtitle#1{#1}\fi
\ifx \bvolume  \undefined \def \bvolume#1{\textbf{#1}}\fi
\ifx \byear  \undefined \def \byear#1{#1}\fi
\ifx \bissue  \undefined \def \bissue#1{#1}\fi
\ifx \bfpage  \undefined \def \bfpage#1{#1}\fi
\ifx \blpage  \undefined \def \blpage #1{#1}\fi
\ifx \burl  \undefined \def \burl#1{\textsf{#1}}\fi
\ifx \doiurl  \undefined \def \doiurl#1{\url{https://doi.org/#1}}\fi
\ifx \betal  \undefined \def \betal{\textit{et al.}}\fi
\ifx \binstitute  \undefined \def \binstitute#1{#1}\fi
\ifx \binstitutionaled  \undefined \def \binstitutionaled#1{#1}\fi
\ifx \bctitle  \undefined \def \bctitle#1{#1}\fi
\ifx \beditor  \undefined \def \beditor#1{#1}\fi
\ifx \bpublisher  \undefined \def \bpublisher#1{#1}\fi
\ifx \bbtitle  \undefined \def \bbtitle#1{#1}\fi
\ifx \bedition  \undefined \def \bedition#1{#1}\fi
\ifx \bseriesno  \undefined \def \bseriesno#1{#1}\fi
\ifx \blocation  \undefined \def \blocation#1{#1}\fi
\ifx \bsertitle  \undefined \def \bsertitle#1{#1}\fi
\ifx \bsnm \undefined \def \bsnm#1{#1}\fi
\ifx \bsuffix \undefined \def \bsuffix#1{#1}\fi
\ifx \bparticle \undefined \def \bparticle#1{#1}\fi
\ifx \barticle \undefined \def \barticle#1{#1}\fi
\bibcommenthead
\ifx \bconfdate \undefined \def \bconfdate #1{#1}\fi
\ifx \botherref \undefined \def \botherref #1{#1}\fi
\ifx \url \undefined \def \url#1{\textsf{#1}}\fi
\ifx \bchapter \undefined \def \bchapter#1{#1}\fi
\ifx \bbook \undefined \def \bbook#1{#1}\fi
\ifx \bcomment \undefined \def \bcomment#1{#1}\fi
\ifx \oauthor \undefined \def \oauthor#1{#1}\fi
\ifx \citeauthoryear \undefined \def \citeauthoryear#1{#1}\fi
\ifx \endbibitem  \undefined \def \endbibitem {}\fi
\ifx \bconflocation  \undefined \def \bconflocation#1{#1}\fi
\ifx \arxivurl  \undefined \def \arxivurl#1{\textsf{#1}}\fi
\csname PreBibitemsHook\endcsname

\bibitem[\protect\citeauthoryear{Huterer and Shafer}{2018}]{Huterer:2017buf}
\begin{barticle}
\bauthor{\bsnm{Huterer}, \binits{D.}},
\bauthor{\bsnm{Shafer}, \binits{D.L.}}:
\batitle{{Dark energy two decades after: Observables, probes, consistency tests}}.
\bjtitle{Rept. Prog. Phys.}
\bvolume{81}(\bissue{1}),
\bfpage{016901}
(\byear{2018})
\doiurl{10.1088/1361-6633/aa997e}
{\href{https://arxiv.org/abs/1709.01091}{{arXiv:1709.01091}}}
{[astro-ph.CO]}
\end{barticle}
\endbibitem

\bibitem[\protect\citeauthoryear{Seife}{2003}]{Seife:2003abc}
\begin{barticle}
\bauthor{\bsnm{Seife}, \binits{C.}}:
\batitle{Illuminating the dark universe}.
\bjtitle{Science}
\bvolume{302}(\bissue{5653}),
\bfpage{2038}--\blpage{2039}
(\byear{2003})
\doiurl{10.1126/science.302.5653.2038}
{\href{https://arxiv.org/abs/https://www.science.org/doi/pdf/10.1126/science.302.5653.2038}{{https://www.science.org/doi/pdf/10.1126/science.302.5653.2038}}}
\end{barticle}
\endbibitem

\bibitem[\protect\citeauthoryear{Glashow}{1961}]{Glashow:1961tr}
\begin{barticle}
\bauthor{\bsnm{Glashow}, \binits{S.L.}}:
\batitle{{Partial Symmetries of Weak Interactions}}.
\bjtitle{Nucl. Phys.}
\bvolume{22},
\bfpage{579}--\blpage{588}
(\byear{1961})
\doiurl{10.1016/0029-5582(61)90469-2}
\end{barticle}
\endbibitem

\bibitem[\protect\citeauthoryear{Weinberg}{1967}]{Weinberg:1967tq}
\begin{barticle}
\bauthor{\bsnm{Weinberg}, \binits{S.}}:
\batitle{{A Model of Leptons}}.
\bjtitle{Phys. Rev. Lett.}
\bvolume{19},
\bfpage{1264}--\blpage{1266}
(\byear{1967})
\doiurl{10.1103/PhysRevLett.19.1264}
\end{barticle}
\endbibitem

\bibitem[\protect\citeauthoryear{Salam}{1968}]{Salam:1968rm}
\begin{barticle}
\bauthor{\bsnm{Salam}, \binits{A.}}:
\batitle{{Weak and Electromagnetic Interactions}}.
\bjtitle{Conf. Proc. C}
\bvolume{680519},
\bfpage{367}--\blpage{377}
(\byear{1968})
\doiurl{10.1142/9789812795915_0034}
\end{barticle}
\endbibitem

\bibitem[\protect\citeauthoryear{Politzer}{1973}]{Politzer:1973fx}
\begin{barticle}
\bauthor{\bsnm{Politzer}, \binits{H.D.}}:
\batitle{{Reliable Perturbative Results for Strong Interactions?}}
\bjtitle{Phys. Rev. Lett.}
\bvolume{30},
\bfpage{1346}--\blpage{1349}
(\byear{1973})
\doiurl{10.1103/PhysRevLett.30.1346}
\end{barticle}
\endbibitem

\bibitem[\protect\citeauthoryear{Gross and Wilczek}{1973}]{Gross:1973id}
\begin{barticle}
\bauthor{\bsnm{Gross}, \binits{D.J.}},
\bauthor{\bsnm{Wilczek}, \binits{F.}}:
\batitle{{Ultraviolet Behavior of Nonabelian Gauge Theories}}.
\bjtitle{Phys. Rev. Lett.}
\bvolume{30},
\bfpage{1343}--\blpage{1346}
(\byear{1973})
\doiurl{10.1103/PhysRevLett.30.1343}
\end{barticle}
\endbibitem

\bibitem[\protect\citeauthoryear{Aoyama et~al.}{2020}]{Aoyama:2020ynm}
\begin{barticle}
\bauthor{\bsnm{Aoyama}, \binits{T.}}, \betal:
\batitle{{The anomalous magnetic moment of the muon in the Standard Model}}.
\bjtitle{Phys. Rept.}
\bvolume{887},
\bfpage{1}--\blpage{166}
(\byear{2020})
\doiurl{10.1016/j.physrep.2020.07.006}
{\href{https://arxiv.org/abs/2006.04822}{{arXiv:2006.04822}}}
{[hep-ph]}
\end{barticle}
\endbibitem

\bibitem[\protect\citeauthoryear{Crivellin and Mellado}{2023}]{Crivellin:2023zui}
\begin{botherref}
\oauthor{\bsnm{Crivellin}, \binits{A.}},
\oauthor{\bsnm{Mellado}, \binits{B.}}:
{Anomalies in Particle Physics}
(2023)
{\href{https://arxiv.org/abs/2309.03870}{{arXiv:2309.03870}}}
{[hep-ph]}
\end{botherref}
\endbibitem

\bibitem[\protect\citeauthoryear{Englert}{2014}]{Englert:2014zpa}
\begin{barticle}
\bauthor{\bsnm{Englert}, \binits{F.}}:
\batitle{{Nobel Lecture: The BEH mechanism and its scalar boson}}.
\bjtitle{Rev. Mod. Phys.}
\bvolume{86}(\bissue{3}),
\bfpage{843}
(\byear{2014})
\doiurl{10.1103/RevModPhys.86.843}
\end{barticle}
\endbibitem

\bibitem[\protect\citeauthoryear{Higgs}{2014}]{Higgs:2014aqa}
\begin{barticle}
\bauthor{\bsnm{Higgs}, \binits{P.W.}}:
\batitle{{Nobel Lecture: Evading the Goldstone theorem}}.
\bjtitle{Rev. Mod. Phys.}
\bvolume{86}(\bissue{3}),
\bfpage{851}
(\byear{2014})
\doiurl{10.1103/RevModPhys.86.851}
\end{barticle}
\endbibitem

\bibitem[\protect\citeauthoryear{Marciano and Pagels}{1979}]{Marciano:1979wa}
\begin{barticle}
\bauthor{\bsnm{Marciano}, \binits{W.J.}},
\bauthor{\bsnm{Pagels}, \binits{H.}}:
\batitle{{QUANTUM CHROMODYNAMICS}}.
\bjtitle{Nature}
\bvolume{279},
\bfpage{479}--\blpage{483}
(\byear{1979})
\doiurl{10.1038/279479a0}
\end{barticle}
\endbibitem

\bibitem[\protect\citeauthoryear{Marciano and Pagels}{1978}]{Marciano:1977su}
\begin{barticle}
\bauthor{\bsnm{Marciano}, \binits{W.J.}},
\bauthor{\bsnm{Pagels}, \binits{H.}}:
\batitle{{Quantum Chromodynamics: A Review}}.
\bjtitle{Phys. Rept.}
\bvolume{36},
\bfpage{137}
(\byear{1978})
\doiurl{10.1016/0370-1573(78)90208-9}
\end{barticle}
\endbibitem

\bibitem[\protect\citeauthoryear{Ding et~al.}{2023}]{Ding:2022ows}
\begin{barticle}
\bauthor{\bsnm{Ding}, \binits{M.}},
\bauthor{\bsnm{Roberts}, \binits{C.D.}},
\bauthor{\bsnm{Schmidt}, \binits{S.M.}}:
\batitle{{Emergence of Hadron Mass and Structure}}.
\bjtitle{Particles}
\bvolume{6},
\bfpage{57}--\blpage{120}
(\byear{2023})
\doiurl{10.3390/particles6010004}
{\href{https://arxiv.org/abs/2211.07763}{{arXiv:2211.07763}}}
{[hep-ph]}
\end{barticle}
\endbibitem

\bibitem[\protect\citeauthoryear{Gao et~al.}{2018}]{Gao:2017uox}
\begin{barticle}
\bauthor{\bsnm{Gao}, \binits{F.}},
\bauthor{\bsnm{Qin}, \binits{S.-X.}},
\bauthor{\bsnm{Roberts}, \binits{C.D.}},
\bauthor{\bsnm{Rodriguez-Quintero}, \binits{J.}}:
\batitle{{Locating the Gribov horizon}}.
\bjtitle{Phys. Rev. D}
\bvolume{97}(\bissue{3}),
\bfpage{034010}
(\byear{2018})
\doiurl{10.1103/PhysRevD.97.034010}
{\href{https://arxiv.org/abs/1706.04681}{{arXiv:1706.04681}}}
{[hep-ph]}
\end{barticle}
\endbibitem

\bibitem[\protect\citeauthoryear{Binosi and Tripolt}{2020}]{Binosi:2019ecz}
\begin{barticle}
\bauthor{\bsnm{Binosi}, \binits{D.}},
\bauthor{\bsnm{Tripolt}, \binits{R.-A.}}:
\batitle{{Spectral functions of confined particles}}.
\bjtitle{Phys. Lett. B}
\bvolume{801},
\bfpage{135171}
(\byear{2020})
\doiurl{10.1016/j.physletb.2019.135171}
{\href{https://arxiv.org/abs/1904.08172}{{arXiv:1904.08172}}}
{[hep-ph]}
\end{barticle}
\endbibitem

\bibitem[\protect\citeauthoryear{Falc\~ao et~al.}{2020}]{Falcao:2020vyr}
\begin{barticle}
\bauthor{\bsnm{Falc\~ao}, \binits{A.F.}},
\bauthor{\bsnm{Oliveira}, \binits{O.}},
\bauthor{\bsnm{Silva}, \binits{P.J.}}:
\batitle{{Analytic structure of the lattice Landau gauge gluon and ghost propagators}}.
\bjtitle{Phys. Rev. D}
\bvolume{102}(\bissue{11}),
\bfpage{114518}
(\byear{2020})
\doiurl{10.1103/PhysRevD.102.114518}
{\href{https://arxiv.org/abs/2008.02614}{{arXiv:2008.02614}}}
{[hep-lat]}
\end{barticle}
\endbibitem

\bibitem[\protect\citeauthoryear{Boito et~al.}{2023}]{Boito:2022rad}
\begin{barticle}
\bauthor{\bsnm{Boito}, \binits{D.}},
\bauthor{\bsnm{Cucchieri}, \binits{A.}},
\bauthor{\bsnm{London}, \binits{C.Y.}},
\bauthor{\bsnm{Mendes}, \binits{T.}}:
\batitle{{Probing the singularities of the Landau-Gauge gluon and ghost propagators with rational approximants}}.
\bjtitle{JHEP}
\bvolume{02},
\bfpage{144}
(\byear{2023})
\doiurl{10.1007/JHEP02(2023)144}
{\href{https://arxiv.org/abs/2210.10490}{{arXiv:2210.10490}}}
{[hep-lat]}
\end{barticle}
\endbibitem

\bibitem[\protect\citeauthoryear{Rutherford}{1919a}]{RutherfordI}
\begin{barticle}
\bauthor{\bsnm{Rutherford}, \binits{E.}}:
\batitle{Collision of alpha particles with light atoms i. hydrogen}.
\bjtitle{Philosophical Magazine}
\bvolume{xxxvii},
\bfpage{537}--\blpage{561}
(\byear{1919})
\end{barticle}
\endbibitem

\bibitem[\protect\citeauthoryear{Rutherford}{1919b}]{RutherfordII}
\begin{barticle}
\bauthor{\bsnm{Rutherford}, \binits{E.}}:
\batitle{Collision of alpha particles with light atoms ii. velocity of the hydrogen atoms}.
\bjtitle{Philosophical Magazine}
\bvolume{xxxvii},
\bfpage{562}--\blpage{571}
(\byear{1919})
\end{barticle}
\endbibitem

\bibitem[\protect\citeauthoryear{Rutherford}{1919c}]{RutherfordIII}
\begin{barticle}
\bauthor{\bsnm{Rutherford}, \binits{E.}}:
\batitle{Collision of alpha particles with light atoms iii. nitrogen and oxygen atoms}.
\bjtitle{Philosophical Magazine}
\bvolume{xxxvii},
\bfpage{571}--\blpage{580}
(\byear{1919})
\end{barticle}
\endbibitem

\bibitem[\protect\citeauthoryear{Rutherford}{1919d}]{RutherfordIV}
\begin{barticle}
\bauthor{\bsnm{Rutherford}, \binits{E.}}:
\batitle{Collision of alpha particles with light atoms iv. an anomalous effect in nitrogen}.
\bjtitle{Philosophical Magazine}
\bvolume{xxxvii},
\bfpage{581}--\blpage{587}
(\byear{1919})
\end{barticle}
\endbibitem

\bibitem[\protect\citeauthoryear{Hofstadter}{1956}]{Hofstadter:1956qs}
\begin{barticle}
\bauthor{\bsnm{Hofstadter}, \binits{R.}}:
\batitle{{Electron scattering and nuclear structure}}.
\bjtitle{Rev. Mod. Phys.}
\bvolume{28},
\bfpage{214}--\blpage{254}
(\byear{1956})
\doiurl{10.1103/RevModPhys.28.214}
\end{barticle}
\endbibitem

\bibitem[\protect\citeauthoryear{Breidenbach et~al.}{1969}]{Breidenbach:1969kd}
\begin{barticle}
\bauthor{\bsnm{Breidenbach}, \binits{M.}},
\bauthor{\bsnm{Friedman}, \binits{J.I.}},
\bauthor{\bsnm{Kendall}, \binits{H.W.}},
\bauthor{\bsnm{Bloom}, \binits{E.D.}},
\bauthor{\bsnm{Coward}, \binits{D.H.}},
\bauthor{\bsnm{DeStaebler}, \binits{H.C.}},
\bauthor{\bsnm{Drees}, \binits{J.}},
\bauthor{\bsnm{Mo}, \binits{L.W.}},
\bauthor{\bsnm{Taylor}, \binits{R.E.}}:
\batitle{{Observed behavior of highly inelastic electron-proton scattering}}.
\bjtitle{Phys. Rev. Lett.}
\bvolume{23},
\bfpage{935}--\blpage{939}
(\byear{1969})
\doiurl{10.1103/PhysRevLett.23.935}
\end{barticle}
\endbibitem

\bibitem[\protect\citeauthoryear{Roberts}{2021}]{Roberts:2021xnz}
\begin{barticle}
\bauthor{\bsnm{Roberts}, \binits{C.D.}}:
\batitle{{On Mass and Matter}}.
\bjtitle{AAPPS Bull.}
\bvolume{31},
\bfpage{6}
(\byear{2021})
\doiurl{10.1007/s43673-021-00005-4}
{\href{https://arxiv.org/abs/2101.08340}{{arXiv:2101.08340}}}
{[hep-ph]}
\end{barticle}
\endbibitem

\bibitem[\protect\citeauthoryear{Roberts}{2023}]{Roberts:2022rxm}
\begin{barticle}
\bauthor{\bsnm{Roberts}, \binits{C.D.}}:
\batitle{{Origin of the Proton Mass}}.
\bjtitle{EPJ Web Conf.}
\bvolume{282},
\bfpage{01006}
(\byear{2023})
\doiurl{10.1051/epjconf/202328201006}
{\href{https://arxiv.org/abs/2211.09905}{{arXiv:2211.09905}}}
{[hep-ph]}
\end{barticle}
\endbibitem

\bibitem[\protect\citeauthoryear{Roberts}{2017}]{Roberts:2016vyn}
\begin{barticle}
\bauthor{\bsnm{Roberts}, \binits{C.D.}}:
\batitle{{Perspective on the origin of hadron masses}}.
\bjtitle{Few Body Syst.}
\bvolume{58}(\bissue{1}),
\bfpage{5}
(\byear{2017})
\doiurl{10.1007/s00601-016-1168-z}
{\href{https://arxiv.org/abs/1606.03909}{{arXiv:1606.03909}}}
{[nucl-th]}
\end{barticle}
\endbibitem

\bibitem[\protect\citeauthoryear{Gao and Vanderhaeghen}{2022}]{Gao:2021sml}
\begin{barticle}
\bauthor{\bsnm{Gao}, \binits{H.}},
\bauthor{\bsnm{Vanderhaeghen}, \binits{M.}}:
\batitle{{The proton charge radius}}.
\bjtitle{Rev. Mod. Phys.}
\bvolume{94}(\bissue{1}),
\bfpage{015002}
(\byear{2022})
\doiurl{10.1103/RevModPhys.94.015002}
{\href{https://arxiv.org/abs/2105.00571}{{arXiv:2105.00571}}}
{[hep-ph]}
\end{barticle}
\endbibitem

\bibitem[\protect\citeauthoryear{Cui et~al.}{2022}]{Cui:2022fyr}
\begin{barticle}
\bauthor{\bsnm{Cui}, \binits{Z.-F.}},
\bauthor{\bsnm{Binosi}, \binits{D.}},
\bauthor{\bsnm{Roberts}, \binits{C.D.}},
\bauthor{\bsnm{Schmidt}, \binits{S.M.}}:
\batitle{{Hadron and light nucleus radii from electron scattering}}.
\bjtitle{Chin. Phys. C}
\bvolume{46}(\bissue{12}),
\bfpage{122001}
(\byear{2022})
\doiurl{10.1088/1674-1137/ac89d0}
{\href{https://arxiv.org/abs/2204.05418}{{arXiv:2204.05418}}}
{[hep-ph]}
\end{barticle}
\endbibitem

\bibitem[\protect\citeauthoryear{Altmannshofer et~al.}{2019}]{Belle-II:2018jsg}
\begin{barticle}
\bauthor{\bsnm{Altmannshofer}, \binits{W.}}, \betal:
\batitle{{The Belle II Physics Book}}.
\bjtitle{PTEP}
\bvolume{2019}(\bissue{12}),
\bfpage{123}--\blpage{01}
(\byear{2019})
\doiurl{10.1093/ptep/ptz106}
{\href{https://arxiv.org/abs/1808.10567}{{arXiv:1808.10567}}}
{[hep-ex]}.
\bcomment{[Erratum: PTEP 2020, 029201 (2020)]}
\end{barticle}
\endbibitem

\bibitem[\protect\citeauthoryear{Aguilar et~al.}{2019}]{Aguilar:2019teb}
\begin{barticle}
\bauthor{\bsnm{Aguilar}, \binits{A.C.}}, \betal:
\batitle{{Pion and Kaon Structure at the Electron-Ion Collider}}.
\bjtitle{Eur. Phys. J. A}
\bvolume{55}(\bissue{10}),
\bfpage{190}
(\byear{2019})
\doiurl{10.1140/epja/i2019-12885-0}
{\href{https://arxiv.org/abs/1907.08218}{{arXiv:1907.08218}}}
{[nucl-ex]}
\end{barticle}
\endbibitem

\bibitem[\protect\citeauthoryear{Yuan and Olsen}{2019}]{Yuan:2019zfo}
\begin{barticle}
\bauthor{\bsnm{Yuan}, \binits{C.-Z.}},
\bauthor{\bsnm{Olsen}, \binits{S.L.}}:
\batitle{{The BESIII physics programme}}.
\bjtitle{Nature Rev. Phys.}
\bvolume{1}(\bissue{8}),
\bfpage{480}--\blpage{494}
(\byear{2019})
\doiurl{10.1038/s42254-019-0082-y}
{\href{https://arxiv.org/abs/2001.01164}{{arXiv:2001.01164}}}
{[hep-ex]}
\end{barticle}
\endbibitem

\bibitem[\protect\citeauthoryear{Brodsky et~al.}{2020}]{Brodsky:2020vco}
\begin{barticle}
\bauthor{\bsnm{Brodsky}, \binits{S.J.}}, \betal:
\batitle{{Strong QCD from Hadron Structure Experiments}: {Newport News, VA, USA, November 4-8, 2019}}.
\bjtitle{Int. J. Mod. Phys. E}
\bvolume{29}(\bissue{08}),
\bfpage{2030006}
(\byear{2020})
\doiurl{10.1142/S0218301320300064}
{\href{https://arxiv.org/abs/2006.06802}{{arXiv:2006.06802}}}
{[hep-ph]}
\end{barticle}
\endbibitem

\bibitem[\protect\citeauthoryear{Chen et~al.}{2020}]{Chen:2020ijn}
\begin{barticle}
\bauthor{\bsnm{Chen}, \binits{X.}},
\bauthor{\bsnm{Guo}, \binits{F.-K.}},
\bauthor{\bsnm{Roberts}, \binits{C.D.}},
\bauthor{\bsnm{Wang}, \binits{R.}}:
\batitle{{Selected Science Opportunities for the EicC}}.
\bjtitle{Few Body Syst.}
\bvolume{61}(\bissue{4}),
\bfpage{43}
(\byear{2020})
\doiurl{10.1007/s00601-020-01574-0}
{\href{https://arxiv.org/abs/2008.00102}{{arXiv:2008.00102}}}
{[hep-ph]}
\end{barticle}
\endbibitem

\bibitem[\protect\citeauthoryear{Anderle et~al.}{2021}]{Anderle:2021wcy}
\begin{barticle}
\bauthor{\bsnm{Anderle}, \binits{D.P.}}, \betal:
\batitle{{Electron-ion collider in China}}.
\bjtitle{Front. Phys. (Beijing)}
\bvolume{16}(\bissue{6}),
\bfpage{64701}
(\byear{2021})
\doiurl{10.1007/s11467-021-1062-0}
{\href{https://arxiv.org/abs/2102.09222}{{arXiv:2102.09222}}}
{[nucl-ex]}
\end{barticle}
\endbibitem

\bibitem[\protect\citeauthoryear{Arrington et~al.}{2021}]{Arrington:2021biu}
\begin{barticle}
\bauthor{\bsnm{Arrington}, \binits{J.}}, \betal:
\batitle{{Revealing the structure of light pseudoscalar mesons at the electron\textendash{}ion collider}}.
\bjtitle{J. Phys. G}
\bvolume{48}(\bissue{7}),
\bfpage{075106}
(\byear{2021})
\doiurl{10.1088/1361-6471/abf5c3}
{\href{https://arxiv.org/abs/2102.11788}{{arXiv:2102.11788}}}
{[nucl-ex]}
\end{barticle}
\endbibitem

\bibitem[\protect\citeauthoryear{Abdul~Khalek et~al.}{2022}]{AbdulKhalek:2021gbh}
\begin{barticle}
\bauthor{\bsnm{Abdul~Khalek}, \binits{R.}}, \betal:
\batitle{{Science Requirements and Detector Concepts for the Electron-Ion Collider}: {EIC Yellow Report}}.
\bjtitle{Nucl. Phys. A}
\bvolume{1026},
\bfpage{122447}
(\byear{2022})
\doiurl{10.1016/j.nuclphysa.2022.122447}
{\href{https://arxiv.org/abs/2103.05419}{{arXiv:2103.05419}}}
{[physics.ins-det]}
\end{barticle}
\endbibitem

\bibitem[\protect\citeauthoryear{Quintans}{2022}]{Quintans:2022utc}
\begin{barticle}
\bauthor{\bsnm{Quintans}, \binits{C.}}:
\batitle{{The New AMBER Experiment at the CERN SPS}}.
\bjtitle{Few Body Syst.}
\bvolume{63}(\bissue{4}),
\bfpage{72}
(\byear{2022})
\doiurl{10.1007/s00601-022-01769-7}
\end{barticle}
\endbibitem

\bibitem[\protect\citeauthoryear{Amoroso et~al.}{2022}]{Amoroso:2022eow}
\begin{barticle}
\bauthor{\bsnm{Amoroso}, \binits{S.}}, \betal:
\batitle{{Snowmass 2021 Whitepaper: Proton Structure at the Precision Frontier}}.
\bjtitle{Acta Phys. Polon. B}
\bvolume{53}(\bissue{12}),
\bfpage{12}--\blpage{1}
(\byear{2022})
\doiurl{10.5506/APhysPolB.53.12-A1}
{\href{https://arxiv.org/abs/2203.13923}{{arXiv:2203.13923}}}
{[hep-ph]}
\end{barticle}
\endbibitem

\bibitem[\protect\citeauthoryear{Carman et~al.}{2023}]{Carman:2023zke}
\begin{barticle}
\bauthor{\bsnm{Carman}, \binits{D.S.}},
\bauthor{\bsnm{Gothe}, \binits{R.W.}},
\bauthor{\bsnm{Mokeev}, \binits{V.I.}},
\bauthor{\bsnm{Roberts}, \binits{C.D.}}:
\batitle{{Nucleon Resonance Electroexcitation Amplitudes and Emergent Hadron Mass}}.
\bjtitle{Particles}
\bvolume{6}(\bissue{1}),
\bfpage{416}--\blpage{439}
(\byear{2023})
\doiurl{10.3390/particles6010023}
{\href{https://arxiv.org/abs/2301.07777}{{arXiv:2301.07777}}}
{[hep-ph]}
\end{barticle}
\endbibitem

\bibitem[\protect\citeauthoryear{Yukawa}{1935}]{Yukawa:1935xg}
\begin{barticle}
\bauthor{\bsnm{Yukawa}, \binits{H.}}:
\batitle{{On the Interaction of Elementary Particles I}}.
\bjtitle{Proc. Phys. Math. Soc. Jap.}
\bvolume{17},
\bfpage{48}--\blpage{57}
(\byear{1935})
\doiurl{10.1143/PTPS.1.1}
\end{barticle}
\endbibitem

\bibitem[\protect\citeauthoryear{Horn and Roberts}{2016}]{Horn:2016rip}
\begin{barticle}
\bauthor{\bsnm{Horn}, \binits{T.}},
\bauthor{\bsnm{Roberts}, \binits{C.D.}}:
\batitle{{The pion: an enigma within the Standard Model}}.
\bjtitle{J. Phys. G}
\bvolume{43}(\bissue{7}),
\bfpage{073001}
(\byear{2016})
\doiurl{10.1088/0954-3899/43/7/073001}
{\href{https://arxiv.org/abs/1602.04016}{{arXiv:1602.04016}}}
{[nucl-th]}
\end{barticle}
\endbibitem

\bibitem[\protect\citeauthoryear{Roberts et~al.}{2021}]{Roberts:2021nhw}
\begin{barticle}
\bauthor{\bsnm{Roberts}, \binits{C.D.}},
\bauthor{\bsnm{Richards}, \binits{D.G.}},
\bauthor{\bsnm{Horn}, \binits{T.}},
\bauthor{\bsnm{Chang}, \binits{L.}}:
\batitle{{Insights into the emergence of mass from studies of pion and kaon structure}}.
\bjtitle{Prog. Part. Nucl. Phys.}
\bvolume{120},
\bfpage{103883}
(\byear{2021})
\doiurl{10.1016/j.ppnp.2021.103883}
{\href{https://arxiv.org/abs/2102.01765}{{arXiv:2102.01765}}}
{[hep-ph]}
\end{barticle}
\endbibitem

\bibitem[\protect\citeauthoryear{Maris et~al.}{1998}]{Maris:1997hd}
\begin{barticle}
\bauthor{\bsnm{Maris}, \binits{P.}},
\bauthor{\bsnm{Roberts}, \binits{C.D.}},
\bauthor{\bsnm{Tandy}, \binits{P.C.}}:
\batitle{{Pion mass and decay constant}}.
\bjtitle{Phys. Lett. B}
\bvolume{420},
\bfpage{267}--\blpage{273}
(\byear{1998})
\doiurl{10.1016/S0370-2693(97)01535-9}
{\href{https://arxiv.org/abs/nucl-th/9707003}{{arXiv:nucl-th/9707003}}}
\end{barticle}
\endbibitem

\bibitem[\protect\citeauthoryear{Maris and Roberts}{1997}]{Maris:1997tm}
\begin{barticle}
\bauthor{\bsnm{Maris}, \binits{P.}},
\bauthor{\bsnm{Roberts}, \binits{C.D.}}:
\batitle{{$\pi$- and $K$-meson Bethe-Salpeter amplitudes}}.
\bjtitle{Phys. Rev. C}
\bvolume{56},
\bfpage{3369}--\blpage{3383}
(\byear{1997})
\doiurl{10.1103/PhysRevC.56.3369}
{\href{https://arxiv.org/abs/nucl-th/9708029}{{arXiv:nucl-th/9708029}}}
\end{barticle}
\endbibitem

\bibitem[\protect\citeauthoryear{Roberts}{2016}]{Roberts:2015lja}
\begin{barticle}
\bauthor{\bsnm{Roberts}, \binits{C.D.}}:
\batitle{{Three Lectures on Hadron Physics}}.
\bjtitle{J. Phys. Conf. Ser.}
\bvolume{706}(\bissue{2}),
\bfpage{022003}
(\byear{2016})
\doiurl{10.1088/1742-6596/706/2/022003}
{\href{https://arxiv.org/abs/1509.02925}{{arXiv:1509.02925}}}
{[nucl-th]}
\end{barticle}
\endbibitem

\bibitem[\protect\citeauthoryear{Eichmann et~al.}{2016}]{Eichmann:2016yit}
\begin{barticle}
\bauthor{\bsnm{Eichmann}, \binits{G.}},
\bauthor{\bsnm{Sanchis-Alepuz}, \binits{H.}},
\bauthor{\bsnm{Williams}, \binits{R.}},
\bauthor{\bsnm{Alkofer}, \binits{R.}},
\bauthor{\bsnm{Fischer}, \binits{C.S.}}:
\batitle{{Baryons as relativistic three-quark bound states}}.
\bjtitle{Prog. Part. Nucl. Phys.}
\bvolume{91},
\bfpage{1}--\blpage{100}
(\byear{2016})
\doiurl{10.1016/j.ppnp.2016.07.001}
{\href{https://arxiv.org/abs/1606.09602}{{arXiv:1606.09602}}}
{[hep-ph]}
\end{barticle}
\endbibitem

\bibitem[\protect\citeauthoryear{Burkert and Roberts}{2019}]{Burkert:2017djo}
\begin{barticle}
\bauthor{\bsnm{Burkert}, \binits{V.D.}},
\bauthor{\bsnm{Roberts}, \binits{C.D.}}:
\batitle{{Colloquium : Roper resonance: Toward a solution to the fifty year puzzle}}.
\bjtitle{Rev. Mod. Phys.}
\bvolume{91}(\bissue{1}),
\bfpage{011003}
(\byear{2019})
\doiurl{10.1103/RevModPhys.91.011003}
{\href{https://arxiv.org/abs/1710.02549}{{arXiv:1710.02549}}}
{[nucl-ex]}
\end{barticle}
\endbibitem

\bibitem[\protect\citeauthoryear{Fischer}{2019}]{Fischer:2018sdj}
\begin{barticle}
\bauthor{\bsnm{Fischer}, \binits{C.S.}}:
\batitle{{QCD at finite temperature and chemical potential from Dyson\textendash{}Schwinger equations}}.
\bjtitle{Prog. Part. Nucl. Phys.}
\bvolume{105},
\bfpage{1}--\blpage{60}
(\byear{2019})
\doiurl{10.1016/j.ppnp.2019.01.002}
{\href{https://arxiv.org/abs/1810.12938}{{arXiv:1810.12938}}}
{[hep-ph]}
\end{barticle}
\endbibitem

\bibitem[\protect\citeauthoryear{Qin and Roberts}{2020}]{Qin:2020rad}
\begin{barticle}
\bauthor{\bsnm{Qin}, \binits{S.-X.}},
\bauthor{\bsnm{Roberts}, \binits{C.D.}}:
\batitle{{Impressions of the Continuum Bound State Problem in QCD}}.
\bjtitle{Chin. Phys. Lett.}
\bvolume{37}(\bissue{12}),
\bfpage{121201}
(\byear{2020})
\doiurl{10.1088/0256-307X/37/12/121201}
{\href{https://arxiv.org/abs/2008.07629}{{arXiv:2008.07629}}}
{[hep-ph]}
\end{barticle}
\endbibitem

\bibitem[\protect\citeauthoryear{Roberts}{2020}]{Roberts:2020hiw}
\begin{barticle}
\bauthor{\bsnm{Roberts}, \binits{C.D.}}:
\batitle{{Empirical Consequences of Emergent Mass}}.
\bjtitle{Symmetry}
\bvolume{12}(\bissue{9}),
\bfpage{1468}
(\byear{2020})
\doiurl{10.3390/sym12091468}
{\href{https://arxiv.org/abs/2009.04011}{{arXiv:2009.04011}}}
{[hep-ph]}
\end{barticle}
\endbibitem

\bibitem[\protect\citeauthoryear{Binosi}{2022}]{Binosi:2022djx}
\begin{barticle}
\bauthor{\bsnm{Binosi}, \binits{D.}}:
\batitle{{Emergent Hadron Mass in Strong Dynamics}}.
\bjtitle{Few Body Syst.}
\bvolume{63}(\bissue{2}),
\bfpage{42}
(\byear{2022})
\doiurl{10.1007/s00601-022-01740-6}
{\href{https://arxiv.org/abs/2203.00942}{{arXiv:2203.00942}}}
{[hep-ph]}
\end{barticle}
\endbibitem

\bibitem[\protect\citeauthoryear{Ferreira and Papavassiliou}{2023}]{Ferreira:2023fva}
\begin{barticle}
\bauthor{\bsnm{Ferreira}, \binits{M.N.}},
\bauthor{\bsnm{Papavassiliou}, \binits{J.}}:
\batitle{{Gauge Sector Dynamics in QCD}}.
\bjtitle{Particles}
\bvolume{6}(\bissue{1}),
\bfpage{312}--\blpage{363}
(\byear{2023})
\doiurl{10.3390/particles6010017}
{\href{https://arxiv.org/abs/2301.02314}{{arXiv:2301.02314}}}
{[hep-ph]}
\end{barticle}
\endbibitem

\bibitem[\protect\citeauthoryear{Deur et~al.}{2024}]{Deur:2023dzc}
\begin{barticle}
\bauthor{\bsnm{Deur}, \binits{A.}},
\bauthor{\bsnm{Brodsky}, \binits{S.J.}},
\bauthor{\bsnm{Roberts}, \binits{C.D.}}:
\batitle{{QCD running couplings and effective charges}}.
\bjtitle{Prog. Part. Nucl. Phys.}
\bvolume{134},
\bfpage{104081}
(\byear{2024})
\doiurl{10.1016/j.ppnp.2023.104081}
{\href{https://arxiv.org/abs/2303.00723}{{arXiv:2303.00723}}}
{[hep-ph]}
\end{barticle}
\endbibitem

\bibitem[\protect\citeauthoryear{Politzer}{2005}]{Politzer:2005kc}
\begin{barticle}
\bauthor{\bsnm{Politzer}, \binits{H.D.}}:
\batitle{{The dilemma of attribution}}.
\bjtitle{Proc. Nat. Acad. Sci.}
\bvolume{102},
\bfpage{7789}--\blpage{7793}
(\byear{2005})
\doiurl{10.1073/pnas.0501644102}
\end{barticle}
\endbibitem

\bibitem[\protect\citeauthoryear{Gross}{2005}]{Gross:2005kv}
\begin{barticle}
\bauthor{\bsnm{Gross}, \binits{D.J.}}:
\batitle{{The discovery of asymptotic freedom and the emergence of QCD}}.
\bjtitle{Proc. Nat. Acad. Sci.}
\bvolume{102},
\bfpage{9099}--\blpage{9108}
(\byear{2005})
\doiurl{10.1073/pnas.0503831102}
\end{barticle}
\endbibitem

\bibitem[\protect\citeauthoryear{Wilczek}{2005}]{Wilczek:2005az}
\begin{barticle}
\bauthor{\bsnm{Wilczek}, \binits{F.}}:
\batitle{{Asymptotic freedom: From paradox to paradigm}}.
\bjtitle{Proc. Nat. Acad. Sci.}
\bvolume{102},
\bfpage{8403}--\blpage{8413}
(\byear{2005})
\doiurl{10.1103/RevModPhys.77.857}
{\href{https://arxiv.org/abs/hep-ph/0502113}{{arXiv:hep-ph/0502113}}}
\end{barticle}
\endbibitem

\bibitem[\protect\citeauthoryear{Cui et~al.}{2020}]{Cui:2019dwv}
\begin{barticle}
\bauthor{\bsnm{Cui}, \binits{Z.-F.}},
\bauthor{\bsnm{Zhang}, \binits{J.-L.}},
\bauthor{\bsnm{Binosi}, \binits{D.}},
\bauthor{\bsnm{Soto}, \binits{F.}},
\bauthor{\bsnm{Mezrag}, \binits{C.}},
\bauthor{\bsnm{Papavassiliou}, \binits{J.}},
\bauthor{\bsnm{Roberts}, \binits{C.D.}},
\bauthor{\bsnm{Rodr\'\i{}guez-Quintero}, \binits{J.}},
\bauthor{\bsnm{Segovia}, \binits{J.}},
\bauthor{\bsnm{Zafeiropoulos}, \binits{S.}}:
\batitle{{Effective charge from lattice QCD}}.
\bjtitle{Chin. Phys. C}
\bvolume{44}(\bissue{8}),
\bfpage{083102}
(\byear{2020})
\doiurl{10.1088/1674-1137/44/8/083102}
{\href{https://arxiv.org/abs/1912.08232}{{arXiv:1912.08232}}}
{[hep-ph]}
\end{barticle}
\endbibitem

\bibitem[\protect\citeauthoryear{Binosi et~al.}{2017}]{Binosi:2016wcx}
\begin{barticle}
\bauthor{\bsnm{Binosi}, \binits{D.}},
\bauthor{\bsnm{Chang}, \binits{L.}},
\bauthor{\bsnm{Papavassiliou}, \binits{J.}},
\bauthor{\bsnm{Qin}, \binits{S.-X.}},
\bauthor{\bsnm{Roberts}, \binits{C.D.}}:
\batitle{{Natural constraints on the gluon-quark vertex}}.
\bjtitle{Phys. Rev. D}
\bvolume{95}(\bissue{3}),
\bfpage{031501}
(\byear{2017})
\doiurl{10.1103/PhysRevD.95.031501}
{\href{https://arxiv.org/abs/1609.02568}{{arXiv:1609.02568}}}
{[nucl-th]}
\end{barticle}
\endbibitem

\bibitem[\protect\citeauthoryear{Sultan et~al.}{2021}]{Sultan:2018tet}
\begin{barticle}
\bauthor{\bsnm{Sultan}, \binits{M.A.}},
\bauthor{\bsnm{Raya}, \binits{K.}},
\bauthor{\bsnm{Akram}, \binits{F.}},
\bauthor{\bsnm{Bashir}, \binits{A.}},
\bauthor{\bsnm{Masud}, \binits{B.}}:
\batitle{{Effect of the quark-gluon vertex on dynamical chiral symmetry breaking}}.
\bjtitle{Phys. Rev. D}
\bvolume{103}(\bissue{5}),
\bfpage{054036}
(\byear{2021})
\doiurl{10.1103/PhysRevD.103.054036}
{\href{https://arxiv.org/abs/1810.01396}{{arXiv:1810.01396}}}
{[nucl-th]}
\end{barticle}
\endbibitem

\bibitem[\protect\citeauthoryear{Gell-Mann and Low}{1954}]{GellMann:1954fq}
\begin{barticle}
\bauthor{\bsnm{Gell-Mann}, \binits{M.}},
\bauthor{\bsnm{Low}, \binits{F.E.}}:
\batitle{{Quantum electrodynamics at small distances}}.
\bjtitle{Phys. Rev.}
\bvolume{95},
\bfpage{1300}--\blpage{1312}
(\byear{1954})
\doiurl{10.1103/PhysRev.95.1300}
\end{barticle}
\endbibitem

\bibitem[\protect\citeauthoryear{Binosi et~al.}{2017}]{Binosi:2016nme}
\begin{barticle}
\bauthor{\bsnm{Binosi}, \binits{D.}},
\bauthor{\bsnm{Mezrag}, \binits{C.}},
\bauthor{\bsnm{Papavassiliou}, \binits{J.}},
\bauthor{\bsnm{Roberts}, \binits{C.D.}},
\bauthor{\bsnm{Rodriguez-Quintero}, \binits{J.}}:
\batitle{{Process-independent strong running coupling}}.
\bjtitle{Phys. Rev. D}
\bvolume{96}(\bissue{5}),
\bfpage{054026}
(\byear{2017})
\doiurl{10.1103/PhysRevD.96.054026}
{\href{https://arxiv.org/abs/1612.04835}{{arXiv:1612.04835}}}
{[nucl-th]}
\end{barticle}
\endbibitem

\bibitem[\protect\citeauthoryear{Deur et~al.}{2022}]{Deur:2022msf}
\begin{barticle}
\bauthor{\bsnm{Deur}, \binits{A.}},
\bauthor{\bsnm{Burkert}, \binits{V.}},
\bauthor{\bsnm{Chen}, \binits{J.P.}},
\bauthor{\bsnm{Korsch}, \binits{W.}}:
\batitle{{Experimental determination of the QCD effective charge $\alpha_{g_1}(Q)$}}.
\bjtitle{Particles}
\bvolume{5}(\bissue{2}),
\bfpage{171}--\blpage{179}
(\byear{2022})
\doiurl{10.3390/particles5020015}
{\href{https://arxiv.org/abs/2205.01169}{{arXiv:2205.01169}}}
{[hep-ph]}
\end{barticle}
\endbibitem

\bibitem[\protect\citeauthoryear{Bjorken}{1970}]{Bjorken:1969mm}
\begin{barticle}
\bauthor{\bsnm{Bjorken}, \binits{J.D.}}:
\batitle{{Inelastic Scattering of Polarized Leptons from Polarized Nucleons}}.
\bjtitle{Phys. Rev. D}
\bvolume{1},
\bfpage{1376}--\blpage{1379}
(\byear{1970})
\doiurl{10.1103/PhysRevD.1.1376}
\end{barticle}
\endbibitem

\bibitem[\protect\citeauthoryear{Cui et~al.}{2021}]{Cui:2020dlm}
\begin{barticle}
\bauthor{\bsnm{Cui}, \binits{Z.-F.}},
\bauthor{\bsnm{Ding}, \binits{M.}},
\bauthor{\bsnm{Gao}, \binits{F.}},
\bauthor{\bsnm{Raya}, \binits{K.}},
\bauthor{\bsnm{Binosi}, \binits{D.}},
\bauthor{\bsnm{Chang}, \binits{L.}},
\bauthor{\bsnm{Roberts}, \binits{C.D.}},
\bauthor{\bsnm{Rodriguez-Quintero}, \binits{J.}},
\bauthor{\bsnm{Schmidt}, \binits{S.M.}}:
\batitle{{Higgs modulation of emergent mass as revealed in kaon and pion parton distributions}}.
\bjtitle{Eur. Phys. J. A}
\bvolume{57}(\bissue{1}),
\bfpage{5}
(\byear{2021})
\doiurl{10.1140/epja/s10050-020-00318-2}
{\href{https://arxiv.org/abs/2006.14075}{{arXiv:2006.14075}}}
{[hep-ph]}
\end{barticle}
\endbibitem

\bibitem[\protect\citeauthoryear{Cui et~al.}{2020}]{Cui:2020tdf}
\begin{barticle}
\bauthor{\bsnm{Cui}, \binits{Z.-F.}},
\bauthor{\bsnm{Ding}, \binits{M.}},
\bauthor{\bsnm{Gao}, \binits{F.}},
\bauthor{\bsnm{Raya}, \binits{K.}},
\bauthor{\bsnm{Binosi}, \binits{D.}},
\bauthor{\bsnm{Chang}, \binits{L.}},
\bauthor{\bsnm{Roberts}, \binits{C.D.}},
\bauthor{\bsnm{Rodr\'\i{}guez-Quintero}, \binits{J.}},
\bauthor{\bsnm{Schmidt}, \binits{S.M.}}:
\batitle{{Kaon and pion parton distributions}}.
\bjtitle{Eur. Phys. J. C}
\bvolume{80}(\bissue{11}),
\bfpage{1064}
(\byear{2020})
\doiurl{10.1140/epjc/s10052-020-08578-4}
\end{barticle}
\endbibitem

\bibitem[\protect\citeauthoryear{Han et~al.}{2021}]{Han:2020vjp}
\begin{barticle}
\bauthor{\bsnm{Han}, \binits{C.}},
\bauthor{\bsnm{Xie}, \binits{G.}},
\bauthor{\bsnm{Wang}, \binits{R.}},
\bauthor{\bsnm{Chen}, \binits{X.}}:
\batitle{{An Analysis of Parton Distribution Functions of the Pion and the Kaon with the Maximum Entropy Input}}.
\bjtitle{Eur. Phys. J. C}
\bvolume{81}(\bissue{4}),
\bfpage{302}
(\byear{2021})
\doiurl{10.1140/epjc/s10052-021-09087-8}
{\href{https://arxiv.org/abs/2010.14284}{{arXiv:2010.14284}}}
{[hep-ph]}
\end{barticle}
\endbibitem

\bibitem[\protect\citeauthoryear{Xie et~al.}{2022}]{Xie:2021ypc}
\begin{barticle}
\bauthor{\bsnm{Xie}, \binits{G.}},
\bauthor{\bsnm{Han}, \binits{C.}},
\bauthor{\bsnm{Wang}, \binits{R.}},
\bauthor{\bsnm{Chen}, \binits{X.}}:
\batitle{{Tackling the kaon structure function at EicC}}.
\bjtitle{Chin. Phys. C}
\bvolume{46}(\bissue{6}),
\bfpage{064107}
(\byear{2022})
\doiurl{10.1088/1674-1137/ac5b0e}
{\href{https://arxiv.org/abs/2109.08483}{{arXiv:2109.08483}}}
{[hep-ph]}
\end{barticle}
\endbibitem

\bibitem[\protect\citeauthoryear{Raya et~al.}{2022}]{Raya:2021zrz}
\begin{barticle}
\bauthor{\bsnm{Raya}, \binits{K.}},
\bauthor{\bsnm{Cui}, \binits{Z.-F.}},
\bauthor{\bsnm{Chang}, \binits{L.}},
\bauthor{\bsnm{Morgado}, \binits{J.-M.}},
\bauthor{\bsnm{Roberts}, \binits{C.D.}},
\bauthor{\bsnm{Rodriguez-Quintero}, \binits{J.}}:
\batitle{{Revealing pion and kaon structure via generalised parton distributions}}.
\bjtitle{Chin. Phys. C}
\bvolume{46}(\bissue{1}),
\bfpage{013105}
(\byear{2022})
\doiurl{10.1088/1674-1137/ac3071}
{\href{https://arxiv.org/abs/2109.11686}{{arXiv:2109.11686}}}
{[hep-ph]}
\end{barticle}
\endbibitem

\bibitem[\protect\citeauthoryear{Cui et~al.}{2022a}]{Cui:2021mom}
\begin{barticle}
\bauthor{\bsnm{Cui}, \binits{Z.-F.}},
\bauthor{\bsnm{Ding}, \binits{M.}},
\bauthor{\bsnm{Morgado}, \binits{J.M.}},
\bauthor{\bsnm{Raya}, \binits{K.}},
\bauthor{\bsnm{Binosi}, \binits{D.}},
\bauthor{\bsnm{Chang}, \binits{L.}},
\bauthor{\bsnm{Papavassiliou}, \binits{J.}},
\bauthor{\bsnm{Roberts}, \binits{C.D.}},
\bauthor{\bsnm{Rodr\'\i{}guez-Quintero}, \binits{J.}},
\bauthor{\bsnm{Schmidt}, \binits{S.M.}}:
\batitle{{Concerning pion parton distributions}}.
\bjtitle{Eur. Phys. J. A}
\bvolume{58}(\bissue{1}),
\bfpage{10}
(\byear{2022})
\doiurl{10.1140/epja/s10050-021-00658-7}
{\href{https://arxiv.org/abs/2112.09210}{{arXiv:2112.09210}}}
{[hep-ph]}
\end{barticle}
\endbibitem

\bibitem[\protect\citeauthoryear{Cui et~al.}{2022b}]{Cui:2022bxn}
\begin{barticle}
\bauthor{\bsnm{Cui}, \binits{Z.-F.}},
\bauthor{\bsnm{Ding}, \binits{M.}},
\bauthor{\bsnm{Morgado}, \binits{J.M.}},
\bauthor{\bsnm{Raya}, \binits{K.}},
\bauthor{\bsnm{Binosi}, \binits{D.}},
\bauthor{\bsnm{Chang}, \binits{L.}},
\bauthor{\bsnm{De~Soto}, \binits{F.}},
\bauthor{\bsnm{Roberts}, \binits{C.D.}},
\bauthor{\bsnm{Rodr\'\i{}guez-Quintero}, \binits{J.}},
\bauthor{\bsnm{Schmidt}, \binits{S.M.}}:
\batitle{{Emergence of pion parton distributions}}.
\bjtitle{Phys. Rev. D}
\bvolume{105}(\bissue{9}),
\bfpage{091502}
(\byear{2022})
\doiurl{10.1103/PhysRevD.105.L091502}
{\href{https://arxiv.org/abs/2201.00884}{{arXiv:2201.00884}}}
{[hep-ph]}
\end{barticle}
\endbibitem

\bibitem[\protect\citeauthoryear{Chang et~al.}{2022}]{Chang:2022jri}
\begin{barticle}
\bauthor{\bsnm{Chang}, \binits{L.}},
\bauthor{\bsnm{Gao}, \binits{F.}},
\bauthor{\bsnm{Roberts}, \binits{C.D.}}:
\batitle{{Parton distributions of light quarks and antiquarks in the proton}}.
\bjtitle{Phys. Lett. B}
\bvolume{829},
\bfpage{137078}
(\byear{2022})
\doiurl{10.1016/j.physletb.2022.137078}
{\href{https://arxiv.org/abs/2201.07870}{{arXiv:2201.07870}}}
{[hep-ph]}
\end{barticle}
\endbibitem

\bibitem[\protect\citeauthoryear{Lu et~al.}{2022}]{Lu:2022cjx}
\begin{barticle}
\bauthor{\bsnm{Lu}, \binits{Y.}},
\bauthor{\bsnm{Chang}, \binits{L.}},
\bauthor{\bsnm{Raya}, \binits{K.}},
\bauthor{\bsnm{Roberts}, \binits{C.D.}},
\bauthor{\bsnm{Rodr\'\i{}guez-Quintero}, \binits{J.}}:
\batitle{{Proton and pion distribution functions in counterpoint}}.
\bjtitle{Phys. Lett. B}
\bvolume{830},
\bfpage{137130}
(\byear{2022})
\doiurl{10.1016/j.physletb.2022.137130}
{\href{https://arxiv.org/abs/2203.00753}{{arXiv:2203.00753}}}
{[hep-ph]}
\end{barticle}
\endbibitem

\bibitem[\protect\citeauthoryear{de~Paula et~al.}{2022}]{dePaula:2022pcb}
\begin{barticle}
\bauthor{\bsnm{Paula}, \binits{W.}},
\bauthor{\bsnm{Ydrefors}, \binits{E.}},
\bauthor{\bsnm{Nogueira~Alvarenga}, \binits{J.H.}},
\bauthor{\bsnm{Frederico}, \binits{T.}},
\bauthor{\bsnm{Salm\`e}, \binits{G.}}:
\batitle{{Parton distribution function in a pion with Minkowskian dynamics}}.
\bjtitle{Phys. Rev. D}
\bvolume{105}(\bissue{7}),
\bfpage{071505}
(\byear{2022})
\doiurl{10.1103/PhysRevD.105.L071505}
{\href{https://arxiv.org/abs/2203.07106}{{arXiv:2203.07106}}}
{[hep-ph]}
\end{barticle}
\endbibitem

\bibitem[\protect\citeauthoryear{Xu et~al.}{2023}]{Xu:2023bwv}
\begin{barticle}
\bauthor{\bsnm{Xu}, \binits{Y.-Z.}},
\bauthor{\bsnm{Raya}, \binits{K.}},
\bauthor{\bsnm{Cui}, \binits{Z.-F.}},
\bauthor{\bsnm{Roberts}, \binits{C.D.}},
\bauthor{\bsnm{Rodr\'\i{}guez-Quintero}, \binits{J.}}:
\batitle{{Empirical Determination of the Pion Mass Distribution}}.
\bjtitle{Chin. Phys. Lett.}
\bvolume{40}(\bissue{4}),
\bfpage{041201}
(\byear{2023})
\doiurl{10.1088/0256-307X/40/4/041201}
{\href{https://arxiv.org/abs/2302.07361}{{arXiv:2302.07361}}}
{[hep-ph]}
\end{barticle}
\endbibitem

\bibitem[\protect\citeauthoryear{Yin et~al.}{2023}]{Yin:2023dbw}
\begin{barticle}
\bauthor{\bsnm{Yin}, \binits{P.-L.}},
\bauthor{\bsnm{Xu}, \binits{Y.-Z.}},
\bauthor{\bsnm{Cui}, \binits{Z.-F.}},
\bauthor{\bsnm{Roberts}, \binits{C.D.}},
\bauthor{\bsnm{Rodr\'\i{}guez-Quintero}, \binits{J.}}:
\batitle{{All-Orders Evolution of Parton Distributions: Principle, Practice, and Predictions}}.
\bjtitle{Chin. Phys. Lett.}
\bvolume{40}(\bissue{9}),
\bfpage{091201}
(\byear{2023})
\doiurl{10.1088/0256-307X/40/9/091201}
{\href{https://arxiv.org/abs/2306.03274}{{arXiv:2306.03274}}}
{[hep-ph]}
\end{barticle}
\endbibitem

\bibitem[\protect\citeauthoryear{Xing et~al.}{2024a}]{Xing:2023wuk}
\begin{barticle}
\bauthor{\bsnm{Xing}, \binits{H.-Y.}},
\bauthor{\bsnm{Ding}, \binits{M.}},
\bauthor{\bsnm{Cui}, \binits{Z.-F.}},
\bauthor{\bsnm{Pimikov}, \binits{A.V.}},
\bauthor{\bsnm{Roberts}, \binits{C.D.}},
\bauthor{\bsnm{Schmidt}, \binits{S.M.}}:
\batitle{{Constraining the pion distribution amplitude using Drell-Yan reactions on a proton}}.
\bjtitle{Phys. Lett. B}
\bvolume{849},
\bfpage{138462}
(\byear{2024})
\doiurl{10.1016/j.physletb.2024.138462}
{\href{https://arxiv.org/abs/2308.13695}{{arXiv:2308.13695}}}
{[hep-ph]}
\end{barticle}
\endbibitem

\bibitem[\protect\citeauthoryear{Xing et~al.}{2024b}]{Xing:2023pms}
\begin{barticle}
\bauthor{\bsnm{Xing}, \binits{H.-Y.}},
\bauthor{\bsnm{Yao}, \binits{Z.-Q.}},
\bauthor{\bsnm{Li}, \binits{B.-L.}},
\bauthor{\bsnm{Binosi}, \binits{D.}},
\bauthor{\bsnm{Cui}, \binits{Z.-F.}},
\bauthor{\bsnm{Roberts}, \binits{C.D.}}:
\batitle{{Developing predictions for pion fragmentation functions}}.
\bjtitle{Eur. Phys. J. C}
\bvolume{84}(\bissue{1}),
\bfpage{82}
(\byear{2024})
\doiurl{10.1140/epjc/s10052-024-12403-7}
{\href{https://arxiv.org/abs/2311.01613}{{arXiv:2311.01613}}}
{[hep-ph]}
\end{barticle}
\endbibitem

\bibitem[\protect\citeauthoryear{Lu et~al.}{2024}]{Lu:2023yna}
\begin{barticle}
\bauthor{\bsnm{Lu}, \binits{Y.}},
\bauthor{\bsnm{Xu}, \binits{Y.-Z.}},
\bauthor{\bsnm{Raya}, \binits{K.}},
\bauthor{\bsnm{Roberts}, \binits{C.D.}},
\bauthor{\bsnm{Rodr\'\i{}guez-Quintero}, \binits{J.}}:
\batitle{{Pion distribution functions from low-order Mellin moments}}.
\bjtitle{Phys. Lett. B}
\bvolume{850},
\bfpage{138534}
(\byear{2024})
\doiurl{10.1016/j.physletb.2024.138534}
{\href{https://arxiv.org/abs/2311.08565}{{arXiv:2311.08565}}}
{[hep-ph]}
\end{barticle}
\endbibitem

\bibitem[\protect\citeauthoryear{Yu et~al.}{2024}]{Yu:2024qsd}
\begin{botherref}
\oauthor{\bsnm{Yu}, \binits{Y.}},
\oauthor{\bsnm{Cheng}, \binits{P.}},
\oauthor{\bsnm{Xing}, \binits{H.-Y.}},
\oauthor{\bsnm{Gao}, \binits{F.}},
\oauthor{\bsnm{Roberts}, \binits{C.D.}}:
{Contact interaction study of proton parton distributions}
(2024)
{\href{https://arxiv.org/abs/2402.06095}{{arXiv:2402.06095}}}
{[hep-ph]}
\end{botherref}
\endbibitem

\bibitem[\protect\citeauthoryear{Bhagwat et~al.}{2007}]{Bhagwat:2006xi}
\begin{barticle}
\bauthor{\bsnm{Bhagwat}, \binits{M.S.}},
\bauthor{\bsnm{Krassnigg}, \binits{A.}},
\bauthor{\bsnm{Maris}, \binits{P.}},
\bauthor{\bsnm{Roberts}, \binits{C.D.}}:
\batitle{{Mind the gap}}.
\bjtitle{Eur. Phys. J. A}
\bvolume{31},
\bfpage{630}--\blpage{637}
(\byear{2007})
\doiurl{10.1140/epja/i2006-10271-9}
{\href{https://arxiv.org/abs/nucl-th/0612027}{{arXiv:nucl-th/0612027}}}
\end{barticle}
\endbibitem

\bibitem[\protect\citeauthoryear{Roberts and Williams}{1994}]{Roberts:1994dr}
\begin{barticle}
\bauthor{\bsnm{Roberts}, \binits{C.D.}},
\bauthor{\bsnm{Williams}, \binits{A.G.}}:
\batitle{{Dyson-Schwinger equations and their application to hadronic physics}}.
\bjtitle{Prog. Part. Nucl. Phys.}
\bvolume{33},
\bfpage{477}--\blpage{575}
(\byear{1994})
\doiurl{10.1016/0146-6410(94)90049-3}
{\href{https://arxiv.org/abs/hep-ph/9403224}{{arXiv:hep-ph/9403224}}}
\end{barticle}
\endbibitem

\bibitem[\protect\citeauthoryear{Munczek}{1995}]{Munczek:1994zz}
\begin{barticle}
\bauthor{\bsnm{Munczek}, \binits{H.J.}}:
\batitle{{Dynamical chiral symmetry breaking, Goldstone's theorem and the consistency of the Schwinger-Dyson and Bethe-Salpeter Equations}}.
\bjtitle{Phys. Rev. D}
\bvolume{52},
\bfpage{4736}--\blpage{4740}
(\byear{1995})
\doiurl{10.1103/PhysRevD.52.4736}
{\href{https://arxiv.org/abs/hep-th/9411239}{{arXiv:hep-th/9411239}}}
\end{barticle}
\endbibitem

\bibitem[\protect\citeauthoryear{Bender et~al.}{1996}]{Bender:1996bb}
\begin{barticle}
\bauthor{\bsnm{Bender}, \binits{A.}},
\bauthor{\bsnm{Roberts}, \binits{C.D.}},
\bauthor{\bsnm{Von~Smekal}, \binits{L.}}:
\batitle{{Goldstone theorem and diquark confinement beyond rainbow ladder approximation}}.
\bjtitle{Phys. Lett. B}
\bvolume{380},
\bfpage{7}--\blpage{12}
(\byear{1996})
\doiurl{10.1016/0370-2693(96)00372-3}
{\href{https://arxiv.org/abs/nucl-th/9602012}{{arXiv:nucl-th/9602012}}}
\end{barticle}
\endbibitem

\bibitem[\protect\citeauthoryear{Roberts}{1996}]{Roberts:1994hh}
\begin{barticle}
\bauthor{\bsnm{Roberts}, \binits{C.D.}}:
\batitle{{Electromagnetic pion form-factor and neutral pion decay width}}.
\bjtitle{Nucl. Phys. A}
\bvolume{605},
\bfpage{475}--\blpage{495}
(\byear{1996})
\doiurl{10.1016/0375-9474(96)00174-1}
{\href{https://arxiv.org/abs/hep-ph/9408233}{{arXiv:hep-ph/9408233}}}
\end{barticle}
\endbibitem

\bibitem[\protect\citeauthoryear{Nambu}{1960}]{Nambu:1960tm}
\begin{barticle}
\bauthor{\bsnm{Nambu}, \binits{Y.}}:
\batitle{{Quasiparticles and Gauge Invariance in the Theory of Superconductivity}}.
\bjtitle{Phys. Rev.}
\bvolume{117},
\bfpage{648}--\blpage{663}
(\byear{1960})
\doiurl{10.1103/PhysRev.117.648}
\end{barticle}
\endbibitem

\bibitem[\protect\citeauthoryear{Goldstone}{1961}]{Goldstone:1961eq}
\begin{barticle}
\bauthor{\bsnm{Goldstone}, \binits{J.}}:
\batitle{{Field Theories with Superconductor Solutions}}.
\bjtitle{Nuovo Cim.}
\bvolume{19},
\bfpage{154}--\blpage{164}
(\byear{1961})
\doiurl{10.1007/BF02812722}
\end{barticle}
\endbibitem

\bibitem[\protect\citeauthoryear{Chang and Roberts}{2009}]{Chang:2009zb}
\begin{barticle}
\bauthor{\bsnm{Chang}, \binits{L.}},
\bauthor{\bsnm{Roberts}, \binits{C.D.}}:
\batitle{{Sketching the Bethe-Salpeter kernel}}.
\bjtitle{Phys. Rev. Lett.}
\bvolume{103},
\bfpage{081601}
(\byear{2009})
\doiurl{10.1103/PhysRevLett.103.081601}
{\href{https://arxiv.org/abs/0903.5461}{{arXiv:0903.5461}}}
{[nucl-th]}
\end{barticle}
\endbibitem

\bibitem[\protect\citeauthoryear{Chang and Roberts}{2012}]{Chang:2011ei}
\begin{barticle}
\bauthor{\bsnm{Chang}, \binits{L.}},
\bauthor{\bsnm{Roberts}, \binits{C.D.}}:
\batitle{{Tracing masses of ground-state light-quark mesons}}.
\bjtitle{Phys. Rev. C}
\bvolume{85},
\bfpage{052201}
(\byear{2012})
\doiurl{10.1103/PhysRevC.85.052201}
{\href{https://arxiv.org/abs/1104.4821}{{arXiv:1104.4821}}}
{[nucl-th]}
\end{barticle}
\endbibitem

\bibitem[\protect\citeauthoryear{Qin et~al.}{2014}]{Qin:2014vya}
\begin{barticle}
\bauthor{\bsnm{Qin}, \binits{S.-X.}},
\bauthor{\bsnm{Roberts}, \binits{C.D.}},
\bauthor{\bsnm{Schmidt}, \binits{S.M.}}:
\batitle{{Ward\textendash{}Green\textendash{}Takahashi identities and the axial-vector vertex}}.
\bjtitle{Phys. Lett. B}
\bvolume{733},
\bfpage{202}--\blpage{208}
(\byear{2014})
\doiurl{10.1016/j.physletb.2014.04.041}
{\href{https://arxiv.org/abs/1402.1176}{{arXiv:1402.1176}}}
{[nucl-th]}
\end{barticle}
\endbibitem

\bibitem[\protect\citeauthoryear{Williams et~al.}{2016}]{Williams:2015cvx}
\begin{barticle}
\bauthor{\bsnm{Williams}, \binits{R.}},
\bauthor{\bsnm{Fischer}, \binits{C.S.}},
\bauthor{\bsnm{Heupel}, \binits{W.}}:
\batitle{{Light mesons in QCD and unquenching effects from the 3PI effective action}}.
\bjtitle{Phys. Rev. D}
\bvolume{93}(\bissue{3}),
\bfpage{034026}
(\byear{2016})
\doiurl{10.1103/PhysRevD.93.034026}
{\href{https://arxiv.org/abs/1512.00455}{{arXiv:1512.00455}}}
{[hep-ph]}
\end{barticle}
\endbibitem

\bibitem[\protect\citeauthoryear{Binosi et~al.}{2016}]{Binosi:2016rxz}
\begin{barticle}
\bauthor{\bsnm{Binosi}, \binits{D.}},
\bauthor{\bsnm{Chang}, \binits{L.}},
\bauthor{\bsnm{Papavassiliou}, \binits{J.}},
\bauthor{\bsnm{Qin}, \binits{S.-X.}},
\bauthor{\bsnm{Roberts}, \binits{C.D.}}:
\batitle{{Symmetry preserving truncations of the gap and Bethe-Salpeter equations}}.
\bjtitle{Phys. Rev. D}
\bvolume{93}(\bissue{9}),
\bfpage{096010}
(\byear{2016})
\doiurl{10.1103/PhysRevD.93.096010}
{\href{https://arxiv.org/abs/1601.05441}{{arXiv:1601.05441}}}
{[nucl-th]}
\end{barticle}
\endbibitem

\bibitem[\protect\citeauthoryear{Qin and Roberts}{2021}]{Qin:2020jig}
\begin{barticle}
\bauthor{\bsnm{Qin}, \binits{S.-X.}},
\bauthor{\bsnm{Roberts}, \binits{C.D.}}:
\batitle{{Resolving the Bethe\textendash{}Salpeter Kernel}}.
\bjtitle{Chin. Phys. Lett.}
\bvolume{38}(\bissue{7}),
\bfpage{071201}
(\byear{2021})
\doiurl{10.1088/0256-307X/38/7/071201}
{\href{https://arxiv.org/abs/2009.13637}{{arXiv:2009.13637}}}
{[hep-ph]}
\end{barticle}
\endbibitem

\bibitem[\protect\citeauthoryear{Xu et~al.}{2023}]{Xu:2022kng}
\begin{barticle}
\bauthor{\bsnm{Xu}, \binits{Z.-N.}},
\bauthor{\bsnm{Yao}, \binits{Z.-Q.}},
\bauthor{\bsnm{Qin}, \binits{S.-X.}},
\bauthor{\bsnm{Cui}, \binits{Z.-F.}},
\bauthor{\bsnm{Roberts}, \binits{C.D.}}:
\batitle{{Bethe\textendash{}Salpeter kernel and properties of strange-quark mesons}}.
\bjtitle{Eur. Phys. J. A}
\bvolume{59}(\bissue{3}),
\bfpage{39}
(\byear{2023})
\doiurl{10.1140/epja/s10050-023-00951-7}
{\href{https://arxiv.org/abs/2208.13903}{{arXiv:2208.13903}}}
{[hep-ph]}
\end{barticle}
\endbibitem

\bibitem[\protect\citeauthoryear{Qin et~al.}{2018}]{Qin:2018dqp}
\begin{barticle}
\bauthor{\bsnm{Qin}, \binits{S.-X.}},
\bauthor{\bsnm{Roberts}, \binits{C.D.}},
\bauthor{\bsnm{Schmidt}, \binits{S.M.}}:
\batitle{{Poincar\'e-covariant analysis of heavy-quark baryons}}.
\bjtitle{Phys. Rev. D}
\bvolume{97}(\bissue{11}),
\bfpage{114017}
(\byear{2018})
\doiurl{10.1103/PhysRevD.97.114017}
{\href{https://arxiv.org/abs/1801.09697}{{arXiv:1801.09697}}}
{[nucl-th]}
\end{barticle}
\endbibitem

\bibitem[\protect\citeauthoryear{Xu et~al.}{2019}]{Xu:2018cor}
\begin{barticle}
\bauthor{\bsnm{Xu}, \binits{S.-S.}},
\bauthor{\bsnm{Cui}, \binits{Z.-F.}},
\bauthor{\bsnm{Chang}, \binits{L.}},
\bauthor{\bsnm{Papavassiliou}, \binits{J.}},
\bauthor{\bsnm{Roberts}, \binits{C.D.}},
\bauthor{\bsnm{Zong}, \binits{H.-S.}}:
\batitle{{New perspective on hybrid mesons}}.
\bjtitle{Eur. Phys. J. A}
\bvolume{55}(\bissue{7}),
\bfpage{113}
(\byear{2019})
\doiurl{10.1140/epja/i2019-12805-4}
{\href{https://arxiv.org/abs/1805.06430}{{arXiv:1805.06430}}}
{[nucl-th]}
\end{barticle}
\endbibitem

\bibitem[\protect\citeauthoryear{Chang et~al.}{2020}]{Chang:2019eob}
\begin{barticle}
\bauthor{\bsnm{Chang}, \binits{L.}},
\bauthor{\bsnm{Chen}, \binits{M.}},
\bauthor{\bsnm{Liu}, \binits{Y.-X.}}:
\batitle{{Excited $B_c$ states via the Dyson-Schwinger equation approach of QCD}}.
\bjtitle{Phys. Rev. D}
\bvolume{102}(\bissue{7}),
\bfpage{074010}
(\byear{2020})
\doiurl{10.1103/PhysRevD.102.074010}
{\href{https://arxiv.org/abs/1904.00399}{{arXiv:1904.00399}}}
{[nucl-th]}
\end{barticle}
\endbibitem

\bibitem[\protect\citeauthoryear{Qin et~al.}{2019}]{Qin:2019hgk}
\begin{barticle}
\bauthor{\bsnm{Qin}, \binits{S.-X.}},
\bauthor{\bsnm{Roberts}, \binits{C.D.}},
\bauthor{\bsnm{Schmidt}, \binits{S.M.}}:
\batitle{{Spectrum of light- and heavy-baryons}}.
\bjtitle{Few Body Syst.}
\bvolume{60}(\bissue{2}),
\bfpage{26}
(\byear{2019})
\doiurl{10.1007/s00601-019-1488-x}
{\href{https://arxiv.org/abs/1902.00026}{{arXiv:1902.00026}}}
{[nucl-th]}
\end{barticle}
\endbibitem

\bibitem[\protect\citeauthoryear{Chang et~al.}{2013}]{Chang:2013pq}
\begin{barticle}
\bauthor{\bsnm{Chang}, \binits{L.}},
\bauthor{\bsnm{Cloet}, \binits{I.C.}},
\bauthor{\bsnm{Cobos-Martinez}, \binits{J.J.}},
\bauthor{\bsnm{Roberts}, \binits{C.D.}},
\bauthor{\bsnm{Schmidt}, \binits{S.M.}},
\bauthor{\bsnm{Tandy}, \binits{P.C.}}:
\batitle{{Imaging dynamical chiral symmetry breaking: pion wave function on the light front}}.
\bjtitle{Phys. Rev. Lett.}
\bvolume{110}(\bissue{13}),
\bfpage{132001}
(\byear{2013})
\doiurl{10.1103/PhysRevLett.110.132001}
{\href{https://arxiv.org/abs/1301.0324}{{arXiv:1301.0324}}}
{[nucl-th]}
\end{barticle}
\endbibitem

\bibitem[\protect\citeauthoryear{Shi et~al.}{2014}]{Shi:2014uwa}
\begin{barticle}
\bauthor{\bsnm{Shi}, \binits{C.}},
\bauthor{\bsnm{Chang}, \binits{L.}},
\bauthor{\bsnm{Roberts}, \binits{C.D.}},
\bauthor{\bsnm{Schmidt}, \binits{S.M.}},
\bauthor{\bsnm{Tandy}, \binits{P.C.}},
\bauthor{\bsnm{Zong}, \binits{H.-S.}}:
\batitle{{Flavour symmetry breaking in the kaon parton distribution amplitude}}.
\bjtitle{Phys. Lett. B}
\bvolume{738},
\bfpage{512}--\blpage{518}
(\byear{2014})
\doiurl{10.1016/j.physletb.2014.07.057}
{\href{https://arxiv.org/abs/1406.3353}{{arXiv:1406.3353}}}
{[nucl-th]}
\end{barticle}
\endbibitem

\bibitem[\protect\citeauthoryear{Binosi et~al.}{2019}]{Binosi:2018rht}
\begin{barticle}
\bauthor{\bsnm{Binosi}, \binits{D.}},
\bauthor{\bsnm{Chang}, \binits{L.}},
\bauthor{\bsnm{Ding}, \binits{M.}},
\bauthor{\bsnm{Gao}, \binits{F.}},
\bauthor{\bsnm{Papavassiliou}, \binits{J.}},
\bauthor{\bsnm{Roberts}, \binits{C.D.}}:
\batitle{{Distribution Amplitudes of Heavy-Light Mesons}}.
\bjtitle{Phys. Lett. B}
\bvolume{790},
\bfpage{257}--\blpage{262}
(\byear{2019})
\doiurl{10.1016/j.physletb.2019.01.033}
{\href{https://arxiv.org/abs/1812.05112}{{arXiv:1812.05112}}}
{[nucl-th]}
\end{barticle}
\endbibitem

\bibitem[\protect\citeauthoryear{Chang et~al.}{2014}]{Chang:2014lva}
\begin{barticle}
\bauthor{\bsnm{Chang}, \binits{L.}},
\bauthor{\bsnm{Mezrag}, \binits{C.}},
\bauthor{\bsnm{Moutarde}, \binits{H.}},
\bauthor{\bsnm{Roberts}, \binits{C.D.}},
\bauthor{\bsnm{Rodr\'\i{}guez-Quintero}, \binits{J.}},
\bauthor{\bsnm{Tandy}, \binits{P.C.}}:
\batitle{{Basic features of the pion valence-quark distribution function}}.
\bjtitle{Phys. Lett. B}
\bvolume{737},
\bfpage{23}--\blpage{29}
(\byear{2014})
\doiurl{10.1016/j.physletb.2014.08.009}
{\href{https://arxiv.org/abs/1406.5450}{{arXiv:1406.5450}}}
{[nucl-th]}
\end{barticle}
\endbibitem

\bibitem[\protect\citeauthoryear{Ding et~al.}{2020a}]{Ding:2019lwe}
\begin{barticle}
\bauthor{\bsnm{Ding}, \binits{M.}},
\bauthor{\bsnm{Raya}, \binits{K.}},
\bauthor{\bsnm{Binosi}, \binits{D.}},
\bauthor{\bsnm{Chang}, \binits{L.}},
\bauthor{\bsnm{Roberts}, \binits{C.D.}},
\bauthor{\bsnm{Schmidt}, \binits{S.M.}}:
\batitle{{Symmetry, symmetry breaking, and pion parton distributions}}.
\bjtitle{Phys. Rev. D}
\bvolume{101}(\bissue{5}),
\bfpage{054014}
(\byear{2020})
\doiurl{10.1103/PhysRevD.101.054014}
{\href{https://arxiv.org/abs/1905.05208}{{arXiv:1905.05208}}}
{[nucl-th]}
\end{barticle}
\endbibitem

\bibitem[\protect\citeauthoryear{Ding et~al.}{2020b}]{Ding:2019qlr}
\begin{barticle}
\bauthor{\bsnm{Ding}, \binits{M.}},
\bauthor{\bsnm{Raya}, \binits{K.}},
\bauthor{\bsnm{Binosi}, \binits{D.}},
\bauthor{\bsnm{Chang}, \binits{L.}},
\bauthor{\bsnm{Roberts}, \binits{C.D.}},
\bauthor{\bsnm{Schmidt}, \binits{S.M.}}:
\batitle{{Drawing insights from pion parton distributions}}.
\bjtitle{Chin. Phys. C}
\bvolume{44}(\bissue{3}),
\bfpage{031002}
(\byear{2020})
\doiurl{10.1088/1674-1137/44/3/031002}
{\href{https://arxiv.org/abs/1912.07529}{{arXiv:1912.07529}}}
{[hep-ph]}
\end{barticle}
\endbibitem

\bibitem[\protect\citeauthoryear{Ding et~al.}{2019}]{Ding:2018xwy}
\begin{barticle}
\bauthor{\bsnm{Ding}, \binits{M.}},
\bauthor{\bsnm{Raya}, \binits{K.}},
\bauthor{\bsnm{Bashir}, \binits{A.}},
\bauthor{\bsnm{Binosi}, \binits{D.}},
\bauthor{\bsnm{Chang}, \binits{L.}},
\bauthor{\bsnm{Chen}, \binits{M.}},
\bauthor{\bsnm{Roberts}, \binits{C.D.}}:
\batitle{{$\gamma^\ast \gamma \to \eta, \eta^\prime$ transition form factors}}.
\bjtitle{Phys. Rev. D}
\bvolume{99}(\bissue{1}),
\bfpage{014014}
(\byear{2019})
\doiurl{10.1103/PhysRevD.99.014014}
{\href{https://arxiv.org/abs/1810.12313}{{arXiv:1810.12313}}}
{[nucl-th]}
\end{barticle}
\endbibitem

\bibitem[\protect\citeauthoryear{Chen et~al.}{2018}]{Chen:2018rwz}
\begin{barticle}
\bauthor{\bsnm{Chen}, \binits{M.}},
\bauthor{\bsnm{Ding}, \binits{M.}},
\bauthor{\bsnm{Chang}, \binits{L.}},
\bauthor{\bsnm{Roberts}, \binits{C.D.}}:
\batitle{{Mass-dependence of pseudoscalar meson elastic form factors}}.
\bjtitle{Phys. Rev. D}
\bvolume{98}(\bissue{9}),
\bfpage{091505}
(\byear{2018})
\doiurl{10.1103/PhysRevD.98.091505}
{\href{https://arxiv.org/abs/1808.09461}{{arXiv:1808.09461}}}
{[nucl-th]}
\end{barticle}
\endbibitem

\bibitem[\protect\citeauthoryear{Chang et~al.}{2013}]{Chang:2013nia}
\begin{barticle}
\bauthor{\bsnm{Chang}, \binits{L.}},
\bauthor{\bsnm{Clo\"et}, \binits{I.C.}},
\bauthor{\bsnm{Roberts}, \binits{C.D.}},
\bauthor{\bsnm{Schmidt}, \binits{S.M.}},
\bauthor{\bsnm{Tandy}, \binits{P.C.}}:
\batitle{{Pion electromagnetic form factor at spacelike momenta}}.
\bjtitle{Phys. Rev. Lett.}
\bvolume{111}(\bissue{14}),
\bfpage{141802}
(\byear{2013})
\doiurl{10.1103/PhysRevLett.111.141802}
{\href{https://arxiv.org/abs/1307.0026}{{arXiv:1307.0026}}}
{[nucl-th]}
\end{barticle}
\endbibitem

\bibitem[\protect\citeauthoryear{Gao et~al.}{2017}]{Gao:2017mmp}
\begin{barticle}
\bauthor{\bsnm{Gao}, \binits{F.}},
\bauthor{\bsnm{Chang}, \binits{L.}},
\bauthor{\bsnm{Liu}, \binits{Y.-X.}},
\bauthor{\bsnm{Roberts}, \binits{C.D.}},
\bauthor{\bsnm{Tandy}, \binits{P.C.}}:
\batitle{{Exposing strangeness: projections for kaon electromagnetic form factors}}.
\bjtitle{Phys. Rev. D}
\bvolume{96}(\bissue{3}),
\bfpage{034024}
(\byear{2017})
\doiurl{10.1103/PhysRevD.96.034024}
{\href{https://arxiv.org/abs/1703.04875}{{arXiv:1703.04875}}}
{[nucl-th]}
\end{barticle}
\endbibitem

\bibitem[\protect\citeauthoryear{Raya et~al.}{2016}]{Raya:2015gva}
\begin{barticle}
\bauthor{\bsnm{Raya}, \binits{K.}},
\bauthor{\bsnm{Chang}, \binits{L.}},
\bauthor{\bsnm{Bashir}, \binits{A.}},
\bauthor{\bsnm{Cobos-Martinez}, \binits{J.J.}},
\bauthor{\bsnm{Guti\'errez-Guerrero}, \binits{L.X.}},
\bauthor{\bsnm{Roberts}, \binits{C.D.}},
\bauthor{\bsnm{Tandy}, \binits{P.C.}}:
\batitle{{Structure of the neutral pion and its electromagnetic transition form factor}}.
\bjtitle{Phys. Rev. D}
\bvolume{93}(\bissue{7}),
\bfpage{074017}
(\byear{2016})
\doiurl{10.1103/PhysRevD.93.074017}
{\href{https://arxiv.org/abs/1510.02799}{{arXiv:1510.02799}}}
{[nucl-th]}
\end{barticle}
\endbibitem

\bibitem[\protect\citeauthoryear{Raya et~al.}{2017}]{Raya:2016yuj}
\begin{barticle}
\bauthor{\bsnm{Raya}, \binits{K.}},
\bauthor{\bsnm{Ding}, \binits{M.}},
\bauthor{\bsnm{Bashir}, \binits{A.}},
\bauthor{\bsnm{Chang}, \binits{L.}},
\bauthor{\bsnm{Roberts}, \binits{C.D.}}:
\batitle{{Partonic structure of neutral pseudoscalars via two photon transition form factors}}.
\bjtitle{Phys. Rev. D}
\bvolume{95}(\bissue{7}),
\bfpage{074014}
(\byear{2017})
\doiurl{10.1103/PhysRevD.95.074014}
{\href{https://arxiv.org/abs/1610.06575}{{arXiv:1610.06575}}}
{[nucl-th]}
\end{barticle}
\endbibitem

\bibitem[\protect\citeauthoryear{Raya et~al.}{2020}]{Raya:2019dnh}
\begin{barticle}
\bauthor{\bsnm{Raya}, \binits{K.}},
\bauthor{\bsnm{Bashir}, \binits{A.}},
\bauthor{\bsnm{Roig}, \binits{P.}}:
\batitle{{Contribution of neutral pseudoscalar mesons to $a_\mu^{HLbL}$ within a Schwinger-Dyson equations approach to QCD}}.
\bjtitle{Phys. Rev. D}
\bvolume{101}(\bissue{7}),
\bfpage{074021}
(\byear{2020})
\doiurl{10.1103/PhysRevD.101.074021}
{\href{https://arxiv.org/abs/1910.05960}{{arXiv:1910.05960}}}
{[hep-ph]}
\end{barticle}
\endbibitem

\bibitem[\protect\citeauthoryear{Eichmann et~al.}{2020}]{Eichmann:2019bqf}
\begin{barticle}
\bauthor{\bsnm{Eichmann}, \binits{G.}},
\bauthor{\bsnm{Fischer}, \binits{C.S.}},
\bauthor{\bsnm{Williams}, \binits{R.}}:
\batitle{{Kaon-box contribution to the anomalous magnetic moment of the muon}}.
\bjtitle{Phys. Rev. D}
\bvolume{101}(\bissue{5}),
\bfpage{054015}
(\byear{2020})
\doiurl{10.1103/PhysRevD.101.054015}
{\href{https://arxiv.org/abs/1910.06795}{{arXiv:1910.06795}}}
{[hep-ph]}
\end{barticle}
\endbibitem

\bibitem[\protect\citeauthoryear{Miramontes et~al.}{2022}]{Miramontes:2021exi}
\begin{barticle}
\bauthor{\bsnm{Miramontes}, \binits{A.}},
\bauthor{\bsnm{Bashir}, \binits{A.}},
\bauthor{\bsnm{Raya}, \binits{K.}},
\bauthor{\bsnm{Roig}, \binits{P.}}:
\batitle{{Pion and Kaon box contribution to a\ensuremath{\mu}HLbL}}.
\bjtitle{Phys. Rev. D}
\bvolume{105}(\bissue{7}),
\bfpage{074013}
(\byear{2022})
\doiurl{10.1103/PhysRevD.105.074013}
{\href{https://arxiv.org/abs/2112.13916}{{arXiv:2112.13916}}}
{[hep-ph]}
\end{barticle}
\endbibitem

\bibitem[\protect\citeauthoryear{Xu et~al.}{2024}]{Xu:2023izo}
\begin{barticle}
\bauthor{\bsnm{Xu}, \binits{Y.-Z.}},
\bauthor{\bsnm{Ding}, \binits{M.}},
\bauthor{\bsnm{Raya}, \binits{K.}},
\bauthor{\bsnm{Roberts}, \binits{C.D.}},
\bauthor{\bsnm{Rodr\'\i{}guez-Quintero}, \binits{J.}},
\bauthor{\bsnm{Schmidt}, \binits{S.M.}}:
\batitle{{Pion and kaon electromagnetic and gravitational form factors}}.
\bjtitle{Eur. Phys. J. C}
\bvolume{84}(\bissue{2}),
\bfpage{191}
(\byear{2024})
\doiurl{10.1140/epjc/s10052-024-12518-x}
{\href{https://arxiv.org/abs/2311.14832}{{arXiv:2311.14832}}}
{[hep-ph]}
\end{barticle}
\endbibitem

\bibitem[\protect\citeauthoryear{Shi et~al.}{2021}]{Shi:2021nvg}
\begin{barticle}
\bauthor{\bsnm{Shi}, \binits{C.}},
\bauthor{\bsnm{Li}, \binits{M.}},
\bauthor{\bsnm{Chen}, \binits{X.}},
\bauthor{\bsnm{Jia}, \binits{W.}}:
\batitle{{Ground state pseudoscalar mesons on the light front: From the light to heavy sector}}.
\bjtitle{Phys. Rev. D}
\bvolume{104}(\bissue{9}),
\bfpage{094016}
(\byear{2021})
\doiurl{10.1103/PhysRevD.104.094016}
{\href{https://arxiv.org/abs/2108.10625}{{arXiv:2108.10625}}}
{[hep-ph]}
\end{barticle}
\endbibitem

\bibitem[\protect\citeauthoryear{Kou et~al.}{2023}]{Kou:2023ady}
\begin{barticle}
\bauthor{\bsnm{Kou}, \binits{W.}},
\bauthor{\bsnm{Shi}, \binits{C.}},
\bauthor{\bsnm{Chen}, \binits{X.}},
\bauthor{\bsnm{Jia}, \binits{W.}}:
\batitle{{Transverse momentum dependent parton distributions of pion at leading twist}}.
\bjtitle{Phys. Rev. D}
\bvolume{108}(\bissue{3}),
\bfpage{036021}
(\byear{2023})
\doiurl{10.1103/PhysRevD.108.036021}
{\href{https://arxiv.org/abs/2304.09814}{{arXiv:2304.09814}}}
{[hep-ph]}
\end{barticle}
\endbibitem

\bibitem[\protect\citeauthoryear{Mezrag}{2023}]{Mezrag:2023nkp}
\begin{barticle}
\bauthor{\bsnm{Mezrag}, \binits{C.}}:
\batitle{{Generalised Parton Distributions in Continuum Schwinger Methods: Progresses, Opportunities and Challenges}}.
\bjtitle{Particles}
\bvolume{6}(\bissue{1}),
\bfpage{262}--\blpage{296}
(\byear{2023})
\doiurl{10.3390/particles6010015}
\end{barticle}
\endbibitem

\bibitem[\protect\citeauthoryear{Brodsky and Lepage}{1989}]{Brodsky:1989pv}
\begin{barticle}
\bauthor{\bsnm{Brodsky}, \binits{S.J.}},
\bauthor{\bsnm{Lepage}, \binits{G.P.}}:
\batitle{{Exclusive Processes in Quantum Chromodynamics}}.
\bjtitle{Adv. Ser. Direct. High Energy Phys.}
\bvolume{5},
\bfpage{93}--\blpage{240}
(\byear{1989})
\doiurl{10.1142/9789814503266_0002}
\end{barticle}
\endbibitem

\bibitem[\protect\citeauthoryear{Lepage and Brodsky}{1979}]{Lepage:1979zb}
\begin{barticle}
\bauthor{\bsnm{Lepage}, \binits{G.P.}},
\bauthor{\bsnm{Brodsky}, \binits{S.J.}}:
\batitle{{Exclusive Processes in Quantum Chromodynamics: Evolution Equations for Hadronic Wave Functions and the Form-Factors of Mesons}}.
\bjtitle{Phys. Lett. B}
\bvolume{87},
\bfpage{359}--\blpage{365}
(\byear{1979})
\doiurl{10.1016/0370-2693(79)90554-9}
\end{barticle}
\endbibitem

\bibitem[\protect\citeauthoryear{Efremov and Radyushkin}{1980}]{Efremov:1979qk}
\begin{barticle}
\bauthor{\bsnm{Efremov}, \binits{A.V.}},
\bauthor{\bsnm{Radyushkin}, \binits{A.V.}}:
\batitle{{Factorization and Asymptotical Behavior of Pion Form-Factor in QCD}}.
\bjtitle{Phys. Lett. B}
\bvolume{94},
\bfpage{245}--\blpage{250}
(\byear{1980})
\doiurl{10.1016/0370-2693(80)90869-2}
\end{barticle}
\endbibitem

\bibitem[\protect\citeauthoryear{Lepage and Brodsky}{1980}]{Lepage:1980fj}
\begin{barticle}
\bauthor{\bsnm{Lepage}, \binits{G.P.}},
\bauthor{\bsnm{Brodsky}, \binits{S.J.}}:
\batitle{{Exclusive Processes in Perturbative Quantum Chromodynamics}}.
\bjtitle{Phys. Rev. D}
\bvolume{22},
\bfpage{2157}
(\byear{1980})
\doiurl{10.1103/PhysRevD.22.2157}
\end{barticle}
\endbibitem

\bibitem[\protect\citeauthoryear{Ding et~al.}{2016}]{Ding:2015rkn}
\begin{barticle}
\bauthor{\bsnm{Ding}, \binits{M.}},
\bauthor{\bsnm{Gao}, \binits{F.}},
\bauthor{\bsnm{Chang}, \binits{L.}},
\bauthor{\bsnm{Liu}, \binits{Y.-X.}},
\bauthor{\bsnm{Roberts}, \binits{C.D.}}:
\batitle{{Leading-twist parton distribution amplitudes of S-wave heavy-quarkonia}}.
\bjtitle{Phys. Lett. B}
\bvolume{753},
\bfpage{330}--\blpage{335}
(\byear{2016})
\doiurl{10.1016/j.physletb.2015.11.075}
{\href{https://arxiv.org/abs/1511.04943}{{arXiv:1511.04943}}}
{[nucl-th]}
\end{barticle}
\endbibitem

\bibitem[\protect\citeauthoryear{Zhang et~al.}{2020}]{Zhang:2020gaj}
\begin{barticle}
\bauthor{\bsnm{Zhang}, \binits{R.}},
\bauthor{\bsnm{Honkala}, \binits{C.}},
\bauthor{\bsnm{Lin}, \binits{H.-W.}},
\bauthor{\bsnm{Chen}, \binits{J.-W.}}:
\batitle{{Pion and kaon distribution amplitudes in the continuum limit}}.
\bjtitle{Phys. Rev. D}
\bvolume{102}(\bissue{9}),
\bfpage{094519}
(\byear{2020})
\doiurl{10.1103/PhysRevD.102.094519}
{\href{https://arxiv.org/abs/2005.13955}{{arXiv:2005.13955}}}
{[hep-lat]}
\end{barticle}
\endbibitem

\bibitem[\protect\citeauthoryear{Neubert}{1994}]{Neubert:1993mb}
\begin{barticle}
\bauthor{\bsnm{Neubert}, \binits{M.}}:
\batitle{{Heavy quark symmetry}}.
\bjtitle{Phys. Rept.}
\bvolume{245},
\bfpage{259}--\blpage{396}
(\byear{1994})
\doiurl{10.1016/0370-1573(94)90091-4}
{\href{https://arxiv.org/abs/hep-ph/9306320}{{arXiv:hep-ph/9306320}}}
\end{barticle}
\endbibitem

\bibitem[\protect\citeauthoryear{Roberts et~al.}{2010}]{Roberts:2010rn}
\begin{barticle}
\bauthor{\bsnm{Roberts}, \binits{H.L.L.}},
\bauthor{\bsnm{Roberts}, \binits{C.D.}},
\bauthor{\bsnm{Bashir}, \binits{A.}},
\bauthor{\bsnm{Gutierrez-Guerrero}, \binits{L.X.}},
\bauthor{\bsnm{Tandy}, \binits{P.C.}}:
\batitle{{Abelian anomaly and neutral pion production}}.
\bjtitle{Phys. Rev. C}
\bvolume{82},
\bfpage{065202}
(\byear{2010})
\doiurl{10.1103/PhysRevC.82.065202}
{\href{https://arxiv.org/abs/1009.0067}{{arXiv:1009.0067}}}
{[nucl-th]}
\end{barticle}
\endbibitem

\bibitem[\protect\citeauthoryear{Holt and Roberts}{2010}]{Holt:2010vj}
\begin{barticle}
\bauthor{\bsnm{Holt}, \binits{R.J.}},
\bauthor{\bsnm{Roberts}, \binits{C.D.}}:
\batitle{{Distribution Functions of the Nucleon and Pion in the Valence Region}}.
\bjtitle{Rev. Mod. Phys.}
\bvolume{82},
\bfpage{2991}--\blpage{3044}
(\byear{2010})
\doiurl{10.1103/RevModPhys.82.2991}
{\href{https://arxiv.org/abs/1002.4666}{{arXiv:1002.4666}}}
{[nucl-th]}
\end{barticle}
\endbibitem

\bibitem[\protect\citeauthoryear{Chang and Roberts}{2021}]{Chang:2021utv}
\begin{barticle}
\bauthor{\bsnm{Chang}, \binits{L.}},
\bauthor{\bsnm{Roberts}, \binits{C.D.}}:
\batitle{{Regarding the Distribution of Glue in the Pion}}.
\bjtitle{Chin. Phys. Lett.}
\bvolume{38}(\bissue{8}),
\bfpage{081101}
(\byear{2021})
\doiurl{10.1088/0256-307X/38/8/081101}
{\href{https://arxiv.org/abs/2106.08451}{{arXiv:2106.08451}}}
{[hep-ph]}
\end{barticle}
\endbibitem

\bibitem[\protect\citeauthoryear{Brodsky et~al.}{1995}]{Brodsky:1994kg}
\begin{barticle}
\bauthor{\bsnm{Brodsky}, \binits{S.J.}},
\bauthor{\bsnm{Burkardt}, \binits{M.}},
\bauthor{\bsnm{Schmidt}, \binits{I.}}:
\batitle{{Perturbative QCD constraints on the shape of polarized quark and gluon distributions}}.
\bjtitle{Nucl. Phys. B}
\bvolume{441},
\bfpage{197}--\blpage{214}
(\byear{1995})
\doiurl{10.1016/0550-3213(95)00009-H}
{\href{https://arxiv.org/abs/hep-ph/9401328}{{arXiv:hep-ph/9401328}}}
\end{barticle}
\endbibitem

\bibitem[\protect\citeauthoryear{Yuan}{2004}]{Yuan:2003fs}
\begin{barticle}
\bauthor{\bsnm{Yuan}, \binits{F.}}:
\batitle{{Generalized parton distributions at x ---\ensuremath{>} 1}}.
\bjtitle{Phys. Rev. D}
\bvolume{69},
\bfpage{051501}
(\byear{2004})
\doiurl{10.1103/PhysRevD.69.051501}
{\href{https://arxiv.org/abs/hep-ph/0311288}{{arXiv:hep-ph/0311288}}}
\end{barticle}
\endbibitem

\bibitem[\protect\citeauthoryear{Barabanov et~al.}{2021}]{Barabanov:2020jvn}
\begin{barticle}
\bauthor{\bsnm{Barabanov}, \binits{M.Y.}}, \betal:
\batitle{{Diquark correlations in hadron physics: Origin, impact and evidence}}.
\bjtitle{Prog. Part. Nucl. Phys.}
\bvolume{116},
\bfpage{103835}
(\byear{2021})
\doiurl{10.1016/j.ppnp.2020.103835}
{\href{https://arxiv.org/abs/2008.07630}{{arXiv:2008.07630}}}
{[hep-ph]}
\end{barticle}
\endbibitem

\bibitem[\protect\citeauthoryear{Dokshitzer}{1977}]{Dokshitzer:1977sg}
\begin{barticle}
\bauthor{\bsnm{Dokshitzer}, \binits{Y.L.}}:
\batitle{{Calculation of the Structure Functions for Deep Inelastic Scattering and e+ e- Annihilation by Perturbation Theory in Quantum Chromodynamics.}}
\bjtitle{Sov. Phys. JETP}
\bvolume{46},
\bfpage{641}--\blpage{653}
(\byear{1977})
\end{barticle}
\endbibitem

\bibitem[\protect\citeauthoryear{Gribov and Lipatov}{1972}]{Gribov:1972ri}
\begin{barticle}
\bauthor{\bsnm{Gribov}, \binits{V.N.}},
\bauthor{\bsnm{Lipatov}, \binits{L.N.}}:
\batitle{{Deep inelastic e p scattering in perturbation theory}}.
\bjtitle{Sov. J. Nucl. Phys.}
\bvolume{15},
\bfpage{438}--\blpage{450}
(\byear{1972})
\end{barticle}
\endbibitem

\bibitem[\protect\citeauthoryear{Lipatov}{1974}]{Lipatov:1974qm}
\begin{barticle}
\bauthor{\bsnm{Lipatov}, \binits{L.N.}}:
\batitle{{The parton model and perturbation theory}}.
\bjtitle{Yad. Fiz.}
\bvolume{20},
\bfpage{181}--\blpage{198}
(\byear{1974})
\end{barticle}
\endbibitem

\bibitem[\protect\citeauthoryear{Altarelli and Parisi}{1977}]{Altarelli:1977zs}
\begin{barticle}
\bauthor{\bsnm{Altarelli}, \binits{G.}},
\bauthor{\bsnm{Parisi}, \binits{G.}}:
\batitle{{Asymptotic Freedom in Parton Language}}.
\bjtitle{Nucl. Phys. B}
\bvolume{126},
\bfpage{298}--\blpage{318}
(\byear{1977})
\doiurl{10.1016/0550-3213(77)90384-4}
\end{barticle}
\endbibitem

\bibitem[\protect\citeauthoryear{Cheng et~al.}{2023}]{Cheng:2023kmt}
\begin{barticle}
\bauthor{\bsnm{Cheng}, \binits{P.}},
\bauthor{\bsnm{Yu}, \binits{Y.}},
\bauthor{\bsnm{Xing}, \binits{H.-Y.}},
\bauthor{\bsnm{Chen}, \binits{C.}},
\bauthor{\bsnm{Cui}, \binits{Z.-F.}},
\bauthor{\bsnm{Roberts}, \binits{C.D.}}:
\batitle{{Perspective on polarised parton distribution functions and proton spin}}.
\bjtitle{Phys. Lett. B}
\bvolume{844},
\bfpage{138074}
(\byear{2023})
\doiurl{10.1016/j.physletb.2023.138074}
{\href{https://arxiv.org/abs/2304.12469}{{arXiv:2304.12469}}}
{[hep-ph]}
\end{barticle}
\endbibitem

\bibitem[\protect\citeauthoryear{Brodsky et~al.}{1980}]{Brodsky:1980pb}
\begin{barticle}
\bauthor{\bsnm{Brodsky}, \binits{S.J.}},
\bauthor{\bsnm{Hoyer}, \binits{P.}},
\bauthor{\bsnm{Peterson}, \binits{C.}},
\bauthor{\bsnm{Sakai}, \binits{N.}}:
\batitle{{The Intrinsic Charm of the Proton}}.
\bjtitle{Phys. Lett. B}
\bvolume{93},
\bfpage{451}--\blpage{455}
(\byear{1980})
\doiurl{10.1016/0370-2693(80)90364-0}
\end{barticle}
\endbibitem

\bibitem[\protect\citeauthoryear{Fan and Lin}{2021}]{Fan:2021bcr}
\begin{barticle}
\bauthor{\bsnm{Fan}, \binits{Z.}},
\bauthor{\bsnm{Lin}, \binits{H.-W.}}:
\batitle{{Gluon parton distribution of the pion from lattice QCD}}.
\bjtitle{Phys. Lett. B}
\bvolume{823},
\bfpage{136778}
(\byear{2021})
\doiurl{10.1016/j.physletb.2021.136778}
{\href{https://arxiv.org/abs/2104.06372}{{arXiv:2104.06372}}}
{[hep-lat]}
\end{barticle}
\endbibitem

\bibitem[\protect\citeauthoryear{Sufian et~al.}{2020}]{Sufian:2020vzb}
\begin{barticle}
\bauthor{\bsnm{Sufian}, \binits{R.S.}},
\bauthor{\bsnm{Egerer}, \binits{C.}},
\bauthor{\bsnm{Karpie}, \binits{J.}},
\bauthor{\bsnm{Edwards}, \binits{R.G.}},
\bauthor{\bsnm{Jo\'o}, \binits{B.}},
\bauthor{\bsnm{Ma}, \binits{Y.-Q.}},
\bauthor{\bsnm{Orginos}, \binits{K.}},
\bauthor{\bsnm{Qiu}, \binits{J.-W.}},
\bauthor{\bsnm{Richards}, \binits{D.G.}}:
\batitle{{Pion Valence Quark Distribution from Current-Current Correlation in Lattice QCD}}.
\bjtitle{Phys. Rev. D}
\bvolume{102}(\bissue{5}),
\bfpage{054508}
(\byear{2020})
\doiurl{10.1103/PhysRevD.102.054508}
{\href{https://arxiv.org/abs/2001.04960}{{arXiv:2001.04960}}}
{[hep-lat]}
\end{barticle}
\endbibitem

\bibitem[\protect\citeauthoryear{Conway et~al.}{1989}]{Conway:1989fs}
\begin{barticle}
\bauthor{\bsnm{Conway}, \binits{J.S.}}, \betal:
\batitle{{Experimental Study of Muon Pairs Produced by 252-GeV Pions on Tungsten}}.
\bjtitle{Phys. Rev. D}
\bvolume{39},
\bfpage{92}--\blpage{122}
(\byear{1989})
\doiurl{10.1103/PhysRevD.39.92}
\end{barticle}
\endbibitem

\bibitem[\protect\citeauthoryear{Aicher et~al.}{2010}]{Aicher:2010cb}
\begin{barticle}
\bauthor{\bsnm{Aicher}, \binits{M.}},
\bauthor{\bsnm{Schafer}, \binits{A.}},
\bauthor{\bsnm{Vogelsang}, \binits{W.}}:
\batitle{{Soft-gluon resummation and the valence parton distribution function of the pion}}.
\bjtitle{Phys. Rev. Lett.}
\bvolume{105},
\bfpage{252003}
(\byear{2010})
\doiurl{10.1103/PhysRevLett.105.252003}
{\href{https://arxiv.org/abs/1009.2481}{{arXiv:1009.2481}}}
{[hep-ph]}
\end{barticle}
\endbibitem

\bibitem[\protect\citeauthoryear{Badier et~al.}{1980}]{Badier:1980fhh}
\begin{barticle}
\bauthor{\bsnm{Badier}, \binits{J.}}, \betal:
\batitle{{Measurement of the $K^- / \pi^-$ Structure Function Ratio Using the {Drell-Yan} Process}}.
\bjtitle{Phys. Lett. B}
\bvolume{93},
\bfpage{354}--\blpage{356}
(\byear{1980})
\doiurl{10.1016/0370-2693(80)90530-4}
\end{barticle}
\endbibitem

\bibitem[\protect\citeauthoryear{Lin et~al.}{2021}]{Lin:2020ssv}
\begin{barticle}
\bauthor{\bsnm{Lin}, \binits{H.-W.}},
\bauthor{\bsnm{Chen}, \binits{J.-W.}},
\bauthor{\bsnm{Fan}, \binits{Z.}},
\bauthor{\bsnm{Zhang}, \binits{J.-H.}},
\bauthor{\bsnm{Zhang}, \binits{R.}}:
\batitle{{Valence-Quark Distribution of the Kaon and Pion from Lattice QCD}}.
\bjtitle{Phys. Rev. D}
\bvolume{103}(\bissue{1}),
\bfpage{014516}
(\byear{2021})
\doiurl{10.1103/PhysRevD.103.014516}
{\href{https://arxiv.org/abs/2003.14128}{{arXiv:2003.14128}}}
{[hep-lat]}
\end{barticle}
\endbibitem

\bibitem[\protect\citeauthoryear{Adler}{1969}]{Adler:1969gk}
\begin{barticle}
\bauthor{\bsnm{Adler}, \binits{S.L.}}:
\batitle{{Axial vector vertex in spinor electrodynamics}}.
\bjtitle{Phys. Rev.}
\bvolume{177},
\bfpage{2426}--\blpage{2438}
(\byear{1969})
\doiurl{10.1103/PhysRev.177.2426}
\end{barticle}
\endbibitem

\bibitem[\protect\citeauthoryear{Bell and Jackiw}{1969}]{Bell:1969ts}
\begin{barticle}
\bauthor{\bsnm{Bell}, \binits{J.S.}},
\bauthor{\bsnm{Jackiw}, \binits{R.}}:
\batitle{{A PCAC puzzle: \mbox{$\pi^0 \to \gamma \gamma$} in the {$\sigma$} model}}.
\bjtitle{Nuovo Cim. A}
\bvolume{60},
\bfpage{47}--\blpage{61}
(\byear{1969})
\doiurl{10.1007/BF02823296}
\end{barticle}
\endbibitem

\bibitem[\protect\citeauthoryear{Adler}{2004}]{Adler:2004ih}
\begin{botherref}
\oauthor{\bsnm{Adler}, \binits{S.L.}}:
{Anomalies}
(2004)
{\href{https://arxiv.org/abs/hep-th/0411038}{{arXiv:hep-th/0411038}}}
\end{botherref}
\endbibitem

\bibitem[\protect\citeauthoryear{Holl et~al.}{2005}]{Holl:2005vu}
\begin{barticle}
\bauthor{\bsnm{Holl}, \binits{A.}},
\bauthor{\bsnm{Krassnigg}, \binits{A.}},
\bauthor{\bsnm{Maris}, \binits{P.}},
\bauthor{\bsnm{Roberts}, \binits{C.D.}},
\bauthor{\bsnm{Wright}, \binits{S.V.}}:
\batitle{{Electromagnetic properties of ground and excited state pseudoscalar mesons}}.
\bjtitle{Phys. Rev. C}
\bvolume{71},
\bfpage{065204}
(\byear{2005})
\doiurl{10.1103/PhysRevC.71.065204}
{\href{https://arxiv.org/abs/nucl-th/0503043}{{arXiv:nucl-th/0503043}}}
\end{barticle}
\endbibitem

\bibitem[\protect\citeauthoryear{Stefanis}{2020}]{Stefanis:2020rnd}
\begin{barticle}
\bauthor{\bsnm{Stefanis}, \binits{N.G.}}:
\batitle{{Pion-photon transition form factor in light cone sum rules and tests of asymptotics}}.
\bjtitle{Phys. Rev. D}
\bvolume{102}(\bissue{3}),
\bfpage{034022}
(\byear{2020})
\doiurl{10.1103/PhysRevD.102.034022}
{\href{https://arxiv.org/abs/2006.10576}{{arXiv:2006.10576}}}
{[hep-ph]}
\end{barticle}
\endbibitem

\bibitem[\protect\citeauthoryear{Christos}{1984}]{Christos:1984tu}
\begin{barticle}
\bauthor{\bsnm{Christos}, \binits{G.A.}}:
\batitle{{Chiral Symmetry and the U(1) Problem}}.
\bjtitle{Phys. Rept.}
\bvolume{116},
\bfpage{251}--\blpage{336}
(\byear{1984})
\doiurl{10.1016/0370-1573(84)90025-5}
\end{barticle}
\endbibitem

\bibitem[\protect\citeauthoryear{Bhagwat et~al.}{2007}]{Bhagwat:2007ha}
\begin{barticle}
\bauthor{\bsnm{Bhagwat}, \binits{M.S.}},
\bauthor{\bsnm{Chang}, \binits{L.}},
\bauthor{\bsnm{Liu}, \binits{Y.-X.}},
\bauthor{\bsnm{Roberts}, \binits{C.D.}},
\bauthor{\bsnm{Tandy}, \binits{P.C.}}:
\batitle{{Flavour symmetry breaking and meson masses}}.
\bjtitle{Phys. Rev. C}
\bvolume{76},
\bfpage{045203}
(\byear{2007})
\doiurl{10.1103/PhysRevC.76.045203}
{\href{https://arxiv.org/abs/0708.1118}{{arXiv:0708.1118}}}
{[nucl-th]}
\end{barticle}
\endbibitem

\bibitem[\protect\citeauthoryear{Maris and Roberts}{2003}]{Maris:2003vk}
\begin{barticle}
\bauthor{\bsnm{Maris}, \binits{P.}},
\bauthor{\bsnm{Roberts}, \binits{C.D.}}:
\batitle{{Dyson-Schwinger equations: A Tool for hadron physics}}.
\bjtitle{Int. J. Mod. Phys. E}
\bvolume{12},
\bfpage{297}--\blpage{365}
(\byear{2003})
\doiurl{10.1142/S0218301303001326}
{\href{https://arxiv.org/abs/nucl-th/0301049}{{arXiv:nucl-th/0301049}}}
\end{barticle}
\endbibitem

\bibitem[\protect\citeauthoryear{Amendolia et~al.}{1986}]{NA7:1986vav}
\begin{barticle}
\bauthor{\bsnm{Amendolia}, \binits{S.R.}}, \betal:
\batitle{{A Measurement of the Space - Like Pion Electromagnetic Form-Factor}}.
\bjtitle{Nucl. Phys. B}
\bvolume{277},
\bfpage{168}
(\byear{1986})
\doiurl{10.1016/0550-3213(86)90437-2}
\end{barticle}
\endbibitem

\bibitem[\protect\citeauthoryear{Horn et~al.}{2008}]{Horn:2007ug}
\begin{barticle}
\bauthor{\bsnm{Horn}, \binits{T.}}, \betal:
\batitle{{Scaling study of the pion electroproduction cross sections and the pion form factor}}.
\bjtitle{Phys. Rev. C}
\bvolume{78},
\bfpage{058201}
(\byear{2008})
\doiurl{10.1103/PhysRevC.78.058201}
{\href{https://arxiv.org/abs/0707.1794}{{arXiv:0707.1794}}}
{[nucl-ex]}
\end{barticle}
\endbibitem

\bibitem[\protect\citeauthoryear{Huber et~al.}{2008}]{JeffersonLab:2008jve}
\begin{barticle}
\bauthor{\bsnm{Huber}, \binits{G.M.}}, \betal:
\batitle{{Charged pion form-factor between Q**2 = 0.60-GeV**2 and 2.45-GeV**2. II. Determination of, and results for, the pion form-factor}}.
\bjtitle{Phys. Rev. C}
\bvolume{78},
\bfpage{045203}
(\byear{2008})
\doiurl{10.1103/PhysRevC.78.045203}
{\href{https://arxiv.org/abs/0809.3052}{{arXiv:0809.3052}}}
{[nucl-ex]}
\end{barticle}
\endbibitem

\bibitem[\protect\citeauthoryear{Dally et~al.}{1980}]{Dally:1980dj}
\begin{barticle}
\bauthor{\bsnm{Dally}, \binits{E.B.}}, \betal:
\batitle{{Direct Measurement of the Negative-Kaon Form Factor}}.
\bjtitle{Phys. Rev. Lett.}
\bvolume{45},
\bfpage{232}--\blpage{235}
(\byear{1980})
\doiurl{10.1103/PhysRevLett.45.232}
\end{barticle}
\endbibitem

\bibitem[\protect\citeauthoryear{Amendolia et~al.}{1986}]{Amendolia:1986ui}
\begin{barticle}
\bauthor{\bsnm{Amendolia}, \binits{S.R.}}, \betal:
\batitle{{A Measurement of the Kaon Charge Radius}}.
\bjtitle{Phys. Lett. B}
\bvolume{178},
\bfpage{435}--\blpage{440}
(\byear{1986})
\doiurl{10.1016/0370-2693(86)91407-3}
\end{barticle}
\endbibitem

\bibitem[\protect\citeauthoryear{Carmignotto et~al.}{2018}]{Carmignotto:2018uqj}
\begin{barticle}
\bauthor{\bsnm{Carmignotto}, \binits{M.}}, \betal:
\batitle{{Separated Kaon Electroproduction Cross Section and the Kaon Form Factor from 6 GeV JLab Data}}.
\bjtitle{Phys. Rev. C}
\bvolume{97}(\bissue{2}),
\bfpage{025204}
(\byear{2018})
\doiurl{10.1103/PhysRevC.97.025204}
{\href{https://arxiv.org/abs/1801.01536}{{arXiv:1801.01536}}}
{[nucl-ex]}
\end{barticle}
\endbibitem

\bibitem[\protect\citeauthoryear{Davies et~al.}{2018}]{Davies:2018zav}
\begin{barticle}
\bauthor{\bsnm{Davies}, \binits{C.T.H.}},
\bauthor{\bsnm{Koponen}, \binits{J.}},
\bauthor{\bsnm{Lepage}, \binits{P.G.}},
\bauthor{\bsnm{Lytle}, \binits{A.T.}},
\bauthor{\bsnm{Zimermmane-Santos}, \binits{A.C.}}:
\batitle{{Meson Electromagnetic Form Factors from Lattice QCD}}.
\bjtitle{PoS}
\bvolume{LATTICE2018},
\bfpage{298}
(\byear{2018})
\doiurl{10.22323/1.334.0298}
{\href{https://arxiv.org/abs/1902.03808}{{arXiv:1902.03808}}}
{[hep-lat]}
\end{barticle}
\endbibitem

\bibitem[\protect\citeauthoryear{Cloet and Roberts}{2008}]{Cloet:2008fw}
\begin{barticle}
\bauthor{\bsnm{Cloet}, \binits{I.C.}},
\bauthor{\bsnm{Roberts}, \binits{C.D.}}:
\batitle{{Form Factors and Dyson-Schwinger Equations}}.
\bjtitle{PoS}
\bvolume{LC2008},
\bfpage{047}
(\byear{2008})
\doiurl{10.22323/1.061.0047}
{\href{https://arxiv.org/abs/0811.2018}{{arXiv:0811.2018}}}
{[nucl-th]}
\end{barticle}
\endbibitem

\bibitem[\protect\citeauthoryear{Cui et~al.}{2021}]{Cui:2021aee}
\begin{barticle}
\bauthor{\bsnm{Cui}, \binits{Z.-F.}},
\bauthor{\bsnm{Binosi}, \binits{D.}},
\bauthor{\bsnm{Roberts}, \binits{C.D.}},
\bauthor{\bsnm{Schmidt}, \binits{S.M.}}:
\batitle{{Pion charge radius from pion+electron elastic scattering data}}.
\bjtitle{Phys. Lett. B}
\bvolume{822},
\bfpage{136631}
(\byear{2021})
\doiurl{10.1016/j.physletb.2021.136631}
{\href{https://arxiv.org/abs/2108.04948}{{arXiv:2108.04948}}}
{[hep-ph]}
\end{barticle}
\endbibitem

\bibitem[\protect\citeauthoryear{Behrend et~al.}{1991}]{CELLO:1990klc}
\begin{barticle}
\bauthor{\bsnm{Behrend}, \binits{H.J.}}, \betal:
\batitle{{A Measurement of the pi0, eta and eta-prime electromagnetic form-factors}}.
\bjtitle{Z. Phys. C}
\bvolume{49},
\bfpage{401}--\blpage{410}
(\byear{1991})
\doiurl{10.1007/BF01549692}
\end{barticle}
\endbibitem

\bibitem[\protect\citeauthoryear{Gronberg et~al.}{1998}]{PhysRevD.57.33}
\begin{barticle}
\bauthor{\bsnm{Gronberg}, \binits{J.}}, \betal:
\batitle{Measurements of the meson-photon transition form factors of light pseudoscalar mesons at large momentum transfer}.
\bjtitle{Phys. Rev. D}
\bvolume{57},
\bfpage{33}--\blpage{54}
(\byear{1998})
\doiurl{10.1103/PhysRevD.57.33}
\end{barticle}
\endbibitem

\bibitem[\protect\citeauthoryear{Aubert et~al.}{2009}]{PhysRevD.80.052002}
\begin{barticle}
\bauthor{\bsnm{Aubert}, \binits{B.}}, \betal:
\batitle{Measurement of the $\ensuremath{\gamma}{\ensuremath{\gamma}}^{*}\ensuremath{\rightarrow}{\ensuremath{\pi}}^{0}$ transition form factor}.
\bjtitle{Phys. Rev. D}
\bvolume{80},
\bfpage{052002}
(\byear{2009})
\doiurl{10.1103/PhysRevD.80.052002}
\end{barticle}
\endbibitem

\bibitem[\protect\citeauthoryear{Uehara et~al.}{2012}]{PhysRevD.86.092007}
\begin{barticle}
\bauthor{\bsnm{Uehara}, \binits{S.}}, \betal:
\batitle{Measurement of $\ensuremath{\gamma}{\ensuremath{\gamma}}^{*}\ensuremath{\rightarrow}{\ensuremath{\pi}}^{0}$ transition form factor at belle}.
\bjtitle{Phys. Rev. D}
\bvolume{86},
\bfpage{092007}
(\byear{2012})
\doiurl{10.1103/PhysRevD.86.092007}
\end{barticle}
\endbibitem

\bibitem[\protect\citeauthoryear{Stefanis et~al.}{2013}]{Stefanis:2012yw}
\begin{barticle}
\bauthor{\bsnm{Stefanis}, \binits{N.G.}},
\bauthor{\bsnm{Bakulev}, \binits{A.P.}},
\bauthor{\bsnm{Mikhailov}, \binits{S.V.}},
\bauthor{\bsnm{Pimikov}, \binits{A.V.}}:
\batitle{{Can We Understand an Auxetic Pion-Photon Transition Form Factor within QCD?}}
\bjtitle{Phys. Rev. D}
\bvolume{87}(\bissue{9}),
\bfpage{094025}
(\byear{2013})
\doiurl{10.1103/PhysRevD.87.094025}
{\href{https://arxiv.org/abs/1202.1781}{{arXiv:1202.1781}}}
{[hep-ph]}
\end{barticle}
\endbibitem

\bibitem[\protect\citeauthoryear{Nedelko and Voronin}{2017}]{Nedelko:2016vpj}
\begin{barticle}
\bauthor{\bsnm{Nedelko}, \binits{S.N.}},
\bauthor{\bsnm{Voronin}, \binits{V.E.}}:
\batitle{{Influence of confining gluon configurations on the $P\to\gamma^*\gamma $ transition form factors}}.
\bjtitle{Phys. Rev. D}
\bvolume{95}(\bissue{7}),
\bfpage{074038}
(\byear{2017})
\doiurl{10.1103/PhysRevD.95.074038}
{\href{https://arxiv.org/abs/1612.02621}{{arXiv:1612.02621}}}
{[hep-ph]}
\end{barticle}
\endbibitem

\bibitem[\protect\citeauthoryear{Eichmann et~al.}{2017}]{Eichmann:2017wil}
\begin{barticle}
\bauthor{\bsnm{Eichmann}, \binits{G.}},
\bauthor{\bsnm{Fischer}, \binits{C.S.}},
\bauthor{\bsnm{Weil}, \binits{E.}},
\bauthor{\bsnm{Williams}, \binits{R.}}:
\batitle{{On the large-$Q^2$ behavior of the pion transition form factor}}.
\bjtitle{Phys. Lett. B}
\bvolume{774},
\bfpage{425}--\blpage{429}
(\byear{2017})
\doiurl{10.1016/j.physletb.2017.10.002}
{\href{https://arxiv.org/abs/1704.05774}{{arXiv:1704.05774}}}
{[hep-ph]}
\end{barticle}
\endbibitem

\bibitem[\protect\citeauthoryear{Choi and Ji}{2020}]{Choi:2020xsr}
\begin{barticle}
\bauthor{\bsnm{Choi}, \binits{H.-M.}},
\bauthor{\bsnm{Ji}, \binits{C.-R.}}:
\batitle{{Chiral anomaly and the pion properties in the light-front quark model}}.
\bjtitle{Phys. Rev. D}
\bvolume{102}(\bissue{3}),
\bfpage{036005}
(\byear{2020})
\doiurl{10.1103/PhysRevD.102.036005}
{\href{https://arxiv.org/abs/2006.08034}{{arXiv:2006.08034}}}
{[hep-ph]}
\end{barticle}
\endbibitem

\bibitem[\protect\citeauthoryear{Zhou et~al.}{2023}]{Zhou:2023ivj}
\begin{barticle}
\bauthor{\bsnm{Zhou}, \binits{H.}},
\bauthor{\bsnm{Yan}, \binits{J.}},
\bauthor{\bsnm{Yu}, \binits{Q.}},
\bauthor{\bsnm{Wu}, \binits{X.-G.}}:
\batitle{{Updated determination of the pion-photon transition form factor}}.
\bjtitle{Phys. Rev. D}
\bvolume{108}(\bissue{7}),
\bfpage{074020}
(\byear{2023})
\doiurl{10.1103/PhysRevD.108.074020}
{\href{https://arxiv.org/abs/2306.10510}{{arXiv:2306.10510}}}
{[hep-ph]}
\end{barticle}
\endbibitem

\bibitem[\protect\citeauthoryear{Bakulev et~al.}{2012}]{Bakulev:2012nh}
\begin{barticle}
\bauthor{\bsnm{Bakulev}, \binits{A.P.}},
\bauthor{\bsnm{Mikhailov}, \binits{S.V.}},
\bauthor{\bsnm{Pimikov}, \binits{A.V.}},
\bauthor{\bsnm{Stefanis}, \binits{N.G.}}:
\batitle{{Comparing antithetic trends of data for the pion-photon transition form factor}}.
\bjtitle{Phys. Rev. D}
\bvolume{86},
\bfpage{031501}
(\byear{2012})
\doiurl{10.1103/PhysRevD.86.031501}
{\href{https://arxiv.org/abs/1205.3770}{{arXiv:1205.3770}}}
{[hep-ph]}
\end{barticle}
\endbibitem

\bibitem[\protect\citeauthoryear{Feng et~al.}{2015}]{Feng:2015uha}
\begin{barticle}
\bauthor{\bsnm{Feng}, \binits{F.}},
\bauthor{\bsnm{Jia}, \binits{Y.}},
\bauthor{\bsnm{Sang}, \binits{W.-L.}}:
\batitle{{Can Nonrelativistic QCD Explain the $\gamma\gamma^* \to \eta_c$ Transition Form Factor Data?}}
\bjtitle{Phys. Rev. Lett.}
\bvolume{115}(\bissue{22}),
\bfpage{222001}
(\byear{2015})
\doiurl{10.1103/PhysRevLett.115.222001}
{\href{https://arxiv.org/abs/1505.02665}{{arXiv:1505.02665}}}
{[hep-ph]}
\end{barticle}
\endbibitem

\bibitem[\protect\citeauthoryear{Lees et~al.}{2010}]{BaBar:2010siw}
\begin{barticle}
\bauthor{\bsnm{Lees}, \binits{J.P.}}, \betal:
\batitle{{Measurement of the $\gamma \gamma* --> \eta_c$ transition form factor}}.
\bjtitle{Phys. Rev. D}
\bvolume{81},
\bfpage{052010}
(\byear{2010})
\doiurl{10.1103/PhysRevD.81.052010}
{\href{https://arxiv.org/abs/1002.3000}{{arXiv:1002.3000}}}
{[hep-ex]}
\end{barticle}
\endbibitem

\bibitem[\protect\citeauthoryear{Polyakov and Schweitzer}{2018}]{Polyakov:2018zvc}
\begin{barticle}
\bauthor{\bsnm{Polyakov}, \binits{M.V.}},
\bauthor{\bsnm{Schweitzer}, \binits{P.}}:
\batitle{{Forces inside hadrons: pressure, surface tension, mechanical radius, and all that}}.
\bjtitle{Int. J. Mod. Phys. A}
\bvolume{33}(\bissue{26}),
\bfpage{1830025}
(\byear{2018})
\doiurl{10.1142/S0217751X18300259}
{\href{https://arxiv.org/abs/1805.06596}{{arXiv:1805.06596}}}
{[hep-ph]}
\end{barticle}
\endbibitem

\bibitem[\protect\citeauthoryear{Mezrag et~al.}{2015}]{Mezrag:2014jka}
\begin{barticle}
\bauthor{\bsnm{Mezrag}, \binits{C.}},
\bauthor{\bsnm{Chang}, \binits{L.}},
\bauthor{\bsnm{Moutarde}, \binits{H.}},
\bauthor{\bsnm{Roberts}, \binits{C.D.}},
\bauthor{\bsnm{Rodr\'\i{}guez-Quintero}, \binits{J.}},
\bauthor{\bsnm{Sabati\'e}, \binits{F.}},
\bauthor{\bsnm{Schmidt}, \binits{S.M.}}:
\batitle{{Sketching the pion's valence-quark generalised parton distribution}}.
\bjtitle{Phys. Lett. B}
\bvolume{741},
\bfpage{190}--\blpage{196}
(\byear{2015})
\doiurl{10.1016/j.physletb.2014.12.027}
{\href{https://arxiv.org/abs/1411.6634}{{arXiv:1411.6634}}}
{[nucl-th]}
\end{barticle}
\endbibitem

\bibitem[\protect\citeauthoryear{Hackett et~al.}{2023}]{Hackett:2023nkr}
\begin{barticle}
\bauthor{\bsnm{Hackett}, \binits{D.C.}},
\bauthor{\bsnm{Oare}, \binits{P.R.}},
\bauthor{\bsnm{Pefkou}, \binits{D.A.}},
\bauthor{\bsnm{Shanahan}, \binits{P.E.}}:
\batitle{{Gravitational form factors of the pion from lattice QCD}}.
\bjtitle{Phys. Rev. D}
\bvolume{108}(\bissue{11}),
\bfpage{114504}
(\byear{2023})
\doiurl{10.1103/PhysRevD.108.114504}
{\href{https://arxiv.org/abs/2307.11707}{{arXiv:2307.11707}}}
{[hep-lat]}
\end{barticle}
\endbibitem

\bibitem[\protect\citeauthoryear{Kumano et~al.}{2018}]{Kumano:2017lhr}
\begin{barticle}
\bauthor{\bsnm{Kumano}, \binits{S.}},
\bauthor{\bsnm{Song}, \binits{Q.-T.}},
\bauthor{\bsnm{Teryaev}, \binits{O.V.}}:
\batitle{{Hadron tomography by generalized distribution amplitudes in pion-pair production process $\gamma^* \gamma \rightarrow \pi^0 \pi^0 $ and gravitational form factors for pion}}.
\bjtitle{Phys. Rev. D}
\bvolume{97}(\bissue{1}),
\bfpage{014020}
(\byear{2018})
\doiurl{10.1103/PhysRevD.97.014020}
{\href{https://arxiv.org/abs/1711.08088}{{arXiv:1711.08088}}}
{[hep-ph]}
\end{barticle}
\endbibitem

\bibitem[\protect\citeauthoryear{Belitsky and Radyushkin}{2005}]{Belitsky:2005qn}
\begin{barticle}
\bauthor{\bsnm{Belitsky}, \binits{A.V.}},
\bauthor{\bsnm{Radyushkin}, \binits{A.V.}}:
\batitle{{Unraveling hadron structure with generalized parton distributions}}.
\bjtitle{Phys. Rept.}
\bvolume{418},
\bfpage{1}--\blpage{387}
(\byear{2005})
\doiurl{10.1016/j.physrep.2005.06.002}
{\href{https://arxiv.org/abs/hep-ph/0504030}{{arXiv:hep-ph/0504030}}}
\end{barticle}
\endbibitem

\bibitem[\protect\citeauthoryear{Mezrag}{2022}]{Mezrag:2022pqk}
\begin{barticle}
\bauthor{\bsnm{Mezrag}, \binits{C.}}:
\batitle{{An Introductory Lecture on Generalised Parton Distributions}}.
\bjtitle{Few Body Syst.}
\bvolume{63}(\bissue{3}),
\bfpage{62}
(\byear{2022})
\doiurl{10.1007/s00601-022-01765-x}
{\href{https://arxiv.org/abs/2207.13584}{{arXiv:2207.13584}}}
{[hep-ph]}
\end{barticle}
\endbibitem

\bibitem[\protect\citeauthoryear{Chouika et~al.}{2018}]{Chouika:2017rzs}
\begin{barticle}
\bauthor{\bsnm{Chouika}, \binits{N.}},
\bauthor{\bsnm{Mezrag}, \binits{C.}},
\bauthor{\bsnm{Moutarde}, \binits{H.}},
\bauthor{\bsnm{Rodr\'\i{}guez-Quintero}, \binits{J.}}:
\batitle{{A Nakanishi-based model illustrating the covariant extension of the pion GPD overlap representation and its ambiguities}}.
\bjtitle{Phys. Lett. B}
\bvolume{780},
\bfpage{287}--\blpage{293}
(\byear{2018})
\doiurl{10.1016/j.physletb.2018.02.070}
{\href{https://arxiv.org/abs/1711.11548}{{arXiv:1711.11548}}}
{[hep-ph]}
\end{barticle}
\endbibitem

\bibitem[\protect\citeauthoryear{\"Ozel and Freire}{2016}]{Ozel:2016oaf}
\begin{barticle}
\bauthor{\bsnm{\"Ozel}, \binits{F.}},
\bauthor{\bsnm{Freire}, \binits{P.}}:
\batitle{{Masses, Radii, and the Equation of State of Neutron Stars}}.
\bjtitle{Ann. Rev. Astron. Astrophys.}
\bvolume{54},
\bfpage{401}--\blpage{440}
(\byear{2016})
\doiurl{10.1146/annurev-astro-081915-023322}
{\href{https://arxiv.org/abs/1603.02698}{{arXiv:1603.02698}}}
{[astro-ph.HE]}
\end{barticle}
\endbibitem

\bibitem[\protect\citeauthoryear{Xu et~al.}{2018}]{Xu:2018eii}
\begin{barticle}
\bauthor{\bsnm{Xu}, \binits{S.-S.}},
\bauthor{\bsnm{Chang}, \binits{L.}},
\bauthor{\bsnm{Roberts}, \binits{C.D.}},
\bauthor{\bsnm{Zong}, \binits{H.-S.}}:
\batitle{{Pion and kaon valence-quark parton quasidistributions}}.
\bjtitle{Phys. Rev. D}
\bvolume{97}(\bissue{9}),
\bfpage{094014}
(\byear{2018})
\doiurl{10.1103/PhysRevD.97.094014}
{\href{https://arxiv.org/abs/1802.09552}{{arXiv:1802.09552}}}
{[nucl-th]}
\end{barticle}
\endbibitem

\bibitem[\protect\citeauthoryear{Raya and Rodr\'\i{}guez-Quintero}{2022}]{Raya:2022eqa}
\begin{barticle}
\bauthor{\bsnm{Raya}, \binits{K.}},
\bauthor{\bsnm{Rodr\'\i{}guez-Quintero}, \binits{J.}}:
\batitle{{Highlights of pion and kaon structure from continuum analyses}}.
\bjtitle{Rev. Mex. Fis. Suppl.}
\bvolume{3}(\bissue{3}),
\bfpage{0308008}
(\byear{2022})
\doiurl{10.31349/SuplRevMexFis.3.0308008}
{\href{https://arxiv.org/abs/2204.01642}{{arXiv:2204.01642}}}
{[hep-ph]}
\end{barticle}
\endbibitem

\bibitem[\protect\citeauthoryear{Albino et~al.}{2022}]{Albino:2022gzs}
\begin{barticle}
\bauthor{\bsnm{Albino}, \binits{L.}},
\bauthor{\bsnm{Higuera-Angulo}, \binits{I.M.}},
\bauthor{\bsnm{Raya}, \binits{K.}},
\bauthor{\bsnm{Bashir}, \binits{A.}}:
\batitle{{Pseudoscalar mesons: Light front wave functions, GPDs, and PDFs}}.
\bjtitle{Phys. Rev. D}
\bvolume{106}(\bissue{3}),
\bfpage{034003}
(\byear{2022})
\doiurl{10.1103/PhysRevD.106.034003}
{\href{https://arxiv.org/abs/2207.06550}{{arXiv:2207.06550}}}
{[hep-ph]}
\end{barticle}
\endbibitem

\bibitem[\protect\citeauthoryear{Almeida-Zamora et~al.}{2024}]{Almeida-Zamora:2023bqb}
\begin{barticle}
\bauthor{\bsnm{Almeida-Zamora}, \binits{B.}},
\bauthor{\bsnm{Cobos-Mart\'\i{}nez}, \binits{J.J.}},
\bauthor{\bsnm{Bashir}, \binits{A.}},
\bauthor{\bsnm{Raya}, \binits{K.}},
\bauthor{\bsnm{Rodr\'\i{}guez-Quintero}, \binits{J.}},
\bauthor{\bsnm{Segovia}, \binits{J.}}:
\batitle{{Algebraic model to study the internal structure of pseudoscalar mesons with heavy-light quark content}}.
\bjtitle{Phys. Rev. D}
\bvolume{109}(\bissue{1}),
\bfpage{014016}
(\byear{2024})
\doiurl{10.1103/PhysRevD.109.014016}
{\href{https://arxiv.org/abs/2309.17282}{{arXiv:2309.17282}}}
{[hep-ph]}
\end{barticle}
\endbibitem

\bibitem[\protect\citeauthoryear{Diehl et~al.}{2001}]{Diehl:2000xz}
\begin{barticle}
\bauthor{\bsnm{Diehl}, \binits{M.}},
\bauthor{\bsnm{Feldmann}, \binits{T.}},
\bauthor{\bsnm{Jakob}, \binits{R.}},
\bauthor{\bsnm{Kroll}, \binits{P.}}:
\batitle{{The overlap representation of skewed quark and gluon distributions}}.
\bjtitle{Nucl. Phys. B}
\bvolume{596},
\bfpage{33}--\blpage{65}
(\byear{2001})
\doiurl{10.1016/S0550-3213(00)00684-2}
{\href{https://arxiv.org/abs/hep-ph/0009255}{{arXiv:hep-ph/0009255}}}.
\bcomment{[Erratum: Nucl.Phys.B 605, 647--647 (2001)]}
\end{barticle}
\endbibitem

\bibitem[\protect\citeauthoryear{Chouika et~al.}{2017}]{Chouika:2017dhe}
\begin{barticle}
\bauthor{\bsnm{Chouika}, \binits{N.}},
\bauthor{\bsnm{Mezrag}, \binits{C.}},
\bauthor{\bsnm{Moutarde}, \binits{H.}},
\bauthor{\bsnm{Rodr\'\i{}guez-Quintero}, \binits{J.}}:
\batitle{{Covariant Extension of the GPD overlap representation at low Fock states}}.
\bjtitle{Eur. Phys. J. C}
\bvolume{77}(\bissue{12}),
\bfpage{906}
(\byear{2017})
\doiurl{10.1140/epjc/s10052-017-5465-6}
{\href{https://arxiv.org/abs/1711.05108}{{arXiv:1711.05108}}}
{[hep-ph]}
\end{barticle}
\endbibitem

\bibitem[\protect\citeauthoryear{Chavez et~al.}{2022}]{Chavez:2021llq}
\begin{barticle}
\bauthor{\bsnm{Chavez}, \binits{J.M.M.}},
\bauthor{\bsnm{Bertone}, \binits{V.}},
\bauthor{\bsnm{De~Soto~Borrero}, \binits{F.}},
\bauthor{\bsnm{Defurne}, \binits{M.}},
\bauthor{\bsnm{Mezrag}, \binits{C.}},
\bauthor{\bsnm{Moutarde}, \binits{H.}},
\bauthor{\bsnm{Rodr\'\i{}guez-Quintero}, \binits{J.}},
\bauthor{\bsnm{Segovia}, \binits{J.}}:
\batitle{{Pion generalized parton distributions: A path toward phenomenology}}.
\bjtitle{Phys. Rev. D}
\bvolume{105}(\bissue{9}),
\bfpage{094012}
(\byear{2022})
\doiurl{10.1103/PhysRevD.105.094012}
{\href{https://arxiv.org/abs/2110.06052}{{arXiv:2110.06052}}}
{[hep-ph]}
\end{barticle}
\endbibitem

\bibitem[\protect\citeauthoryear{Dall'Olio et~al.}{2024}]{DallOlio:2024vjv}
\begin{botherref}
\oauthor{\bsnm{Dall'Olio}, \binits{P.}},
\oauthor{\bsnm{De~Soto}, \binits{F.}},
\oauthor{\bsnm{Mezrag}, \binits{C.}},
\oauthor{\bsnm{Morgado~Ch\'avez}, \binits{J.M.}},
\oauthor{\bsnm{Moutarde}, \binits{H.}},
\oauthor{\bsnm{Rodr\'\i{}guez-Quintero}, \binits{J.}},
\oauthor{\bsnm{Sznajder}, \binits{P.}},
\oauthor{\bsnm{Segovia}, \binits{J.}}:
{Unraveling Generalized Parton Distributions Through Lorentz Symmetry and Partial DGLAP Knowledge}
(2024)
{\href{https://arxiv.org/abs/2401.12013}{{arXiv:2401.12013}}}
{[hep-ph]}
\end{botherref}
\endbibitem

\bibitem[\protect\citeauthoryear{Jo\'o et~al.}{2019}]{Joo:2019bzr}
\begin{barticle}
\bauthor{\bsnm{Jo\'o}, \binits{B.}},
\bauthor{\bsnm{Karpie}, \binits{J.}},
\bauthor{\bsnm{Orginos}, \binits{K.}},
\bauthor{\bsnm{Radyushkin}, \binits{A.V.}},
\bauthor{\bsnm{Richards}, \binits{D.G.}},
\bauthor{\bsnm{Sufian}, \binits{R.S.}},
\bauthor{\bsnm{Zafeiropoulos}, \binits{S.}}:
\batitle{{Pion valence structure from Ioffe-time parton pseudodistribution functions}}.
\bjtitle{Phys. Rev. D}
\bvolume{100}(\bissue{11}),
\bfpage{114512}
(\byear{2019})
\doiurl{10.1103/PhysRevD.100.114512}
{\href{https://arxiv.org/abs/1909.08517}{{arXiv:1909.08517}}}
{[hep-lat]}
\end{barticle}
\endbibitem

\bibitem[\protect\citeauthoryear{Sufian et~al.}{2019}]{Sufian:2019bol}
\begin{barticle}
\bauthor{\bsnm{Sufian}, \binits{R.S.}},
\bauthor{\bsnm{Karpie}, \binits{J.}},
\bauthor{\bsnm{Egerer}, \binits{C.}},
\bauthor{\bsnm{Orginos}, \binits{K.}},
\bauthor{\bsnm{Qiu}, \binits{J.-W.}},
\bauthor{\bsnm{Richards}, \binits{D.G.}}:
\batitle{{Pion Valence Quark Distribution from Matrix Element Calculated in Lattice QCD}}.
\bjtitle{Phys. Rev. D}
\bvolume{99}(\bissue{7}),
\bfpage{074507}
(\byear{2019})
\doiurl{10.1103/PhysRevD.99.074507}
{\href{https://arxiv.org/abs/1901.03921}{{arXiv:1901.03921}}}
{[hep-lat]}
\end{barticle}
\endbibitem

\bibitem[\protect\citeauthoryear{Alexandrou et~al.}{2021}]{Alexandrou:2021mmi}
\begin{barticle}
\bauthor{\bsnm{Alexandrou}, \binits{C.}},
\bauthor{\bsnm{Bacchio}, \binits{S.}},
\bauthor{\bsnm{Cloet}, \binits{I.}},
\bauthor{\bsnm{Constantinou}, \binits{M.}},
\bauthor{\bsnm{Hadjiyiannakou}, \binits{K.}},
\bauthor{\bsnm{Koutsou}, \binits{G.}},
\bauthor{\bsnm{Lauer}, \binits{C.}}:
\batitle{{Pion and kaon $\langle x^3 \rangle$ from lattice QCD and PDF reconstruction from Mellin moments}}.
\bjtitle{Phys. Rev. D}
\bvolume{104}(\bissue{5}),
\bfpage{054504}
(\byear{2021})
\doiurl{10.1103/PhysRevD.104.054504}
{\href{https://arxiv.org/abs/2104.02247}{{arXiv:2104.02247}}}
{[hep-lat]}
\end{barticle}
\endbibitem

\end{thebibliography}

\end{document}